\documentclass[3p,preprint,review,12pt]{elsarticle} 
\makeatletter\if@twocolumn\PassOptionsToPackage{switch}{lineno}\else\fi\makeatother
\usepackage{tabulary,xcolor}
\usepackage{amsfonts,amsmath,amssymb}
\usepackage[T1]{fontenc}

\usepackage{ifluatex}
\ifluatex
\usepackage{fontspec}
\defaultfontfeatures{Ligatures=TeX}
\usepackage[]{unicode-math}
\unimathsetup{math-style=TeX}
\else 
\usepackage[utf8]{inputenc}
\fi 
\ifluatex\else\usepackage{stmaryrd}\fi

\usepackage{url,multirow,morefloats,floatflt,cancel,tfrupee}
\makeatletter

\AtBeginDocument{\@ifpackageloaded{textcomp}{}{\usepackage{textcomp}}}
\makeatother
\usepackage{colortbl}
\usepackage{pifont}
\usepackage[nointegrals]{wasysym}
\makeatletter

\def\mcWidth#1{\csname TY@F#1\endcsname+\tabcolsep}

\def\cAlignHack{\rightskip\@flushglue\leftskip\@flushglue\parindent\z@\parfillskip\z@skip}
\def\rAlignHack{\rightskip\z@skip\leftskip\@flushglue \parindent\z@\parfillskip\z@skip}

\@ifundefined{etal}{}{}

\usepackage{ifxetex}
\ifxetex\else\if@twocolumn\@ifpackageloaded{stfloats}{}{\usepackage{dblfloatfix}}\fi\fi

\AtBeginDocument{
\expandafter\ifx\csname eqalign\endcsname\relax
\def\eqalign#1{\null\vcenter{\def\\{\cr}\openup\jot\m@th
  \ialign{\strut$\displaystyle{##}$\hfil&$\displaystyle{{}##}$\hfil
      \crcr#1\crcr}}\,}
\fi
}

\AtBeginDocument{%
  \@ifpackageloaded{endfloat}%
  {\renewcommand\efloat@iwrite[1]{\immediate\expandafter\protected@write\csname efloat@post#1\endcsname{}}}{\newif\ifefloat@tables}%
}%

\def\BreakURLText#1{\@tfor\brk@tempa:=#1\do{\brk@tempa\hskip0pt}}
\let\lt=<
\let\gt=>
\def\processVert{\ifmmode|\else\textbar\fi}

\@ifundefined{subparagraph}{
\def\subparagraph{\@startsection{paragraph}{5}{2\parindent}{0ex plus 0.1ex minus 0.1ex}%
{0ex}{\normalfont\small\itshape}}%
}{}

\newcommand\role[1]{\unskip}
\newcommand\aucollab[1]{\unskip}
  
\@ifundefined{tsGraphicsScaleX}{\gdef\tsGraphicsScaleX{1}}{}
\@ifundefined{tsGraphicsScaleY}{\gdef\tsGraphicsScaleY{.9}}{}
\def\checkGraphicsWidth{\ifdim\Gin@nat@width>\linewidth
	\tsGraphicsScaleX\linewidth\else\Gin@nat@width\fi}

\def\checkGraphicsHeight{\ifdim\Gin@nat@height>.9\textheight
	\tsGraphicsScaleY\textheight\else\Gin@nat@height\fi}

\def\fixFloatSize#1{}%\@ifundefined{processdelayedfloats}{\setbox0=\hbox{\includegraphics{#1}}\ifnum\wd0<\columnwidth\relax\renewenvironment{figure*}{\begin{figure}}{\end{figure}}\fi}{}}
\let\ts@includegraphics\includegraphics

\def\inlinegraphic[#1]#2{{\edef\@tempa{#1}\edef\baseline@shift{\ifx\@tempa\@empty0\else#1\fi}\edef\tempZ{\the\numexpr(\numexpr(\baseline@shift*\f@size/100))}\protect\raisebox{\tempZ pt}{\ts@includegraphics{#2}}}}

\AtBeginDocument{\def\includegraphics{\@ifnextchar[{\ts@includegraphics}{\ts@includegraphics[width=\checkGraphicsWidth,height=\checkGraphicsHeight,keepaspectratio]}}}

\DeclareMathAlphabet{\mathpzc}{OT1}{pzc}{m}{it}

\def\URL#1#2{\@ifundefined{href}{#2}{\href{#1}{#2}}}

\def\UrlOrds{\do\*\do\-\do\~\do\'\do\"\do\-}%
\g@addto@macro{\UrlBreaks}{\UrlOrds}

\edef\fntEncoding{\f@encoding}

\makeatother

\newif\ifmultipleabstract\multipleabstractfalse%
\newcommand{\estA}{\hat{\matr{A}}_t}
\let\vec\mathbf

\newcommand{\matr}[1]{\mathbf{#1}}

\newcommand{\vx}{\Vec{x}}
\newcommand{\vf}{\Vec{f}}

\DeclareMathOperator*{\argmin}{arg\;min}

\newcommand{\MA}{\mathbf{A}}

\newcommand{\MB}{\mathbf{B}}
\newcommand{\MBt}{\mathbf{B}'}
\newcommand{\MY}{\mathbf{Y}}
\newcommand{\MX}{\mathbf{X}}
\newcommand{\MH}{\mathbf{H}}

\newcommand{\MTh}{\mathbf{\Theta}}

\newcommand{\MQ}{\vec{Q}}

\newcommand{\sumoT}{\sum_{t=1}^{T}}
\newcommand{\oTsumoT}{\frac{1}{T}\sum_{t=1}^{T}}
\newcommand{\fot}{\frac{1}{T}}

\newcommand{\Tyt}{\mathbf{\tilde{y}}_t}
\newcommand{\Txt}{\mathbf{\tilde{x}}_t}

\newcommand{\ft}{\mathbf{f}_t}
\newcommand{\xt}{\vec{x}_t}
\newcommand{\yt}{\vec{y}_t}

\newcommand{\ut}{\vec{u}_t}
\newcommand{\et}{\boldsymbol{\epsilon}_t}

\newcommand{\fxt}{\vec{f}_{x,t}}

\newcommand{\vv}{\Vec{v}}
\newcommand{\vu}{\Vec{u}}
\newcommand{\Sd}[1]{\mathcal{S}^{#1-1}}

\newcommand{\psit}{{\psi_2}}

\newcommand{\hA}{\hat{\mathbf{A}}}

\newcommand{\hB}{\hat{\mathbf{B}}}
\newcommand{\hBt}{\hat{\mathbf{B}}'}

\newcommand{\hft}{\hat{\mathbf{f}}_t}

\newcommand{\hCOV}{\widehat{\vec{\Sigma}}}

\newcommand{\taot}{\tau_t}
\newcommand{\taut}{\tau_t}

\newcommand{\gammastart}{\boldsymbol{\gamma}_t^*}

\usepackage{natbib}
\biboptions{authoryear}
\usepackage{mathtools}
\usepackage{lineno}
\usepackage[symbol]{footmisc}

\usepackage{makecell}
\usepackage{booktabs}

\usepackage{soul}
\usepackage{lipsum}
\usepackage{xcolor}
\usepackage{amsmath}
\usepackage{graphicx,psfrag,epsf}
\usepackage{enumerate}
\usepackage{url} % not crucial - just used below for the URL 
\usepackage{amsmath,amsfonts,bm}
\usepackage[ruled,vlined]{algorithm2e}

\usepackage{array}
\usepackage{multirow}
\usepackage{amsthm}

\newtheorem{theorem}{Theorem}%[section]

\newtheorem{proposition}{Proposition}
\newtheorem{asmp}{Assumption}
\newtheorem{condition}{Condition}

\newtheorem{remark}{Remark}

 \newlength{\leftstackrelawd}
\newlength{\leftstackrelbwd}
\def\leftstackrel#1#2{\settowidth{\leftstackrelawd}%
{${{}^{#1}}$}\settowidth{\leftstackrelbwd}{$#2$}%
\addtolength{\leftstackrelawd}{-\leftstackrelbwd}%
\leavevmode\ifthenelse{\lengthtest{\leftstackrelawd>0pt}}%
{\kern-.5\leftstackrelawd}{}\mathrel{\mathop{#2}\limits^{#1}}}

\newcommand{\ratetheoremo}{    
\fot + 
{\frac{1}{T^{2\eta}N^2} }
+ 
\frac{u_2^2}{T} + 
{\frac{u_2^4}{K^2T^{2\eta}}}
    + \frac{u_3^2}{K} 
    }
    
\newcommand{\ratetheoremt}{   
\frac{K}{N} 
+ \frac{1}{T^2}
+  {\frac{1}{T^{2\eta}N^2}}
+ \frac{u_2^2}{T} 
+ {\frac{u_2^4}{K^2T^{2\eta}}}
+ \frac{u_3^2}{K}
}
\DeclareMathOperator{\cov}{cov}

\usepackage{dirtytalk}
\usepackage{xcite}

\usepackage{xr-hyper}
\usepackage{hyperref}       % hyperlinks

\makeatletter
\newcommand*{\addFileDependency}[1]{% argument=file name and extension
  \typeout{(#1)}
  \@addtofilelist{#1}
  \IfFileExists{#1}{}{\typeout{No file #1.}}
}
\makeatother
\newcommand*{\myexternaldocument}[1]{%
    \externaldocument{#1}%
    \addFileDependency{#1.tex}%
    \addFileDependency{#1.aux}%
}

\myexternaldocument{0_CAFE_supplement}

\usepackage[utf8]{inputenc}
\begin{document}

\begin{frontmatter}
	
\title{\mbox{}
Mining the Factor Zoo: \\ 
Estimation of Latent Factor Models with Sufficient Proxies
% Candidates
}
% Efficient
\renewcommand{\thefootnote}{\fnsymbol{footnote}}

\author[]{Runzhe Wan\footnote[1]{Department of Statistics, North Carolina State University, Raleigh, NC, USA. Email: rwan@ncsu.edu.}
, 
Yingying Li\footnote[2]{Department of ISOM and Department of Finance, Hong Kong University of Science and Technology, Clear Water Bay, Kowloon, Hong Kong.  Email: yyli@ust.hk. Research is supported in part by grants NSFC71922902(NSFC19BM03), RGC GRF16503419, T31-604/18-N and T31-603/21-N.}
, 
Wenbin Lu \footnote[3]{Department of Statistics, North Carolina State University, Raleigh, NC, USA. Email: wlu4@ncsu.edu.
}
and 
Rui Song 
\footnote[4]{Department of Statistics, North Carolina State University, Raleigh, NC, USA. Email: rsong@ncsu.edu.
}
\footnote[5]{Rui Song is the Corresponding Author. Runzhe Wan was a Graduate Student when the paper was finished. Author names for Professors Yingying Li, Wenbin Lu, and Rui Song are listed in alphabetical order. 
We thank the guest editor and two anonymous referees for their thoughtful and constructive suggestions.
}
}

\begin{abstract}
Latent factor model estimation typically relies on either using domain knowledge to manually pick several observed covariates as factor proxies, or purely conducting multivariate analysis such as principal component analysis. 
However, the former approach may suffer from the bias while the latter can not incorporate additional information.  
We propose to bridge these two approaches while allowing the number of factor proxies to diverge, 
and hence make the latent factor model estimation robust, flexible, and statistically more accurate. 
As a bonus, the number of factors is also allowed to grow. 
At the heart of our method is a penalized reduced rank regression to combine information. 
To further deal with heavy-tailed data, a computationally attractive penalized robust reduced rank regression method is proposed. 
We establish faster rates of convergence compared with the benchmark. 
Extensive simulations and real examples are used to illustrate the advantages.
\end{abstract}

\end{frontmatter}
\noindent%

% already modified
\noindent%
\vfill
\newpage

% is proposed to improve the estimation efficiency by 
% Specifically, to extract information from both the multivariate observations $\yt$ of main interest and a large number of factor candidates $\xt$, 
% we first run a to combine information and estimate the  loading matrix, and then project $\yt$ onto the estimated loadings to recover the latent factors.

\section{Introduction}\label{sec:intro}
Factor model is a useful tool for modeling common dependence among multivariate outputs and it is gaining popularity with
the emergence of high-dimensional data. Suppose we observe a sample of $N$-dimensional multivariate data $\{\yt\}_{t = 1} ^T$, where $\yt=(y_{1,t},...,y_{N,t})^T$. 
Assume the covariance structure of $\yt$ is 
largely driven by a small number of common factors
$\ft = (f_{t,1},...,f_{t,K})^T$. The factor model is 
\begin{equation}\label{model_1}
    \yt = \MA \ft + \et, t =1,...,T, 
\end{equation}
where $\MA$ is an $N \times K$ loading matrix and $\et$ is the idiosyncratic error. 

Since the true factors are typically unknown, there are two approaches being commonly adopted.
One is to use domain knowledge to manually pick several observed covariates  as factor proxies, treat them as $\ft$ in model (\ref{model_1}),  and then estimate $\MA$ and $\et$. 
We refer to this approach as the \textit{factor proxy} approach. 
A well-known example is the Fama-French three-factor model \citep{fama1993common} in finance. 
However, the subjective and static nature of this approach raises many concerns. For example, in finance, there is a long-lasting debate on which covariates are the best set of factor proxies \citep{harvey2019census}, and emerging evidence shows that some commonly used proxies are only weakly correlated with stock returns \citep{anatolyev2018factor, gospodinov2019too}, part of the factor structure remains  unexplained \citep{kleibergen2015unexplained, fan2016augmented}, 
and the true factors change over time and across different markets \citep{li2017determining}. 

The second approach is to treat the factors as latent and estimate the whole model via statistical methods, such as principal components-based methods \citep{stock2002forecasting} or maximum likelihood estimation \citep{doz2012quasi}. 
We refer to this approach as the \textit{proxy-agnostic} approach. 
Such an approach only utilizes the information in $\yt$ and cannot incorporate information from other variables, and hence may lead to loss of efficiency when there is useful auxiliary information, such as the factor proxies. 
% increasing number of factors
% not adaptive. 

To combine the merits of these two approaches and mitigate their drawbacks, 
\cite{fan2016augmented} took the first step and proposed to bridge them by approximating the latent factors with the nonlinear transformation of several observed covariates $\xt^*$. 
Specifically, they consider the model $\ft = \boldsymbol{g}(\xt^*) + \gammastart$, where $\boldsymbol{g}(\xt^*) = \mathbb{E}(\ft|\xt^*)$ is a nonlinear function and $\gammastart$ represents the approximation error. 
To (implicitly) utilize this model, they first run non-parametric regression of $\{y_{i,t}\}_{t=1}^T$ onto $\{\xt^*\}_{t=1}^T$ to obtain fitted values $\{\hat{y}_{i,t}\}_{t=1}^T$, for each $i \in \{1, \dots, N\}$ separately, 
then apply principal component analysis (PCA) to $\{(\hat{y}_{1,t}, \dots, \hat{y}_{N,t})'\}_{t=1}^T$ to estimate $\MA$ as $\hat{\MA}$, and finally estimate $\ft$ as $N^{-1} \hat{\MA}' \yt$. 
The approximation power of the covariates plays a vital role in their method. 
Let $||\cdot||$ denote the matrix operator norm. 
When $||\cov(\gammastart)|| \rightarrow 0$ as $N$ or $T$ goes to infinity, that is, when the approximation error becomes  negligible, we can establish improved convergence rates for $\hat{\MA}$ and $\{\hat{\ft}\}$ over PCA; 
while when the explanatory power of $\xt^*$ on $\ft$ is weak, the convergence rates can be slower than PCA. 
Therefore, it shares similar limitations with the factor proxy approach as it is not easy to guarantee the approximation quality with only several manually picked  covariates. 
In addition, the condition  $||\cov(\gammastart)|| \rightarrow 0$ naturally requires the approximation power of $\xt^*$ to increase, which may not hold when the dimension of $\xt^*$ is fixed as in \cite{fan2016augmented}. 
Therefore, restricting $\xt^*$ to be low-dimensional can be risky and may cause loss of efficiency. 
In this paper, we aim to relax this restriction by allowing the number of covariates to grow.  
%%%%%%%%%%%%%%%%%%%%%%
% not adaptively choose; adaptive to the dataset
% 1. more factors, misspecified, bad proxies, unexplained, varies
% robust, efficient, adaptive
% hesitated: covariates or factor proxies.
%%%%%%%%%%%%%%%%%%%%%%

One specific motivating example of this work is the \say{\textit{factor zoo}} problem arising in finance. 
As summarized in \cite{harvey2019census}, over $400$ covariates have been proposed in the literature and claimed to have approximation power on the latent factors of  stock returns. 
This chaos raises a lot of concerns and discussions. 
The existing works mainly focus on the factor selection problem \citep{feng2017taming, hwang2018searching}. 
% carefully selecting several good proxies from the large pool of candidates
% select them to use as proxies in the first approach
% combination of factors more stable
Instead, we approach the problem from a different angle, by considering directly improving the latent factor model estimation accuracy via utilizing the information contained in these covariates. 
Motivated by this example, 
in this paper, we refer to the observed covariates $\xt$ as \textit{factor proxies} to emphasize their explanatory power on $\ft$. 
However, the scope of our method is not limited to this application.   
Some other examples of such factor proxies include the numerous macroeconomic variables recorded in econometrics and the covariates collected for each individual in health studies. 
% high-dimensional

%  \cite{li2016supervised} is not an example.  low-dimensional

% Nowadays, in many applications of factor models, we have a large number of auxiliary variables available, denoted by the $p$-dimensional vector \(\xt\), which after certain  transformations and combinations can provide good approximation to $\ft$.

Our target is to utilize the information from both the multivariate observations of main interest $\yt$ and a large set of factor proxies $\xt$ to improve the estimation accuracy of the latent factor model. 
Specifically, we consider approximating the latent factors by some linear transformation of $\xt$ and regard the following model as our working model: 
\begin{equation}\label{model_2}
    \ft = \MB' \xt + \ut, 
\end{equation}
% (possibly sparse) no sparsity -> theory? 
where $\vec{B}$ is a $p\times K$ matrix and $\ut$ is the mean-zero approximation error. 
The effect of potential misspecification of the linear form will be analyzed both theoretically and numerically. 
We assume $K < \min(T,N,p)$ throughout this paper.  
% more motivations here! 
% sparse
% PCA; linear combinations
% Linear transformation is a reasonable approximation in this setting. 
% For illustration, among $99$ stock risk factor proxies survied in \cite{feng2017taming}, $16$ of them are about different aspects of Value-versus-Growth and $8$ are about Momentum, and hence it is natural to assume a weighted version of these candidates might be better proxies. 
% Strong multicolinearity
% We do not require $\ut$ to be uncorrelated with $\xt$, and hence allow mild misspecification of the linear form. 
% because, asymptotic. 

% augmenting covariates are sufficient for approximating $\ft$,
% explicitly utilize the approximation model (\ref{model_2}), high-dim, linear form. 
To combine information by integrating model (\ref{model_1}) and (\ref{model_2}), we propose to fit a reduced rank regression (RRR) with $\yt$ onto $\xt$ to estimate the loading matrix, and then recover the latent factors by projecting $\yt$ onto the estimated loadings. 
The RRR model is a multivariate regression model with a low-rank 
coefficient matrix \citep{velu2013multivariate}, which naturally links model (\ref{model_1}) and (\ref{model_2}) together. 
Suppose the information provided by $\xt$ is useful, that is, the variance of $\ut$ is relatively small, 
the RRR step can be regarded as a denoising procedure with the guidance of  $\xt$ and with the low-rank structure in mind. 
We can expect that, compared with the noisy data $\yt$, 
the fitted value $\hat{\Vec{y}}_t$ can help us recover the latent factor structure more accurately. 
In addition, such a procedure allows our estimator to yield the desired \say{exact identification} property, that is, the loading matrix is identifiable and can be consistently estimated even when $N$ is fixed. 
To guarantee the approximation power, we allow the number of proxies $p$ to increase so that one can incorporate a large number of proxies,  % with $T$
and we replace the vanilla RRR with a penalized variant to handle the high dimensionality. 
As a bonus, the number of factors $K$ is then naturally allowed to grow.
The resulting method is named as \textit{Factor Model estimation with Sufficient Proxies (FMSP)}. 
% \textit{Candidates-Augmented Factor model Estimation (CAFE)}. 
Finally, in some applications of factor models, the error terms usually have heavy-tailed distributions and hence non-robust procedures may fail  \citep{calzolari2018estimating}. 
To accommodate these scenarios, we further propose a novel penalized robust RRR method as a subroutine to extend FMSP to the heavy-tailed situations.

\subsection{Related literature}
%%%%%%%%%%%%%%%%%%%%%%%% Intro: Literature %%%%%%%%%%%%%%%%%%%%%%%%%%%%%%
There is a rich literature on the estimation of latent factor models. 
For the proxy-agnostic approach, some popular methods include
the principal components-based methods \citep{stock2002forecasting, bai2017principal}, 
the generalized principal components-based methods \citep{choi2012efficient, bai2013statistical}, and the 
maximum likelihood-based methods \citep{bai2012statistical, bai2016efficient}. 
The factor proxy approach is also widely used in finance \citep{fama1993common}, economics \citep{cox1985intertemporal} and other areas \citep{mulaik2009foundations}. 
Our work bridges these two approaches and we generalize the method proposed in \cite{fan2016augmented}.
Compared with \cite{fan2016augmented}, our approach has advantages in three aspects: 
(i) our approach allows the number of covariates to diverge, and hence can better ensure the approximation power from these covariates  and the resulting estimation accuracy; 
(ii) their method requires fitting $N$ separate non-parametric regressions and hence could be computationally intense, while our estimator yields a quasi-closed form; 
(iii) our approach allows the number of factors $K$ to increase, a feature receiving increasing attention in recent years % with $N$ and $T$
in order to accommodate the growing data size and potential structural breaks  \citep{jurado2015measuring, li2017determining, luo2017inverse, yu2020nonparametric}.

With the emergence of high-dimensional and complex datasets, in recent years, a lot of works focus on 
the low-rank structure \citep{bickel2008regularized, negahban2011estimation}, 
the penalization methods \citep{fan2001variable, zou2005regularization}, 
the heavy-tailed problem \citep{fan2016shrinkage, sun2019adaptive}, 
and their combinations \citep{candes2011robust, chen2018robust}.
Our approach contributes to their intersection area. 
Specifically, our regression step is closely related to the literature on RRR, especially the penalized and robust variants  (e.g., \cite{chen2012sparse}, \cite{she2017robust}, and \cite{tan2018distributionally}). 
We note that some RRR literature refers to the transformed predictors as factors, in a way to offer certain interpretability \citep{bernardini2015macroeconomic, kargin2015estimation, she2017selective}. 
However, both the setting and the results of this paper differ from those in the RRR literature. 
A RRR model is a regression model, hence the focus is typically on the accuracy of the coefficient matrix estimation or the prediction. 
It is also typically assumed that the error term has independent entries. 
In contrast, we use RRR as an intermediate step to facilitate the factor model estimation, and the focus is on the latent factor structure recovery. % only
Besides, our problem is formulated under the general factor model setup, and hence in the RRR step, we allow the residuals to have correlated entries or even some factor structures. 
The estimation procedure as well as the theoretical and numerical results are developed accordingly and are different from the RRR literature. 
As a by-product, a computationally efficient penalized robust RRR method is proposed  
% for our robust factor model estimator.
as a subroutine of our robust extension. 
Finally, \citet{freyberger2020dissecting} considers using high-dimensional
stock-specific characteristics to predict the expected excess return, while we aim to describe the covariance structure of stock returns using a set of shared risk factors. 
% However, we only use RRR as an intermediate step and our focus is on the latent factor structure recovery instead of the regression problem itself. 
% As a by-product, a computationally efficient penalized robust  RRR method is proposed for our robust factor model estimator. 

% to our knowledge, first time, penalized RRR + FM. 

%%%%%%%%%%%%%%%%%%%%% 1. guided DR
Our work is also related to research on supervised dimension reduction, which studies the dimension reduction for one dataset with the help of auxiliary information. 
Popular methods include  supervised principal components \citep{bair2006prediction}, 
principal fitted components \citep{cook2008principal}, and sufficient dimension reduction \citep{adragni2009sufficient}. See \cite{chao2019recent} for a recent survey. %envelope model
Most methods study the regression problem and seek the best dimension reduction for the regressors so as to better predict the responses. 
In contrast, our paper focuses on recovering the low-dimensional structure of the multivariate data with the guidance of some covariates, which are used as regressors in our first step. 
Therefore, the objectives are totally different. 
The most related works are \citet{huang2019shrinking} and \citet{li2016supervised}. 
\citet{huang2019shrinking} empirically studies the performance of the vanilla RRR model in asset pricing, while our work focuses on the latent  factor model estimation problem itself. 
To better suit this problem, we extend the RRR model by introducing an unexplained factor term, studying its penalized version, and also proposing a robust extension. 
In addition, we provide systematic simulation experiments and theoretical analysis, while \citet{huang2019shrinking} do not. 
\citet{li2016supervised} also studies better recovering the low-rank structure of $\yt$ with auxiliary information $\xt$. 
However, their work is motivated by a problem different from the factor model estimation and hence has fundamental differences with ours in many aspects. 
For example, they consider the setup where $\xt$ is low-dimensional, variables have normal distributions, and noise entries are independent.  Their likelihood-based estimator and its theoretical property are also both developed under these conditions, which do not hold in our setting. 

% which also studies approximating the low-rank structure of $\yt$ with some linear transformations of $\xt$. 

In the asset pricing literature, there is a surge of interest in the \say{factor zoo} problem. 
    To address the high dimensionality issue, existing efforts can be roughly divided as the \textit{selection} approaches and the \textit{aggregation} approaches. 
    Among the selection approaches, 
\citet{feng2017taming} designs a model selection procedure to estimate the incremental contribution of a newly proposed factor, 
and \citet{hwang2018searching} applies Bayesian variable selection to choose among many candidate pricing factors. 
% Shrinking the Cross Section
Among the aggregation approaches, 
\citet{kozak2020shrinking} and \citet{bryzgalova2019bayesian} aim to aggregate  information from the factor zoo to learn a better factor pricing model, with either a ridge-like prior or a \say{spike-and-slab} prior. 
The empirical findings suggest that the aggregation approach seems to outperform the selection approach \citep{bryzgalova2019bayesian}, and that the true model seems to be sparse in the linear span of the observable proxies instead of in the  raw  proxies. 
These findings are consistent with our approach. 
Instead of targeting asset pricing and risk premia estimation, 
% of observable factors (under the linear factor asset pricing model assumption), 
% large number of factor proxies
our work is the first paper to leverage the rich information in the factor zoo to improve the estimation of the latent factor model. 
% (i.e., maximizing the time-series $R^2$). 
Finally, to handle the high dimensionality issue or to incorporate side information, various shrinkage methods (such as the $L1$ penalty, the $L2$ penalty and their Bayesian counterparts) are commonly adopted in the aforementioned large factor model literature or factor zoo literature.  
Our work contributes to this line of research. 
% and we use the ridge penalty due to its 

{
In asset pricing, 
one possible issue of using manually picked observable factors is that some true pricing factors might be omitted in the manually specified model, which belongs to model misspecification \citep{bryzgalova2015spurious, gospodinov2014misspecification} and may lead to invalid estimation. 
For example, \citet{giglio2021asset} studies how to obtain an valid risk premia estimation even in this case. % In particular
Although our objective is not the inference tasks in asset pricing, the similar consideration motivates us to incorporate a sufficient pool of factor proxies and to explicitly model the unexplained factor component. 
% On one hand, this motivates us to consider incorporating a large number of factor proxies to improve the robustness to manual proxy selection; 
% on the other hand, we also explicitly model and account for the part that can not be explained by factor proxies in our methodology design and theoretical analysis. 
A reduced rank representation is also commonly adopted in the linear asset pricing literature \citep{adrian2015regression}, where the assumption of no arbitrage leads to that the cross-sectional expected excess return and the time-series variations can be explained by the same loadings, which induces a reduced rank structure in the whole model. 
This is different from our approach of using RRR to aggregate information in a large number of factor proxies. 
}

% empirical. Not high-dimensional. Robustness. Pricing. Variance. Heavytail. Empirical. no real data. No additional term 
% Some previous empirical evidence has been found in XXX. 
% Should argue more difference? not sure what

% ASMP, methods, applications, theories
% underlying patterns of genetic variation among tumors
% However, we note that the two models arise in the analysis of different problems, and that they have different applications and interpretations. The

% I can argue the target is to explain the covariance, but still just regression? 
% They aim to better predict the responses, so does ours?

% How Fan cite that Econ paper?
% No additional term

%%%%%%%%%%%%%%%%%%%%% 2. Factor zoo %%%%%%%%%%%%%%%%%%%%% 
% The focus of most existing methods is to find a dimension reduced version of X that keeps all the information about Y. This is different from the scope of the current paper. Here our primary goal is to identify low rank structure of X, whether or not the structure is related to the auxiliary information Y. T
% supervised DR
% Sufficient Dimension Reduction
% PCR
% li, 2016; dashan huang; other related papers?

% discuss more about factor seletion? those econ paper? confused. factor
% zoo? How to do review here?
 %%%%%%%%%%%%%%%%%%%%%  %%%%%%%%%%%%%%%%%%%%%  %%%%%%%%%%%%%%%%%%%%% 

%%%%%%%%%%%%%%%%%%%%%%%% Intro: Contributions %%%%%%%%%%%%%%%%%%%%%%%%%%%%%%
\subsection{Contributions}
The main contributions of this paper can be summarized as follows. 
Methodologically, we propose a novel latent factor model estimation approach to bridge the factor proxy approach and the proxy-agnostic approach, 
and it allows incorporating a large number of factor proxies. 
As such, the proposed estimator FMSP is robust, flexible, and statistically more accurate: 
it is robust as it avoids the factor selection task and hence mitigates the  potential resulting bias, compared with the factor proxy approach and SPCA; 
it is flexible as it can adapt to different datasets and dimensionality, and also allows the number of factors to diverge; 
it is more accurate as it yields improved convergence rates compared with baselines such as PCA, and can even achieve the \say{oracle} rate for the loading matrix estimation as if we can observe the true factors. 
% the true factors may vary among different applications and datasets, their methods still can not avoid the potential insufficiency of $\xt$. %and bias.
% change after structural break and across different markets

Theoretically, we establish convergence rates for both the estimator of the loading matrix and that of the latent factors. 
The asymptotic results are established under a general setting: $N$ and $K$ can either grow or stay fixed, and mild model misspecification is allowed. 
The convergence rates under different conditions are discussed and also compared with those of several baselines, which provides insights into the expected performance of the proposed method. 

Empirically, we conduct extensive simulations to study the finite-sample performance of the proposed method under different conditions. 
% Robust version 
Two real applications on estimating the factor structure of monthly stock returns are also presented. 
The numerical performance supports our theoretical findings and illustrates the advantages of FMSP. 
% , and confirms the necessity of including a sufficient number of candidates. 
% 99. findings

%%%%%%%%%%%%%%%%%%%%%%%% Intro: Organization %%%%%%%%%%%%%%%%%%%%%%%%%%%%%%

The rest of the paper is organized as follows. 
In Section 2, we discuss and formally define our estimation procedure. 
The rates of convergence are presented and discussed in Section 3. 
In Section 4 and 5, we illustrate the performance of our method using  simulated datasets and real data examples, respectively. 
An extension for handling heavy-tailed data is introduced in Section 6. 
Technical details, proofs, and additional numerical results can be found in the appendix.  

In the paper, for a matrix $\MA$, we denote its minimum, maximum and $i$-th largest singular value as $\sigma_{min}({\MA})$, $\sigma_{max}({\MA})$ and
 $\sigma_i({\MA})$, respectively. The rank of $\MA$ is denoted as $r(\MA)$. 
 We define $||\MA||_F = tr^{1/2}(\MA^{T}\MA)$ and $||\MA|| = \sigma_{max}^{1/2}(\MA^{T}\MA)$.
Let $\MA^{+}$ represent the Moore–Penrose inverse of $\MA$ and $\MA'$ be the transpose of $\MA$. 
For a vector $\vv$, we define $||\vv||$ to be its $\ell_2$ norm and denote the set of  $p$-dimensional vector $\vv$ satisfying $||\vv||=1$ as $\Sd{p}$. 
For a $p$-dimensional random vector $\vec{x}$, 
define its sub-Gaussian norm $||\vec{x}||_{\psi_2} := sup_{\vv \in \Sd{p}, s\ge 1}(E|\vv^T \Vec{\xt}|^s)^{\frac{1}{s}}/\sqrt{s}$. 
We use $\Vec{I}$ to denote an identity matrix and $\textbf{1}$ to denote an all-one vector, the  dimensions of which can be inferred from the context. 
For two variables $a$ and $b$, we denote the smaller one as $a \wedge b$ and the larger one as $a \vee b$. 
For two sequences $\{a_i\}$ and $\{b_i\}$, we write $a_i \asymp b_i$ if $a_i = O(b_i)$ and $b_i = O(a_i)$.

\section{Methodology}\label{sec:meth}
% Estimator
% \subsection{The estimator}\label{method:non-robust}

Combining model (1) and (2), we are considering the following working model: 
\begin{equation}\label{eqn:full_model}
    \yt =  \MA \ft + \et = \MA(\MBt \xt + \ut) + \et. 
\end{equation}
It is well known that additional identification conditions are needed for latent factor models: for any $K \times K$ invertible matrix $\mathbf{H}$, (\ref{eqn:full_model}) is unchanged when we replace $(\MA, \MBt, \{\ft\})$ with $(\MA\MH, \MH^{-1}\MBt, \{\MH^{-1}\ft\})$. 
For ease of presentation, we assume $\MA'\MA = NK^{-1} \Vec{I}$ throught this paper.

The first and most crucial question is how to utilize information from both $\yt$ and $\xt$ to estimate the factor model. 
Suppose the variance of $\ut$ is small. 
We can expect that, by regressing $\yt$ onto $\xt$, 
the fitted value $\hat{\Vec{y}}_t$ can be regarded as a denoised version of $\yt$ with the guidance of $\xt$,  
and compared with the noisy data $\yt$, $\hat{\Vec{y}}_t$ can help us recover the latent factor structure more accurately. 
We formalize this property with convergence rates in Section \ref{theory}. 
In addition, similar to Theorem 2.1 in \cite{fan2016augmented}, we can establish the following desired \say{exact identification} property for the loading matrix. %$\MA$. 

\begin{proposition}[Exact identification]\label{proposition: exact_identification}
Suppose $\sigma_{min}(E(E(\ft|\xt) E'(\ft|\xt))/K) >  0$ and $E(\et|\xt) = 0$. 
There is an invertible $K \times K$ matrix $\MH$ such that: 
$E(E(\yt|\xt) E'(\yt|\xt))$ has rank $K$, and the eigenvectors of  $E(E(\yt|\xt) E'(\yt|\xt))$ corresponding to its non-zero eigenvalues 
are the columns of $\MA\MH$.  
\end{proposition}

% together with $\MA'\MA = NK^{-1} \Vec{I}$
% $K$ factors are \textit{strong}. 
% The condition $\sigma_{min}(E(E(\ft|\xt) E'(\ft|\xt))/K) > 0$ implies the 
The condition $\sigma_{min}(E(E(\ft|\xt) E'(\ft|\xt))/K) > 0$ implies that $\xt$ is correlated with every factor. 
Proposition \ref{proposition: exact_identification} states that, by incorporating the covariates $\xt$, the loading matrix is identifiable even when $N$ is fixed. 
In contrast, for the proxy-agnostic approach, $N \rightarrow \infty$ is generally required. 
A similar consistency result is established in Theorem \ref{rate:loading}. 

% no cov requirement
% with penalty
% so as to reduce the effect of $\et$ with diversification
% according to $E(\yt\yt') = \MA cov(\ft) \MA' + cov(\et)$, 

% The identification of $\ft$ follows from the relationship $\ft = N^{-1}K\MA'(\yt - \et)$. 
% with an optional hyperparameter $\tau$. 
Motivated by the above discussion, 
we consider the following regression problem  to estimate the loading matrix
% following
% such a regression can be formularized as the following problem: 
\begin{equation}\label{prob:motivation}
    \min_{\MA^{T}\MA = NK^{-1}\vec{I}, \;\MB}\;  \sumoT \mathcal{L}(\yt, \MA\MBt\xt), 
\end{equation}
where \(\mathcal{L}\) is some loss function. 
We set $\mathcal{L}$ as the least square loss in this section. 
The heavy-tailed scenario and the corresponding robust loss functions are considered in Section \ref{sec: robust FMSP}.

Problem (\ref{prob:motivation}) shares a similar form with the reduced rank regression (RRR) problem \citep{velu2013multivariate} 
\begin{equation*}
    \yt = \MTh'\xt + \tilde{\boldsymbol{\epsilon}}_t, r(\MTh) \le K, 
\end{equation*}
where $\MTh$ is a low-rank coefficient matrix and $\tilde{\boldsymbol{\epsilon}}_t$ is the error term which is typically assumed to have independent entries. 
Our formulation is different from the standard RRR model as we allow $\tilde{\boldsymbol{\epsilon}}_t$ to have correlated entries or even some factor structures (note that $\tilde{\boldsymbol{\epsilon}}_t = \MA\ut + \et$). 
% and we require the rank to be equal to the number of factors
In addition, the regression is only used as an intermediate step in our procedure to facilitate the factor model estimation.  
% ($\tilde{\boldsymbol{\epsilon}}_t = \MA\ut + \et$)

In the aforementioned approach, what plays a vital role is the signal quality of $\xt$, which may not be easy to guarantee when $p$ is small. 
As an extreme example, when $p=K$, the method reduces to the factor proxy approach. 
Luckily, as introduced in Section \ref{sec:intro}, nowadays, we usually have a large number of auxiliary variables which are believed to have explanatory power on the latent factors. % may
By allowing incorporating a large number of factor proxies, it is expected that the potential bias (and the resulting efficiency loss) from manually selecting proxies can be mitigated. 
As a bonus, the number of factors $K$ is allowed to grow. 
% * One additional desired property of FMSP is that our method naturally allows the number of factors to diverge with $T$, and the theory still holds. 

Based on the above discussion, in this paper, we allow the number of proxies $p$ to grow with $T$. 
To account for the high dimensionality of $\xt$, we focus on a penalized version of  (\ref{prob:motivation}) as the first step of our estimation method: 
\begin{equation}\label{general optimization equation}
    \min_{\MA^{T}\MA= NK^{-1}\vec{I}, \; \MB}\; \sumoT \mathcal{L}(\yt, \MA\MBt\xt) +  \mathcal{J}(\vec{B}; \lambda). 
\end{equation}
Here, $\mathcal{J}(\cdot; \lambda)$ is some penalization function with a positive penalty parameter $\lambda$. 

For a general penalty function $\mathcal{J}$, even with the condition $\MA \MA' = NK^{-1}\vec{I}$, 
the optimization problem (\ref{general optimization equation}) may still be ill-posed: for an arbitrary rotation matrix $\vec{Q}$ satisfying $\MQ \MQ'=\vec{I}$, 
model (\ref{eqn:full_model}) is unchanged when we replace $(\MA, \MB, \{\ft\})$ with $(\MA\MQ, \MB\MQ, \{\MQ^{-1}\ft\})$, while it is not necessary that  $\mathcal{J}(\MB,\lambda)$ and $\mathcal{J}(\MB \vec{Q},\lambda)$ are equal. 
% make it clearer about the group lasso here. 
Among commonly used penalty functions, the group lasso \citep{friedman2010note} and ridge-type penalty both can avoid this issue, while lasso \citep{tibshirani1996regression}, elastic net \citep{zou2005regularization}, and SCAD \citep{fan2001variable} cannot. 

In this paper, we focus on the ridge-type penalty $\mathcal{J}(\MB,\lambda) =\lambda ||\MB||_F^2$ given that the resulting estimator is generally computationally efficient and that the estimator is less sensitive to multicollinearity, which usually exists in a large pool of factor proxies. 
In particular, with $\mathcal{J}(\MB,\lambda) =\lambda ||\MB||_F^2$, we can obtain a quasi-closed-form solution of (\ref{general optimization equation}). 
Specifically, 
let $\MY = (\vec{y}_1,...,\vec{y}_T)'$, $\MX = (\vec{x}_1,...,\vec{x}_T)'$, and $\hat{\MY}^* = (\MX ',\sqrt{\lambda/N}\Vec{I})'(\MX '\MX +
\lambda/N\vec{I})^{-1}\MX '\MY$. 
Let $\sum_{i=1}^{N \wedge p}\sigma_i \vu_i \vv_i'$  be the singular value decomposition of $\hat{\MY}^*$. 
Then, up to a rotation, we have that the loading matrix estimator obtained by solving \eqref{general optimization equation} is 
\begin{equation}\label{eqn:A_hat}
    \hA = \sqrt{N/K}(\vv_1,...,\vv_K). 
\end{equation}
Formally, we have the following result. 
% oroposition. 
% is formalized in Proposition \ref{proof: Derivation}. 
% , where the condition on singular values is for sake of simplicity and can be satisfied by most real datasets. 
% zhu ji. proposition
\begin{proposition}[Quasi-closed form of the loading matrix estimator]\label{proof: Derivation}
% Assume the largest $K$  singular values of  
% $\hat{\MY}^*$ are non-zero and  distinct. 
Suppose $\mathcal{J}(\MB,\lambda) =\lambda ||\MB||_F^2$ and the largest $K$  singular values of  $\hat{\MY}^*$ are non-zero and  distinct. 
Let $\hA = \sqrt{N/K}(\vv_1,...,\vv_K)$, then there is a matrix $\hB$ such that $(\hA,\hB)$ is a solution of \eqref{general optimization equation}. 
Conversely, for any solution of (\ref{general optimization equation}) denoted by $(\hA_1,\hB_1)$, there exists an orthonormal matrix $\MQ$ such that $\hA = \hA_1\MQ$ and $\hB = \hB_1\MQ$. 
\end{proposition} 
The derivation for ridge RRR has been studied in \citet{mukherjee2011reduced}, under the regression setting. 
We note that the proposed factor model estimation framework can be extended to other penalty function options. For example, when explicit factor proxy selection is preferred, the group lasso penalty can be adopted and the iterative algorithm developed in \cite{chen2012sparse} can be used to solve (\ref{general optimization equation}).
% The form of B is defered to XX. 

Once we solve the optimization problem (\ref{general optimization equation}) and obtain ($\hA$, $\hB$) following Proposition \ref{proof: Derivation}, it remains to estimate the latent factors. 
When $\ut$ can be ignored, that is, when the linear form is correctly specified and the approximation power of $\xt$ is sufficient, $\hBt \xt$ seems to be a straightforward estimator.  % desired
However, when either of these two conditions does not hold, there could be a significant bias and the estimator may not be consistent. 
Instead, motivated by the relationship $\ft = N^{-1}K\MA'(\yt - \et)$, 
we propose to estimate the latent factors by $\hft = N^{-1}K\hA' \yt$, which is robust to either model misspecification or insufficiency of $\xt$'s approximation power. 
% , while the price to pay is little when $\ut$ is negligible. 

The proposed estimation procedure, FMSP, can be summarized as the following two steps:
\begin{enumerate}
    \item Solve the penalized reduced rank regression problem to estimate the loading matrix: 
    \begin{equation*}%\label{eqn:step2}
    \hA,\hB = \argmin_{\MA \in \mathcal{R}^{N\times K}, \MB \in \mathcal{R}^{p\times K}, \MA'\MA=NK^{-1}\mathbf{I}}\quad ||\MY - \MX\vec{B}\MA'||_F^2 +  \lambda ||\vec{B}||_F^2.
    \end{equation*}
    The quasi-closed form of the final estimator $\hA$ is given in \eqref{eqn:A_hat}.  
    \item Project $\yt$ on the estimated loading matrix to recover the latent factors: 
    \begin{equation*}
        \hft = N^{-1}K\hA' \yt, \;\; t = 1, \dots, T. 
    \end{equation*}
\end{enumerate}
In this paper, we focus on the factor model estimation problem and treat the number of factors $K$ as  pre-specified. 
In practice, $K$ can be consistently estimated by many methods in the 
literature, such as the eigenvalue-ratio method proposed in \cite{lam2012factor}  when $K$ is fixed or the information criteria-based method studied in \cite{li2017determining} when $K$ diverges. 
Alternatively, we can consider utilizing the information in both $\yt$ and $\xt$ to select $K$, via using either cross-validation or certain information criteria in step 1. 
Such a supervised estimation is expected to be more accurate and is left for future investigation. 
% Compared with most existing methods 
% , and our  when $K = o(T)$. 

%%%%%%%%%%%%%%%%%%    
%  Assumptions    
%%%%%%%%%%%%%%%%
\section{Asymptotic Properties}\label{theory}
\renewcommand{\theenumi}{\roman{enumi}}

In this section, we present and discuss the convergence rates of the proposed estimators. 
Recall that, so far, we are dealing with the working model 
$\ft = \MBt\xt+\ut$, where $E(\ut|\xt) = \mathbf{0}$. 
To understand how the potential misspecification of the linear form affects the estimation, 
in this section, we consider a general form of the data generation model as $\ft = \mathbf{g}(\xt) + \ut$, where $\mathbf{g}(\xt) = E(\ft|\xt)$ and $\mathbf{g}(\cdot)$ is some nonparametric function. 
To ensure identifiability, we define 
$\MB = \argmin_{\Tilde{\MB}} ||\ft - \Tilde{\MBt}\xt||$. 
In such a setup, $E(\ut|\xt) = \mathbf{0}$ may not hold any longer, hence $\ut$ and $\xt$ can be correlated.

% $\yt = \MA(\MBt\xt+\ut)+\et$

\subsection{Assumptions}\label{assumptions}
%%%%%%%%%%%%%%%%%%    Assumptions 1   %%%%%%%%%%%%%%%%
We first state some assumptions. 
% Recall that our full model is $\yt = \MA(\MBt\xt+\ut)+\et$. 
For simplicity of notations, 
denote $\fxt = \MBt\xt$.
% We assume $\yot,\xt,\ut,\et$ are all centered.

\begin{asmp}[Tail distributions]\label{asmp:tail}
There are constants $M_0, M_1, M_2 >0$, such that
$||\fxt||_{\psi_2} \le M_0$, 
$||\xt||_{\psi_2} \le \sqrt{p} M_1$, 
and 
$||\et||_{\psi_2} \le M_2$.
\end{asmp}
% Not an issue; ||\xt||_{\psi_2} = ||cov(\xt)||, bounded (???)
We assume $\fxt$ and $\xt$ have sub-Gaussian tails and allow the factor proxies in $\xt$ to have correlations. 
The sub-Gaussian condition on $\et$ is required for our analysis of RRR with random matrix tools. 
Roughly speaking, 
it suffices for $\et$ to have entry-wise sub-Gaussian tails and weak dependency among entries, which is typically imposed in the approximate factor model literature \citep{fan2011high, bai2016efficient}. 
% citation here? 
For example, the condition that $\et$ follows a multivariate normal distribution with $||E(\et \et')||_1 \le M_2'$
can lead to $||\et||_\psit \le \sqrt{M_2'/2}$.

% Moreover, for a symmetric matrix $\MA_1$, we have $||\MA_1|| \le ||\MA_1||_1$ (\cite{van1983matrix}). Thus, following Assumption3.(\romannumeral1)  we have $||E(\et  \et')|| \le ||E\et \et'|| \le M_2$.

% [explanation on the sub-gaussian of et? how?]
% [sparsity?]

% $max_iE\boldsymbol{\epsilon}_{it}^4 \le M_1$
%$max_{i} E(E(\boldsymbol{\epsilon}_{it}^2|\xt,\ut))^k \le M_1, k > 1$% how to write?
% do we need this?
 % $E(X_i) = 0$% \vspace{-1cm}

%%%%%%%%%%%%%%%%%%    Assumptions 2  %%%%%%%%%%%%%%%%
% The eigenvalues of $KN^{-1} \MA' \MA$ are bounded away from zero and infinity; 
\begin{asmp}[Signal strengths and dimensions]\label{asmp:signal}
\hfill
\begin{enumerate}
    \item $ 0 < m_3 \le \sigma_{min}(E(\ft\ft') / K)\le \sigma_{max}(E(\ft\ft')/K) \le M_3 < \infty$ for some positive constants $m_3$ and $M_3$;
    \item $\sigma_{min}(E(\fxt\fxt')/K) \ge m_4 $, $\sigma_{min} (E(\xt\xt')) \ge m_5 $ and $u_2 \coloneqq ||\ut||_{\psi_2}
     \le K^{1/2} M_4$, where $m_4, M_4$, and $m_5$ are all positive constants;
    \item $K = o(T)$ and {$p = O(T^{1-\eta})$ for some $\eta \in (0,1]$.}
    % , $p = O( T^{1/2+2\delo})$ and $p = O(\sqrt{T}N)$
    % {[we do not need these any longer]}
\end{enumerate}
\end{asmp}
When $K$ is fixed, the first condition is standard in the factor model literature \citep{fan2011high, bai2016efficient}. 
It assumes that each factor impacts a non-negligible portion of $\yt$ with non-degenerating strength. 
The additional dependency on $K$ is because we allow $K$ to diverge, and it ensures that $||N^{-1}\MA E(\ft \ft') \MA'||$, the magnitude of the systematic part, is bounded as $K \rightarrow \infty$. 
The second condition implies two things: the unexplained part $\ut$ has sub-Gaussian tails and the magnitude of the approximation error is bounded. 
The latter is a reasonable assumption under Condition \textit{\romannumeral1}. 
% given that the latent factor 
% power of $\xt$ does not decay, with $u_2$ representing the strength of the unexplained part in $\ft$. 
% When strong  multicollinearity exists among the factor proxies, prepossessing procedures such as PCA can be applied to $\xt$ in advance. 
{
The large pool of factor proxies usually comes with a strong multicollinearity,  given that some of them have been observed as explaining almost the same exposures. 
In practice, we assume some preprocessing procedures have been applied to avoid a nearly perfect multicollinearity, and hence we can assume the least eigenvalue is bounded away from $0$. 
}
The condition $K = o(T)$ is standard in the literature \citep{li2017determining}. 
% \wedge N
% selection or

% , and $K^{-1}u_2^2$ is assumed to be bounded as $K \rightarrow \infty$. 
%may degenerate with $p$. 
% comment on the dimension assumptions

%%%%%%%%%%%%%%%%%%    Assumptions 3  %%%%%%%%%%%%%%%%

\begin{asmp}[Weak dependences] \label{asmp:dependences}
\hfill 
\begin{enumerate}
%     \item There is a constant $M_3 >0$ so that 
% \begin{equation*}
%     sup_{\vec{x},\vec{u}} max_i \sum_{j=1}^{N}|E(\boldsymbol{\epsilon}_{it} \boldsymbol{\epsilon}_{jt}|\xt=\vec{x},\ut=\vec{u})| < M_3
% \end{equation*}
\item  $||E(\xt\et')|| = 0$ and $||E(\ut\et')|| = 0$;
\item $u_3 \coloneqq max(||E(||\xt||_{\psit}^{-1}\xt \ut')||, ||E(\fxt \ut')||) = o(K^{1/2})$; 
\item $\{(\xt,\ut,\et)\}_{1 \le t \le T}$ are independent and identically distributed.
\end{enumerate}   
\end{asmp}
% \hl{EGs; corr; cov(ut); xt?}

The first condition is standard in the literature. 
For the second condition, 
when the linear form of \eqref{model_2} is correctly specified, or at least when $\ut$ and $\xt$ are uncorrelated, we have $||E(\ut\xt')|| = 0$ and then the condition is automatically satisfied. 
In general, we do not require $\ut$ and $\xt$ to be uncorrelated and hence leave room for mild misspecification of the linear form. 
To achieve consistency, it suffices to have $K^{-1}u_3^2$ to degenerate. 
Note that $u_3$ depends on both $u_2$, the magnitude of the unexplained part, 
and $||E(\frac{\xt}{||\xt||_{\psit}}  \frac{\ut'}{||\ut||_{\psit}})||$, the strength of correlation  between $\ut$ and $\xt$. 
Under Assumption 2.\textit{\romannumeral2} that $u_2 = O(K^{1/2})$, it suffices to have   $||E(\frac{\xt}{||\xt||_{\psit}}  \frac{\ut'}{||\ut||_{\psit}})|| = o(1)$ and 
$||E(\frac{\fxt}{||\fxt||_{\psit}}  \frac{\ut'}{||\ut||_{\psit}})|| = o(1)$, that is, the correlation degenerates. 
The last condition requires serial independence. 
We admit that this might be restrictive in some applications. 
Our proof relies on several random matrix results, for which the series-dependent versions have not been established in the literature yet. 
However, the idea of approximating latent factors with observed proxies is still applicable, and our numerical results support the proposed method in the presence of serial correlations. 
Similar conditions are also imposed in the literature \citep{fan2016augmented}. 
% the superiority of 

% technical difficulties because
% heavily
% It is generally reasonable to assume $u_3 = O(u_2)$. 
% For the third condition, it is possible to extend our theory to series-dependent cases, 
% while that is  not the main focus of this paper and 
% will add many technical difficulties because our proof relies heavily on the random matrix theory for independent data. 

% Assumption 4.3 (i) requires serial independence, and we admit that it can be restrictive
% in applications. Allowing for serial dependence is technically difficult due to the non-
% smooth Huber’s loss. To obtan the Bahadur representation of the estimated eigenvectors,
% we rely on the symmetrization and contraction theorems (e.g., van der Vaart and Wellner
% (1996)), which requires the data be independently distributed. Nevertheless, the idea of
% using covariates would still be applicable for serial dependent data. For instance, it is not
% difficult to allow for weak serial correlations when the data are not heavy-tailed, by using 
% the sieve least squares estimator Σ (introduced in Section 3.1) in place of the Huber’s estimator Σ��. We conduct numerical studies when the data are serially correlated in the simulations, and find that the proposed methods continue to perform well in the presence of serial correlations.
\subsection{Rates of convergence }\label{rates}

%%%%%%%%%%%%%%%%%%    Theorem 1  %%%%%%%%%%%%%%%%
We are now ready to present the rates of convergence for our estimators. 
Let $(\hA, \{\hft\})$ be the estimates obtained with FMSP. 
For the loading matrix estimation, we have the following result. 
% $(N+p)log(N) = O(T)$
\begin{theorem}[Loadings]\label{rate:loading}
Under Assumptions 1-3, when we choose $\lambda = O(N/\sqrt{T+N})$, there is an invertible matrix $\MH$, such that 
as $T \rightarrow \infty$, and $N$ and $K$ either grow or stay fixed, we have 
% {these p ASMPs should be put where? should be shared by both N=o(T) and; both factors and loadings}
% p = o(T) and p = O(N), where N  can either be fixed or increase with T, 
\begin{equation}\label{theorem_1_1}
        \frac{1}{N}||\hA -\MA\MH||_F^2  = O_p\big(\ratetheoremo\big).
\end{equation}
\end{theorem}

\begin{remark}\label{remart:1}
In Theorem \ref{rate:loading}, we can see that the first term $T^{-1}$ is the unavoidable variance caused by the error term, the second term {$T^{-2\eta}N^{-2}$} is mainly concerned with the number of covariates $p$  relative to $N$ and $T$, the third and the fourth terms are about the relative magnitude of the unexplained part, and the last term describes the impact of misspecification of the linear form. 
When there is no misspecification, the last term automatically disappears. 
\end{remark}

%%%%%%%%%%%%%%%%%%    Theorem 2  %%%%%%%%%%%%%%%%

As for the estimation of latent factors, we have the following rate. 
\begin{theorem}[Factors]\label{rate:factors}
Suppose Assumptions 1-3 hold. 
% For simplicity of expression, we assume $p=O(\sqrt{T}N)$, $u_2^2 = O(1 \wedge  T^{1/2}p^{-1}K)$ and $K^{-1}u_3^2 = O(T^{-1})$. 
By choosing $\lambda = O(N/\sqrt{T+N})$, 
for the same matrix $\MH$ in Theorem 1, 
as $T \rightarrow \infty$, and $N$ and $K$ either grow or stay fixed, we have 
% as $T, N \rightarrow \infty$, and $K$ either grows or stays fixed, 
% we have that 
\begin{equation}\label{theorem_2}
    \frac{1}{K}\oTsumoT ||\hft - \MH^{-1}\ft ||^2 = 
            O_p\big(\ratetheoremt\big). 
\end{equation}
\end{theorem}

% \hl{not strong and clear; try derivations again} 
\begin{remark} 
% The additional conditions are for simplicity of expression. 
% As argue in xxx, they are not strong. 
% The full form is provided in the appendix with similar implications. 
% and not necessary to derive a consistent estimator. 
% Without these conditions, we can still 
Reviewing our proof of Theorem $\ref{rate:factors}$, the  $O_p( KN^{-1})$ term in (\ref{theorem_2})
comes from the unavoidable variance caused by the error term $\et$; 
the $O_p(T^{-2})$ term is caused by the estimation error of $\MA$; 
%  in Theorem $\ref{rate:loading}$
the remaining  terms mainly depend on the dimension of $\xt$, the decay rate of $||\ut||_{\psi_2}$, and the decay rate of the correlation between $\ut$ and $\xt$. 
We require $N$ grows and $K = o(N)$ for the estimator to be consistent. 
The same condition is also assumed in the literature \citep{li2017determining}. 
% dimension, signal, and correlation
\end{remark}

% OP: (X’TX’)X’TX beta. Not consistent
\subsection{Discussions on the convergence rates}
In this section, we compare the convergence rates of FMSP with two  baselines. 
We first review the asymptotic properties of PCA and SPCA. 
According to \cite{stock2002forecasting}, when $K$ is fixed,  for the PCA-based factor model estimator $(\hA_{PCA}, \{\hat{\mathbf{f}}_{t, PCA}\})$, there is a rotation matrix $\MH_{PCA}$ such that 
\begin{equation}\label{rate:PCA}
    \frac{1}{N}||\hA_{PCA} -\MA\MH_{PCA}||_F^2  = O_p\big(\frac{1}{T} + \frac{1}{N}\big),\quad \oTsumoT ||\hat{\vf}_{t,PCA} - \MH_{PCA}^{-1}\ft ||^2 = O_p\big(\frac{1}{T} + \frac{1}{N}\big). 
\end{equation}

For SPCA, recall the model is $\ft = \boldsymbol{g}(\xt^*) + \gammastart$, where $\boldsymbol{g}(\cdot)$ is a nonlinear function and $\gammastart$ is the approximation error. 
Let $\chi_N = ||E\ft\ft' - \cov(\gammastart)||$, which represents the approximation power of $\xt$. 
$\chi_N$ is bounded from above. 
According to \cite{fan2016augmented}, when $K$ is fixed,  for the SPCA-based factor model estimator $(\hA_{SPCA}, \{\hat{\mathbf{f}}_{t, SPCA}\})$, there is a rotation matrix $\MH_{SPCA}$, such that under certain assumptions, when 
$T \rightarrow \infty$ and $N$ is either fixed or diverging, 
it (approximately) holds that 
\begin{equation}\label{rate:SPCA}
    \begin{split}
            \frac{1}{N}||\hA_{SPCA} -\MA\MH_{SPCA}||_F^2  
    &= O_p\big(\frac{1}{T} \chi_N^{-1}\big), \\
     \oTsumoT ||\hat{\vf}_{t,SPCA} - \MH_{SPCA}^{-1}\ft ||^2 
    &= O_p\big(\frac{||\cov(\gammastart)|| + \chi_N^{-1}}{T} + \frac{1}{N} + \frac{log^2 N}{T^2}\big). 
    \end{split}
\end{equation}

% The rate (\ref{theorem_1_1}) mainly depends on the signal strength and dimensions.
We first remark two desired properties of FMSP. 
First, our loading matrix estimator is consistent even when $N$ is fixed, while PCA requires both $N$ and $T$ to go to infinity. 
Such a property is consistent with the exact identification property discussed in Proposition \ref{proposition: exact_identification}. 
Besides, 
different from competing methods such as SPCA, FMSP allows $K$ to grow,  and the estimator is consistent as long as $K = o(N)$. 
% $N$ grows and 
% In the following, we compare when K is fixed
% This property comes from utilizing additional information from $\xt$ to approximate $\ft$, instead of solely relying on $\yt$. 
% Compared with SPCA
% when the model is correctly specified

In terms of convergence rates, 
similar to SPCA, 
the approximation power of $\xt$ on $\ft$ is central to the performance of FMSP. 
Below, we make comparisons under various regimes. 
We first note that, when $K$ is fixed, the convergence rate of our loading matrix estimator and that of the latent factor estimator can be bounded as $O_p(T^{-1} + N^{-2} + {u_2^4T^{-2\eta}} + u_3^2)$ and $O_p(N^{-1} + u_2^2T^{-1} + {u_2^4T^{-2\eta}} + u_3^2)$, respectively. 
Therefore, our estimators converge at least as fast as PCA when $u_2^4 = {O\big(N^{-1}T^{2\eta} + T^{2\eta - 1}\big)}$ and $\mu_3 = O\big(N^{-1/2} + T^{-1/2}\big)$, according to \eqref{rate:PCA}. 

% To see this relationship more clearly

In addition, 
when the approximation error decreases at a faster rate, FMSP can achieve improved convergence rates. 
To formalize this statement, we introduce the following two sets of conditions. 
\begin{condition}[Approximation errors] \label{asmp:approximation}
\hfill 
\begin{enumerate}
    \item $u_2^2 = O(1 \wedge {T^{\eta-1/2}}K)$ and $K^{-1}u_3^2 = O(T^{-1})$. 
    \item $u_2^2 = o(1 \wedge {T^{\eta-1/2}}K)$ and $K^{-1}u_3^2 = o(T^{-1})$. 
\end{enumerate}
\end{condition}
We first focus on discussing Condition \ref{asmp:approximation}.\textit{\romannumeral1}, and discussion for Condition \ref{asmp:approximation}.\textit{\romannumeral2} is similar. 
The condition $u_2^2 = O(1 \wedge {T^{\eta-1/2}}K)$ essentially requires the approximation errors  to diminish at a moderate speed with $p$, relative to $T$ and $K$.  
When $K$ is fixed, under Assumption \ref{asmp:signal}, $u_2 = O(p^{-1/4})$ or $p = O(T^{1/2})$ is sufficient; 
when $K$ is diverging, we provide two sufficient conditions below as examples: 
% ut is mean-zero?
% [label=\arabic*)]
% not increasing is enough. 

% {More on $p = O(T^{1-\eta})$ below! Where should we begin to use $\eta$? $\eta$ is just an upper bound, so why does it help? because see the rate dependency more clearly?
% Or, it is a fixed constant now??
% }

\begin{enumerate}
    \item $\ut$ follows a multivariate normal distribution with $||\cov(\ut)|| = O(Kp^{-1/2} \wedge 1 )$, then $u_2 = ||\ut||_{\psi_2} = ||\cov(\ut)|| = O(Kp^{-1/2} \wedge 1 ) = O(1 \wedge {T^{\eta-1/2}}K)$; 
    \item Let $\ut = (u_{1,t},\dots,u_{K,t})'$ with independent coordinates and $\max_{1 \le i \le K} ||u_{i,t}||_\psit = O(Kp^{-1/2} \wedge 1 )$, 
    then according to Lemma 3.4.2 in \cite{vershynin2018high}, 
    we have $u_2 \le c_1 \max_{i \le K}||u_{i,t}||_\psit$, where $c_1$ is some absolute constant.
\end{enumerate}
For the second condition in Assumption \ref{asmp:approximation}, note that we have $u_3 \le ||E(\frac{\ut}{||\ut||} \frac{\xt^T}{||\xt||} )||  u_2$. 
Therefore, given the decay rate of $u_2$, a mild degeneration rate of the correlation between $\xt$  and $\ut$ would be sufficient to guarantee $u_3^2 = O(K/T)$. 
Again, $u_3$ is zero when $E(\ft|\xt = \vx)$ is a linear function of $\vx$. 
% Roughly speaking, this would hold when the true factors (or their linear transformations) are included in the factor candidates or are certain linear combinations
% $ = O(\sqrt{K/T})$
% some mild additional requirements on the degeneration
% Again, for instance, correctly specified. 
% and holds
% {More on $p = O(T^{1-\eta})$ above!}

Under Condition \ref{asmp:approximation}.\textit{\romannumeral1}, 
when $p=O(\sqrt{T}N)$, 
the convergence rate of the loading matrix estimator \eqref{theorem_1_1} reduces to 
% $KN^{-1}||\hA -\MA\MH||_F^2  = O(T^{-1})$. 
\begin{equation*}%\label{theorem_1_2}
      \frac{1}{N}||\hA -\MA\MH||_F^2  = O(\fot). 
\end{equation*}
Therefore, when $N = o(T)$, we get a faster convergence rate  compared with PCA. 
Moreover, note that based on the classical least square theory, the "oracle" rate one can achieve when the true latent factors $\ft$ are observable is ${N}^{-1}||\hA -\MA\MH||_F^2  = O_p(T^{-1})$.  
This implied that, by incorporating a large number of factor proxies of satisfactory qualities, we can achieve the same rate as if we can observe the true  factors.  
Finally, SPCA achieves the same rate when the explanatory power of $\boldsymbol{g}(\xt^*)$ is strong, in the sense that $\chi_N$ is bounded away from zero. 
Compared with SPCA, 
although the condition for either method essentially requires a desired approximation power of the covariates, by incorporating an increasing number of factor proxies, FMSP is expected to provide additional protection for this condition and hence for the resulting convergence rate. 

As for the estimation of latent factors, to make comparisons with PCA and SPCA, we focus on the case where  $K$ is fixed. 
Then, the rate (\ref{rate:factors}) becomes
\begin{equation}\label{theorem_2_2}
  \oTsumoT ||\hft - \MH^{-1}\ft ||^2 = 
  O_p(
\frac{1}{N} 
+ \frac{1}{T^2}
+ \frac{u_2^2}{T} 
+ {\frac{u_2^4}{T^{2\eta}}}
+ u_3^2
). 
\end{equation}
% Besides, under the assumptions, we have that (\ref{theorem_2_2}) implies $(KT)^{-1} ||\hft - \MH^{-1}\ft ||^2 = O(N^{-1} + T^{-1})$, and hence the estimator converges at least at the same rate as PCA. 
Compared with SPCA, 
the term $O(N^{-1} + T^{-2})$ is shared, and the remaining terms in either (\ref{rate:SPCA}) or   (\ref{theorem_2_2}) essentially depend on the approximation power of the covariates. 
Note that in SPCA, the number of covariates is fixed, hence it is generally reasonable to consider that $||\cov(\gammastart)||$ does not decrease (i.e., the approximation does not improve without more covariates), which (approximately) implies that 
$ N^{-1}||\hA_{SPCA} -\MA\MH_{SPCA}||_F^2  = O_p(T^{-1} + N^{-1})$. 
In this case, under 
Condition \ref{asmp:approximation}.\textit{\romannumeral2},  
our convergence rate is faster than SPCA when 
 $T = o(N)$. 
% , under a slightly stronger version of :   $u_2^2 = o(1 \wedge T^{1/2}p^{-1}K)$ and $K^{-1}u_3^2 = o(T^{-1})$. 
Our convergence rate is also faster than PCA under the same conditions.

To conclude, when the approximation error of $\xt$ on $\ft$ is small, FMSP can achieve improved convergence rates compared with baselines such as PCA under appropriate regimes. 
Compared with SPCA, by incorporating a sufficient number of factor proxies, FMSP provides additional protection for the required regime and hence the desired rates. 
% Finally, it is clear that, for the factor proxy approach, when the manually picked proxies are not equal to the true latent factors (up to a rotation), the estimator will not be consistent. 
% FMSP provides 
% but if the linear form is mis-specified? then also not consistent?

% $O_p( T^{-1}(||\cov(\gammastart)|| + \chi_N^{-1}) )$,
%   O_p(
% \frac{u_2^2}{T} + u_3^2
% + \frac{p^{3/2}}{T^3} 
%     + \frac{p^2u_2^4}{T^2} 
% +  \frac{p^2u_2^2}{T^3}(\sqrt{p} \wedge \frac{N}{T})
% )
% any more insights?

% {
% \begin{remark}
%     In this section, we discuss that our estimators enjoy improved convergence rates compared with baselines,  when the approximation error diminishes at a moderate speed with $p$. 
%     In reality, it is possible that, the approximation error will never be zero even with a huge number of proxies. 
%     % 1. State this limitation 
%     % $\ut$ 
%     % 2. Repeat our arguments
%     As standard in the literature (e.g., \cite{fan2016augmented}), we regard our asymptotic statement as a tool to study the finite-sample performance. 
%     For example, these convergence rates help us understand the interplay between the number of proxies $p$, the magnitude of the approximation error, and the estimation accuracy. 
%     This further facilities our understanding on when the proposed method might have improved accuracy, even with a finite sample.  
%     Alternatively, 
%     a finite-sample error bound that explicitly involves the non-zero approximate error is of interest and this is a meaningful future direction.  % certainly; great
% \end{remark}
% }

\subsection{Extension to accommodate weak factors}
{
In observable factor-based pricing model, it is noticed that some observable factors might be weakly related to the returns and hence the risk premium inference might be invalid \citep{gospodinov2019too}. 
In contrast, our method aims to aggregate information from observable proxies to approximate latent factors, and hence we need no assumption on how strong every single proxy needs to be. 
% : invalid inference with weak observable pricing factors. 
}

{
However, to focus on our main contribution, we have assumed that the latent factors are strong. 
There is empirical evidence \citep{uematsu2022estimation} that suggests that  some latent factors might be weak, in the sense that the number of affected assets scales with rate $N^{\alpha}$ where $\alpha \in (0, 1)$. 
For simplicity, by replacing the condition  $\MA'\MA = NK^{-1} \Vec{I}$ with  $\MA'\MA = N^{\alpha}K^{-1} \Vec{I}$ and closely following our proof, we can similarly derive the following results. 
When $N^{1-\alpha} = o(T)$, we have 
    %%%%%%%%%%%%%%%%%%%%%%%%%%%%%%%%%%%%%%%%%%%%%
    \begin{align*}
        \frac{1}{{N^{\alpha}}}||\hA -\MA\MH||_F^2 &= O_p\Big(
        \frac{1}{T} {N^{1-\alpha}} + 
        \frac{1}{T^{2\eta} N^2} {N^{2-2\alpha}} + 
        \frac{u_2^2}{T} + 
        \frac{u_2^4}{K^2T^{2\eta}} + 
        \frac{u_3^2}{K}  
        \Big),\\ 
    \frac{1}{K}\oTsumoT ||\hft - \MH^{-1}\ft ||^2 
    &=O_p\Big(
    \frac{K}{N} {N^{1-\alpha}}
        + \frac{1}{T^2} {N^{2-2\alpha}}
        + \frac{1}{T^{2\eta} N^2} {N^{2-2\alpha}}
        + \frac{u_2^4}{T^2}
        + \frac{u_2^4}{K^2T^{2\eta}} 
        + \frac{u_3^2}{K} {N^{1-\alpha}}\\
        &\quad\quad\quad+ 
        \frac{{N^{2-2\alpha}}}{TK} \times 
        (\frac{{N^{1-\alpha}}}{T} % +  \frac{u_3^2}{K}
        + \frac{p^2}{T^2} {N^{-2\alpha}} + \frac{p^2u_2^4}{T^2K^2})
        + \frac{u_2^2}{T}{N^{1-\alpha}}
        + \frac{u_3^4}{K^2}
        \Big). 
    \end{align*}
    %%%%%%%%%%%%%%%%%%%%%%%%%%%%%%%%%%%%%%%%%%%%%
Compared with the rates for PCA with weak factors derived in \citet{uematsu2022estimation}, our main advantages still hold, such as that the loading matrix estimator is consistent with a fixed $N$ and that the rates are faster  when the approximation error is relatively low (and $\alpha$ is not too small). 
}

%%%%%%%%%%%%%%%%%%%%%%%%%%%%%%%%%%%%%%%%%%%%%%%%%%%%%%
%%%%%%%%%%%%%%%%%%    Simulation    %%%%%%%%%%%%%%%%%%
%%%%%%%%%%%%%%%%%%%%%%%%%%%%%%%%%%%%%%%%%%%%%%%%%%%%%%
\section{Simulation}\label{sec: simulation}
%%%%%%%%%%%%%%%%%%  Settings  %%%%%%%%%%%%%%%%

% its non-projection variant FMSP-NP (estimate $\ft$ by $\hft = \hB \xt$), 
% several related estimators: 
In this section, we use simulated datasets to compare our proposed method FMSP with the principle component-based estimator (PCA), the smoothed PCA (SPCA)
proposed in \citet{fan2016augmented}, and  factor models with observed proxies (OP). 
Recall that, with $\xt^*$ as a low-dimensional vector of manually picked proxies, OP assumes $\ft = \xt^*$ and runs a regression to estimate $\MA$. 
In addition to these methods, we also would like to investigate the role of our RRR component.  % necessity
Towards this end, we replace the RRR step in FMSP with two other regression methods which are commonly compared with RRR: principal component regression (PCR) and partial least squares (PLS). 
Both of these two estimators utilize the information in $\xt$ and $\yt$, but do not directly utilize the structure of  \eqref{eqn:full_model}. %  in the non-heavy tail settings 
%  in certain ways
Throughout this paper, the tuning parameters for all these methods are chosen by $5$-fold cross-validation.
% , SPCA assumes $\ft = \Vec{h}(\xt^*) + \ut$, where $\Vec{h}$ is a nonlinear function and $\ut$ is the approximation error
For SPCA, following \citet{fan2016augmented}, we use additive polynomial basis as the sieve basis for its non-parametric regressions step and use the least square version. 
% FMSP is not sensitive to lambda. 

We design a general data generation process in a similar way with \citet{fan2016augmented} to investigate the performance of different methods under various conditions: 
\begin{equation*}
    \vec{y}_t = \matr{A}\vec{f}_t +  \et \mbox{   and   } \vec{f}_t = \sqrt{w_2} \mbox{ } c_1 \vec{g}(\MBt \vec{x}_t) + \sqrt{ 1-w_2} \ut,
\end{equation*}
% \frac{1}{\sqrt{3+w_1^2}} \mbox{ }
where $\MA$ is the $N \times K$  loading matrix, $\xt$ is the $p$-dimensional vector of factor proxies, and the other terms are specified below.  
Our parameter specifications largely follow \citet{fan2016augmented}, except for $\MB$ and $\xt$ since we have a large number of factor proxies. 
% We follow \citet{fan2016augmented} to specify the parameters
Specifically, we generate the entries of $\MA$ and $\ut$ independently from $\mathcal{N}(0,1)$. 
The entries of $\et$ are independently drawn from $\mathcal{N}(0, 3K)$. 

The factor proxy part is calibrated from the real dataset that we study in Section \ref{real-data}. 
In that dataset, we have $99$ factor proxies, and hence $p$ is set as $99$ in our simulation study. 
For $\xt$, 
we consider both the independent setting, which is consistent with the theory established in Section \ref{theory}, as well as the serially dependent setting, which further supports the validity of FMSP when serial dependence is present. 
In the independent setting, $\{\xt\}_{t=1}^T$ are generated independently from a mean-zero multivariate normal distribution, the covariance matrix of which is fitted from the real data; 
in the serially dependent setting, $\{\xt\}_{t=1}^T$ are generated from a vector autoregression model of lag $1$, the parameters of which are estimated from the real dataset via the penalized vector autoregression \citep{basu2015regularized}. 
% a real dataset (see Section \ref{real-data} for details). 
% $\xt$ is drawn from the multivariate normal distribution $\mathcal{N}(\Vec{0}, \boldsymbol\Theta)$, where $\boldsymbol\Theta_{jk} = 0.9^{|j - k|}$. 
% \hl{satisfying that assumptions}
% Fama, french

Our specification of the matrix $\MB$ is motivated by the following two points. 
First, to allow a fair comparison with SPCA and OP, we need to make sure there are several factor proxies in $\xt$ yielding  relatively strong approximation power, which is difficult to ensure by fitting the noisy real data. 
Second, as found in the literature \citep{gu2018empirical}, usually only a proportion of the factor proxies are strongly correlated with $\yt$. 
% in generally, in the factor zoo asset returns 
Therefore, we specify $\MB$ in the following manner: 
we first set $\vec{B}_{ii} = w_1 >0$ for $i=1,...,K$, 
then for each column of $\MB$, we randomly choose $s$ other entries to have values independently sampled from $\mathcal{N}(0,1)$, 
and finally set the remaining entries of $\MB$ as zero. 
In our experiment, we set $\xt^*$ as $(\vx_{t1},...,\vx_{tK})'$ for OP and SPCA. 
In the real dataset, these factor proxies correspond to the Fama-French factors. 
The hyper-parameter $w_1$ controls the relative importance of $\xt^*$ in $\xt$. 

We consider two forms of  $\vec{g}(\cdot)$: $\vec{g}(\vec{z}_t) = \vec{z}_t$ and $\vec{g}(\vec{z}_t) = 0.5 (sin(\frac{1}{2}\pi\vec{z}_t) + \vec{z}_t)$. 
The latter one corresponds to the situation where there exists mild model misspecification. 
The constant $c_1$ is chosen such that $c_1 \vec{g}(\MBt \vec{x}_t)$ has unit variance. 
Finally, we introduce the hyperparameter $w_2 \in [0,1]$ to control the ratio between the explained and unexplained parts. 
% percent of 
% total approximation power of $\xt$ on $\ft$, and . 
% These hyper-parameters allow us to investigate the adaptiveness of our method. 
% To address different signal-noise regimes

% Two motivations
% 5 * 6 = 30

%%%%%%%%%%%%%%%%%%  Results  %%%%%%%%%%%%%%%%

% a table of std in the appendix % but not easy for the ratios??

In this section, we present our results under the setting where $N = 100, T= 90$, and $K=5$. 
This is consistent with our real data experiments. 
$s$ is set as 5. 
Additional results for other combinations can be found in the appendix and the findings are similar. 
For each setting, $200$ replications are conducted. 
Since the factors and loadings are only identifiable up to a rotation, 
following \citet{bai2016efficient}, for both the factors and the loadings, 
we compare the performance by the average of the smallest canonical coefficient between the estimated values and the true ones. 
% A canonical coefficient is a value between $0$ and $1$, and 
A larger canonical coefficient implies a more accurate estimation, up to a rotation. 
We report the relative accuracy of the other methods with FMSP as the denominator, so that it is easier the see the trends. 

% \begin{table}[h]
% \centering
% \begin{tabular}{cccccccccccc}
%     \toprule
%     {} & {} & \multicolumn{5}{c}{Factors} & \multicolumn{5}{c}{Factors}\\
%     \cmidrule{1-12}
%     %  \cmidrule{5-5} 
%     $\omega_2$ & $\omega_1$ & PCA & PLS & PCR & SPCA & OP  & PCA  & PLS & PCR   & SPCA  & OP \\
%     \midrule
%     \multirow{2}{*}{0.7} & 1          &  0.69 & 0.56 & 0.979 & 0.639 & 0.201 & 0.612 & 0.511 & 0.982 & 0.66 & 0.403  \\
%     {}& 2          &  0.69 & 0.56 & 0.979 & 0.639 & 0.201 & 0.612 & 0.511 & 0.982 & 0.66 & 0.403  \\
%     \midrule
%     \multirow{2}{*}{0.7} & 1          &  0.69 & 0.56 & 0.979 & 0.639 & 0.201 & 0.612 & 0.511 & 0.982 & 0.66 & 0.403  \\
%     {}& 2          &  0.69 & 0.56 & 0.979 & 0.639 & 0.201 & 0.612 & 0.511 & 0.982 & 0.66 & 0.403  \\
%     \bottomrule
% \end{tabular}
% \end{table}

\begin{table}[h!]
\centering
\caption{Relative values of the smallest canonical correlations with FMSP as the denominator, when $\vec{g}(\vec{z}_t) = \vec{z}_t$ and under the \textit{i.i.d.} setting. 
A value lower than 1 implies the estimation is less accurate than FMSP.}
\begin{tabular}{|c|c|ccccc|ccccc|}
\hline
                      &            & \multicolumn{5}{c|}{Factors} & \multicolumn{5}{c|}{Loading} \\ \hline
$\omega_2$            & $\omega_1$ &  PLS & PCR & PCA & SPCA & OP  & PLS & PCR & PCA & SPCA & OP \\ \hline
\multirow{3}{*}{0.7} 
 &  1  &  0.66 & 0.55 & 1.01 & 0.56 & 0.11 & 0.49 & 0.38 & 1.02 & 0.46 & 0.21 \\ 

 &  2  &  0.38 & 0.33 & 1.00 & 0.63 & 0.20 & 0.29 & 0.23 & 1.01 & 0.54 & 0.36 \\ 

 &  5  &  0.67 & 0.51 & 1.00 & 0.85 & 0.54 & 0.51 & 0.36 & 1.00 & 0.74 & 0.74 \\ 

\hline \multirow{3}{*}{0.95}
 &  1  &  1.21 & 1.00 & 0.66 & 0.53 & 0.22 & 1.01 & 0.80 & 0.72 & 0.50 & 0.37 \\ 

 &  2  &  0.44 & 0.40 & 0.71 & 0.67 & 0.43 & 0.38 & 0.33 & 0.75 & 0.64 & 0.67 \\ 

 &  5  &  0.85 & 0.63 & 0.76 & 1.02 & 0.97 & 0.72 & 0.50 & 0.80 & 0.95 & 1.04 \\ 
  \hline
\end{tabular}\label{tab:non_robust_IID}
\caption{Relative values of the smallest canonical correlations with FMSP as the denominator,  when $\vec{g}(\vec{z}_t) = 0.5 ( sin(\frac{1}{2}\vec{z}_t) +  \vec{z}_t)$ and under the \textit{i.i.d.} setting. A value lower than 1 implies the estimation is less accurate than FMSP. }
\begin{tabular}{|c|c|ccccc|ccccc|}
\hline
                      &            & \multicolumn{5}{c|}{Factors} & \multicolumn{5}{c|}{Loading} \\ \hline
$\omega_2$            & $\omega_1$ & PLS & PCR & PCA & SPCA & OP  & PLS & PCR & PCA & SPCA & OP \\ \hline
\multirow{3}{*}{0.7}   
 &  1  &  0.50 & 0.43 & 1.01 & 0.47 & 0.10 & 0.37 & 0.29 & 1.02 & 0.39 & 0.19 \\ 

 &  2  &  0.38 & 0.32 & 1.01 & 0.52 & 0.16 & 0.27 & 0.22 & 1.02 & 0.44 & 0.28 \\ 

 &  5  &  0.52 & 0.41 & 1.01 & 0.72 & 0.41 & 0.38 & 0.29 & 1.02 & 0.61 & 0.58 \\ 

\hline \multirow{3}{*}{0.95}
 &  1  &  1.54 & 1.31 & 0.81 & 0.67 & 0.34 & 1.19 & 0.97 & 0.89 & 0.63 & 0.50 \\ 

 &  2  &  0.55 & 0.52 & 0.82 & 0.72 & 0.55 & 0.48 & 0.43 & 0.87 & 0.68 & 0.75 \\ 

 &  5  &  0.86 & 0.68 & 0.75 & 1.03 & 1.11 & 0.70 & 0.54 & 0.79 & 0.96 & 1.16 \\ 
                      \hline
\end{tabular}\label{tab:non_robust_sin_IID}
\end{table}

%%%%%%%%%%%%%%%%%%%%%%%%%%%%%%%%%%%%%%%%%%%%%%%%%%%%%%%%%%%%%%%%%%%%%%%%%%%%%%%%%%%%%%%%%%%%%%%%%%%%%%%%%
%%%%%%%%%%%%%%%%%%%%%%%%%%%%%%%%%%%%%%%%%%%%%%%%%%%%%%%%%%%%%%%%%%%%%%%%%%%%%%%%%%%%%%%%%%%%%%%%%%%%%%%%%
\begin{table}[h!]
\centering
\caption{Relative values of the smallest canonical correlations with FMSP as the denominator, when $\vec{g}(\vec{z}_t) = \vec{z}_t$ and under the \textit{serially dependent} setting. 
A value lower than 1 implies the estimation is less accurate than FMSP.}
\begin{tabular}{|c|c|ccccc|ccccc|}
\hline
                      &            & \multicolumn{5}{c|}{Factors} & \multicolumn{5}{c|}{Loading} \\ \hline
$\omega_2$            & $\omega_1$ &  PLS & PCR & PCA & SPCA & OP  & PLS & PCR & PCA & SPCA & OP \\ \hline
\multirow{3}{*}{0.7}  
 &  1  &  0.69 & 0.50 & 1.01 & 0.58 & 0.13 & 0.51 & 0.34 & 1.01 & 0.47 & 0.23 \\ 

 &  2  &  0.50 & 0.39 & 0.99 & 0.68 & 0.23 & 0.37 & 0.27 & 1.00 & 0.57 & 0.39 \\ 

 &  5  &  0.56 & 0.42 & 0.99 & 0.90 & 0.64 & 0.42 & 0.28 & 0.99 & 0.81 & 0.81 \\ 

\hline \multirow{3}{*}{0.95}
 &  1  &  1.00 & 0.69 & 0.77 & 0.56 & 0.18 & 0.82 & 0.53 & 0.82 & 0.50 & 0.32 \\ 

 &  2  &  0.57 & 0.47 & 0.81 & 0.71 & 0.43 & 0.47 & 0.36 & 0.84 & 0.64 & 0.62 \\ 

 &  5  &  0.69 & 0.51 & 0.85 & 1.10 & 1.06 & 0.55 & 0.38 & 0.87 & 1.02 & 1.06 \\ 
\hline
\end{tabular}\label{tab:non_robust_TS}
\caption{Relative values of the smallest canonical correlations with FMSP as the denominator,  when $\vec{g}(\vec{z}_t) = 0.5 ( sin(\frac{1}{2}\vec{z}_t) +  \vec{z}_t)$ and under the \textit{serially dependent} setting. A value lower than 1 implies the estimation is less accurate than FMSP. }
\begin{tabular}{|c|c|ccccc|ccccc|}
\hline
                      &            & \multicolumn{5}{c|}{Factors} & \multicolumn{5}{c|}{Loading} \\ \hline
$\omega_2$            & $\omega_1$ & PLS & PCR & PCA & SPCA & OP  & PLS & PCR & PCA & SPCA & OP \\ \hline
\multirow{3}{*}{0.7} 
 &  1  &  0.51 & 0.41 & 1.01 & 0.43 & 0.12 & 0.38 & 0.26 & 1.01 & 0.35 & 0.21 \\ 

 &  2  &  0.47 & 0.32 & 0.99 & 0.56 & 0.19 & 0.34 & 0.23 & 1.00 & 0.47 & 0.31 \\ 

 &  5  &  0.45 & 0.34 & 1.00 & 0.81 & 0.51 & 0.35 & 0.24 & 1.01 & 0.71 & 0.66 \\ 

\hline \multirow{3}{*}{0.95}
 &  1  &  1.12 & 0.83 & 0.80 & 0.53 & 0.25 & 0.88 & 0.60 & 0.81 & 0.45 & 0.40 \\ 

 &  2  &  0.65 & 0.52 & 0.80 & 0.65 & 0.53 & 0.53 & 0.39 & 0.81 & 0.58 & 0.65 \\ 

 &  5  &  0.67 & 0.54 & 0.85 & 1.20 & 1.23 & 0.53 & 0.41 & 0.88 & 1.07 & 1.20 \\ 
                      \hline
\end{tabular}\label{tab:non_robust_sin_TS}
\end{table}

Results for $\vec{g}(\vec{z}_t) = \vec{z}_t$ and $\vec{g}(\vec{z}_t) = 0.5 (sin(\frac{1}{2}\pi\vec{z}_t) + \vec{z}_t)$ under the i.i.d. setting 
are shown in Table \ref{tab:non_robust_IID}
and \ref{tab:non_robust_sin_IID}, respectively. 
Results under the serially dependent setting are shown in Table \ref{tab:non_robust_TS}
and \ref{tab:non_robust_sin_TS}. 
% shown in Table \ref{tab: light_linear }
% and \ref{tab: light_sin }, respectively. 
% Note that FMSP-NP and FMSP share the same loading matrix estimator and hence yield the same performance on its estimation. 
The proposed approach FMSP shows superior performance under various settings. We summarize our findings as the following aspects: 
\begin{enumerate}
    \item Compared with PCA, when $\xt$ yields good approximation power ($w_2$ is large), the performance of FMSP is significantly better; when $w_2$ is small, the performance is comparable. 
    \item FMSP is more robust than SPCA and OP, which rely on several manually picked proxies $\xt^*$. Only when $\xt^*$ provides an almost perfect approximation to the latent factors ($w_1 = 5$ and $w_2 = 0.95$), their performance becomes comparable. 
    % when $w_2 = 0.95$, 
    \item FMSP shows robustness to mild model misspecification. 
    Comparing Table \ref{tab:non_robust_IID} and \ref{tab:non_robust_sin_IID}, as expected, 
    the performance improvement of FMSP over the other methods slightly decreases under most settings when there exists model misspecification. 
    Nevertheless, FMSP generally still shows the highest accuracy and the findings above continue to hold. 
    \item When there exists serial dependency in the factor proxies, the performance improvement of FMSP over the other methods slightly decreases. However, FMSP generally still outperforms the other methods and shows robustness. 
    \item In general, replacing the RRR step of FMSP with either PLS or PCR significantly affects the accuracy. 
    This is due to that PLS and PCR do not appropriately utilize the structural information between $\xt$ and $\yt$. 
    % \item Compared with FMSP-NP, for factor estimation, FMSP gains more robustness
    % % to both model misspecification (Table \ref{tab: light_sin }) and insufficiency of $\xt$'s approximation power ($w_2$ is low). 
    % Only when the linear combination of observed covariates almost perfectly explains the latent factors ($w_2 = 0.95$ and $\vec{g}(\vec{z}_t) = \vec{z}_t$), there is slight loss of efficiency. 
    % \item Compared with FMSP-NP, for factor estimation, FMSP gains more robustness. 
    % % to both model misspecification (Table \ref{tab: light_sin }) and insufficiency of $\xt$'s approximation power ($w_2$ is low). 
    % Only when the linear combination of observed covariates almost perfectly explains the latent factors ($w_2 = 0.95$ and $\vec{g}(\vec{z}_t) = \vec{z}_t$), there is slight loss of efficiency. 
\end{enumerate}

%%%%%%%%%%%%%%%%%%%%%%%%%%%%%%%%%%%%%%%%%%%%%%%%%%%%%%
%%%%%%%%%%%%%%%%%%    REAL DATA     %%%%%%%%%%%%%%%%%%
%%%%%%%%%%%%%%%%%%%%%%%%%%%%%%%%%%%%%%%%%%%%%%%%%%%%%%

\section{Real Data}\label{real-data}
Factor models have broad applications in finance, including asset pricing, portfolio allocation, and risk assessment. 
In this section, we apply the proposed method to estimate the factor structure of monthly stock returns. 
We first introduce the datasets in Section \ref{sec:dataset}. % we use
The estimation accuracy is evaluated in Section \ref{sec:Estimation accuracy} and an application to optimal portfolio allocation is presented in Section \ref{sec: real_port}.  

% risk
\subsection{Datasets}\label{sec:dataset}
We use the dataset created in \cite{feng2017taming} as our  factor proxies. 
The dataset consists of the monthly values of $99$ proxies from July $1980$ to December $2016$. 
These factor proxies are all proposed in the literature and claimed to have approximation power on the latent factors of stock returns. 
{
The list of proxies contains $17$ publicly available factors, including the Fama-French factors \citep{fama2015five}, the q-factors \citep{hou2015digesting}, the liquidity risk \citep{pastor2003liquidity}, and $82$ long-short factors using firm characteristics  \citep{green2017characteristics}. 
The set of proxies covers many risk sources, including Momentum, Value-versus-Growth, Investment, Profitability, Intangibles, and Trading Frictions. 
}
% We thank Dr. Dacheng Xiu for providing us the dataset.
Among these proxies, we use the Fama-French three \citep{fama1993common} or five \citep{fama2015five} factors as $\xt^*$ for OP and SPCA when $K$ is $3$ or $5$, respectively. % are used
% The Fama-French factors have already been included in the $99$ factor proxies.
We use the $567$ stocks in the U.S. market with less than $5$ missing records from July $1980$ to December $2016$ in the CRSP (The Center for Research in Security Prices) database as our stock universe, 
and the sample mean of each is used to impute the missing records.
The total number of months is $438$. 

{
The datasets have two features that may need attention when applying our estimators.
% and may call for further extensions. 
First, there exists serial dependency. 
For example, based on the partial autocorrelation function, at the significance level of $0.05$, among the $99$ factor proxies, we observe that $23$ have lag $1$, $1$ has lag $2$, and $1$ has lag $3$. 
Higher-order dependence (such as conditional heteroskedasticity) may also be present. 
Our method can still apply to this case, however our theory is established under the i.i.d. case due to the lack of appropriate technical tools. 
We validate our estimators via numerical experiments, and leave the theoretical extension and possible methodology improvement for future study. 
Second, as in most factor model applications, certain homogeneity conditions may not hold. For example, even in a fixed time window, the loadings may not be static. 
Therefore, extensions such as dynamic factor models that aim to handle the heterogeneity would be a meaningful future direction. 
}

% stock2011dynamic

% An issue that needs to be acknowledged more forcefully in the factor model literature, more generally, is that certain homogeneity conditions are unlikely to be satisfied in economic and financial data. 
% For example, the Stock-Watson series that are routinely used in macroeconomics and are a potential application for this paper as well, are very heterogeneous in terms of their dynamic properties (even though they are all transformed to induce stationarity). It would be useful to know the robustness properties of the method with respect to potential data heterogeneity across the factor candidates. 
% A short comment on this would suffice. 

%%%%%%%%%%%%%%%%%%%%%%%%%%%%%%%%%%%%%%%%%%%%%%%%%%%%%%
%%%%%%%%%%%%%%%%%%    REAL DATA 1   %%%%%%%%%%%%%%%%%%
%%%%%%%%%%%%%%%%%%%%%%%%%%%%%%%%%%%%%%%%%%%%%%%%%%%%%%

\subsection{Estimation accuracy}\label{sec:Estimation accuracy}
We evaluate the estimation accuracy based on a rolling-window scheme. 
Let $\yt$ be the vector of stock returns at month $t$ and $\estA$ be the  estimated loading matrix using all available data from month 
$t - 89$ to $t$, for $t = 90, \dots, 426$. 
Following \cite{wang2019factor}, we evaluate the accuracy of an estimated factor model by the out-of-sample residual sum of squares for stock returns from $t+1$ to $t+12$, defined as
\begin{equation*}
SSE_t = \sum_{i=1}^{12} ||\vec{y}_{t+i}-\frac{K}{N}\hat{\MA}_t\hat{\MA}_t'\vec{y}_{t+i}||_2^2. 
\end{equation*}

% more discussion here
% To illustrate the necessity of including a sufficient number of candidates, we randomly choose $10$, $20$ and $30$ candidates from the $99$ ones and use them in the proposed estimation procedure. 
% For fairness, the Fama-French five factors are always included. 
% The resulting estimators are denoted as FMSP-$10$, FMSP-$20$, and FMSP-$30$, respectively. 
% FMSP with the whole set of $99$ candidates is denoted as FMSP. 
Besides conducting experiments with all the $567$ stocks, we also randomly sample $N = 50$ and $300$ stocks from the universe and repeat the same experiment, to investigate the effect of the dimension $N$. 
The average of $SSE_t$  from $t = 90$ to $426$ is reported in Table \ref{real: loading}, for $K = 3, 5,$ and $7$. 
In addition, we test the statistical significance of the differences in prediction accuracy between FMSP and the other methods by applying the Diebold-Mariano test \citep{diebold2002comparing, harvey1997testing} to the hypotheses 

\begin{equation}\label{formula: test}
    H_0: \mathbb{E} \big(SSE^0_t - SSE^1_t\big) \ge 0  
    \;\;\;\;\; \text{v.s.} \;\;\;\;\;
    H_1: \mathbb{E} \big(SSE^0_t - SSE^1_t\big) < 0, 
\end{equation}
where $SSE^0_t$ and $SSE^1_t$ correspond to the residual sum of squares of FMSP and an alternative method, respectively. 

% Note that this evaluation metric mainly focuses on the . loading matrix estimator
In all situations, FMSP shows higher accuracy than the competing methods, 
and in most cases, this finding is statistically significant. 
The results demonstrate the efficiency and robustness of FMSP in loading matrix estimation. 
Consistent with our theoretical results in Section \ref{theory}, we notice the advantage of FMSP, in terms of the loading matrix estimation, is more significant when $N$ is moderate. 
We can also find that, consistent with our simulation results, replacing the RRR step of FMSP with either PLS or PCR harms the accuracy. 
Finally, when $K = 7$, there is no standard choice of factor proxies and hence we do not evaluate OP and SPCA. 
In contrast, FMSP naturally allows $K$ to grow.
% The relative SSEs of different methods with FMSP as the benchmark are 
% With $N$ increasing, the relative performance of OP becomes worse, which implies the need for  more complex factors to explain the larger set of stocks. 
% and avoids this issue. 
% When the number of candidates $p$ grows, the performance of FMSP gradually improves, which supports our motivation of incorporating a sufficient number of candidates. 
% When K increases from 3 to 5; but the following story may not hold

\vspace{.2cm}

%%%%%%%%%%%%%%%%%%%%%%%%%%%%%%%%%%%%%%%%%%%%%%%%%%%%%%%%%%%%%%%%%%%%%%%%%%%%%%%%%%%%%%%%%%%%%%%%%%%%%%%%%
%%%%%%%%%%%%%%%%%%%%%%%%%%%%%%%%%%%%%%%%%%%%%%%%%%%%%%%%%%%%%%%%%%%%%%%%%%%%%%%%%%%%%%%%%%%%%%%%%%%%%%%%%
% and their standard errors (in parentheses) 
% relative SSE with FMSP-$99$ as the denominator. 
% A value lower than 1 implies the estimation is less accurate than FMSP-99. 
% Larger than 1 means worse than FMSP-99. 
\begin{table}[h!]
\centering
\caption{ Estimation accuracy: 
average out-of-sample SSEs for different factor model estimators. 
No experiment results for OP and SPCA when $K = 7$ due to the lack of a standard choice of proxies. 
SST stands for the total sum of squares for the evaluation periods. 
Bold indicates the best. 
The symbols $^{\ast\ast\ast}$, $^{\ast\ast}$, and $^{\ast}$ indicate statistical significance for testing \eqref{formula: test} at levels $0.01$, $0.05$, and $0.1$, respectively. 
}
\label{real: loading}
\begin{tabular}{|l|lll|lll|lll|} % if c, then the numbers can not be aligned
\hline
N & \multicolumn{3}{c|}{50} & \multicolumn{3}{c|}{300} & \multicolumn{3}{c|}{567}\\ \hline
K   & 3 & 5  & 7 & 3 & 5  & 7 & 3 & 5  & 7 \\ \hline

SST & $6.50^{\ast\ast\ast}$ & $6.50^{\ast\ast\ast}$ & $6.50^{\ast\ast\ast}$ & $36.41^{\ast\ast\ast}$ & $36.41^{\ast\ast\ast}$ & $36.41^{\ast\ast\ast}$ & $67.71^{\ast\ast\ast}$ & $67.71^{\ast\ast\ast}$ & $67.71^{\ast\ast\ast}$\\

PCA & $4.98^{\ast\ast\ast}$ & $4.26$ & $3.83^{\ast\ast}$ & $31.28^{\ast\ast\ast}$ & $30.13^{\ast\ast\ast}$ & $29.46^{\ast\ast\ast}$ & $58.77^{\ast\ast\ast}$ & $57.30^{\ast\ast\ast}$ & $56.29^{\ast\ast\ast}$\\

OP & $5.22^{\ast}$ & $4.74^{\ast\ast}$ & $\backslash$ & $30.87^{\ast\ast\ast}$ & $30.17^{\ast\ast\ast}$ & $\backslash$ & $58.39^{\ast\ast\ast}$ & $57.39^{\ast\ast\ast}$ & $\backslash$\\

% SPCA & $5.03^{\ast}$ & $4.45^{\ast\ast\ast}$ & $\backslash$ & $30.81^{\ast\ast}$ & $29.93^{\ast\ast\ast}$ & $\backslash$ & $58.33^{\ast\ast\ast}$ & $57.07^{\ast\ast\ast}$ & $\backslash$\\

SPCA & $5.12^{\ast}$ & $4.45^{\ast\ast\ast}$ & $\backslash$ & $30.94^{\ast\ast\ast}$ & $30.13^{\ast\ast\ast}$ & $\backslash$ & $58.56^{\ast\ast\ast}$ & $57.38^{\ast\ast\ast}$ & $\backslash$\\

PCR & $5.26^{\ast}$ & $4.80^{\ast\ast}$ & $4.36^{\ast\ast}$ & $31.46^{\ast\ast\ast}$ & $30.33^{\ast\ast\ast}$ & $29.61^{\ast\ast\ast}$ & $59.44^{\ast\ast\ast}$ & $57.75^{\ast\ast\ast}$ & $56.63^{\ast\ast\ast}$\\

PLS & $5.14$ & $4.72^{\ast\ast}$ & $4.23^{\ast\ast}$ & $30.88^{\ast\ast\ast}$ & $30.00^{\ast\ast\ast}$ & $29.39^{\ast\ast\ast}$ & $58.47^{\ast\ast\ast}$ & $57.10^{\ast\ast\ast}$ & $56.20^{\ast\ast\ast}$\\

FMSP & \textbf{4.81} & \textbf{4.24} & \textbf{3.74} & \textbf{30.57} & \textbf{29.67} & \textbf{29.02} & \textbf{57.84} & \textbf{56.50} & \textbf{55.66}\\

\hline
\end{tabular}
\end{table}

% [['***', '**', 0, 0, 0, 0],
%  ['***', 0, '*', '*', '.', '.'],
%  ['***', '*', 0, 0, '.', '.'],
%  ['**', '***', '*', '.', '***', '*'],
%  ['***', '***', '**', '***', '**', '**'],
%  ['***', '***', 0, 0, '***', '***'],
%  ['***', '***', '***', '***', '***', '**'],
%  ['***', '***', '***', '***', '**', '***'],
%  ['***', '***', 0, 0, '**', '**']]

%%%%%%%%%%%%%%%%%%%%%%%%%%%%%%%%%%%%%%%%%%%%%%%%%%%%%%
%%%%%%%%%%%%%%%%%%    REAL DATA 2   %%%%%%%%%%%%%%%%%%
%%%%%%%%%%%%%%%%%%%%%%%%%%%%%%%%%%%%%%%%%%%%%%%%%%%%%%

\subsection{Optimal portfolio allocation}\label{sec: real_port}
% with observed factor proxies 
% not a convincing story?
% how to tell such a story?

\newcommand{\vw}{\vec{w}}
\newcommand{\hS}{\vec{\widehat{\Sigma}}}
\newcommand{\COV}{\vec{\Sigma}}

One important application of factor models is portfolio optimization. 
For a set of $N$ stocks, we denote the covariance matrix of their returns as $\COV$. 
The global minimum variance portfolio (GMV) $\vw_{GMV}$ is defined as 
\begin{equation}\label{formula: GMV}
    \vw_{GMV} = \argmin_{\vw} \vec{w}'\COV\vec{w}, \; \text{s.t.} \; \vw'\Vec{1} = 1.
\end{equation}
% $\sqrt{\vec{w}'\COV\vec{w}}$
The GMV yields the lowest risk  among all portfolios and it is widely used in portfolio management because of its good finite-sample performance \citep{DING2020}. %cite???
To estimate the GMV with data, we need to replace $\COV$ in (\ref{formula: GMV}) with a covariance matrix estimator, denote as $\hS$. 
As widely acknowledged \citep{fan2012vast, DING2020}, when $N$ is relatively large, the sample covariance matrix leads to poor portfolio performance due to error accumulation. 
Instead, the factor model-based covariance matrix estimator is popular. 
Suppose we have an estimated factor model ($\hat{\MA}$, $\{\hft\}$) and denote the sample covariance matrix of the residuals $\{\yt - \hA\hft\}$ as $\widehat{\vec{\Sigma}}_\epsilon$. 
Existing methods \citep{fan2013large, fan2011high} first threshold $\widehat{\vec{\Sigma}}_\epsilon$ in a certain way with some threshold hyperparameter $\eta$ to obtain $\widehat{\vec{\Sigma}}_{\epsilon,\eta}$, and then construct the covariance matrix estimator as $\hCOV_\eta = \hA \widehat{\cov}(\ft) \hA' + \widehat{\vec{\Sigma}}_{\epsilon,\eta}$ (refer to page 295, \cite{fan2017elements} for the detailed description). 
Under appropriate sparsity and low-rank assumptions, $\hCOV_\eta$ has been shown to be consistent even when $N \asymp T$  \citep{fan2013large, fan2011high}. 
The superior performance of plugging in such a covariance estimator in GMV estimation has been illustrated in \cite{fan2013large}, and in general, the performance should rely on the quality of the estimated factor structures. 
In this section, we show that, by capturing the factor structure more accurately and hence getting better covariance matrix estimation, FMSP can improve the performance of the estimated GMVs in terms of yielding lower out-of-sample variances. 
% require there is no omitted factor structure in the residual part, [theory?]
% which has been challenged by some empirical work \citep{kleibergen2015unexplained, lettau2001resurrecting}. 

% with a given covariance matrix estimator, 
We construct portfolios based on a rolling-window scheme. 
At the beginning of month $t$, for $t=91, \dots, 438$, we use the data of the $90$ preceding months to estimate a covariance matrix,  and then calculate the estimated GMV as $\widehat{\vw}_{GMV, t}$. 
% Following \citet{bodnar2018estimation} and \citet{demiguel2009generalized}, 
We measure the performance of the estimated GMVs by the out-of-sample variance 
\begin{equation*}
    \hat{\sigma}^2 = \frac{1}{347} 
    \sum_{t = 91}^{438} (\widehat{\vw}_{GMV, t}' \Vec{y}_{t} - \hat{\mu})^2, 
\end{equation*}
where $\hat{\mu} = \frac{1}{347} \sum_{t=91}^{438} \widehat{\vw}_{GMV, t}' \vec{y}_{t}$ is the sample mean. 

% how to say it is better?
In addition to the sample covariance matrix  $\Vec{S}$, 
we also consider the factor model-based covariance matrix estimators with the factor structure estimated via methods introduced in Section \ref{sec: simulation}. 
Specifically, we consider OP, PCA, SPCA, PLS, PCR, and FMSP. 
% We consider the following covariance matrix estimators: 
% the sample covariance matrix  $\Vec{S}$, 
% OP-based estimator $\vec{\widehat{\Sigma}}_{OP, \eta}$,  
% PCA-based estimator $\vec{\widehat{\Sigma}}_{PCA, \eta}$, 
% PLS-based estimator $\vec{\widehat{\Sigma}}_{PLS, \eta}$, 
% PCR-based estimator $\vec{\widehat{\Sigma}}_{PCR, \eta}$, 
% SPCA-based estimator $\vec{\widehat{\Sigma}}_{SPCA, \eta}$, 
% and FMSP-based estimator $\vec{\widehat{\Sigma}}_{FMSP, \eta}$. 
% Except for the sample covariance matrix, the other estimators are all factor model-based covariance matrix estimators with threshold $\eta$ (refer to page 295, \cite{fan2017elements} for the detailed description), where the factor models are estimated with corresponding methods as described in Section \ref{sec: simulation}. 
The threshold $\eta$ is tuned separately for the best performance of each estimator. 
% No point to use absolute 
% growing k?
As a baseline, we also compare with the performance of the equally weighted portfolio (EW) $\widehat{\vw}_{EW, t} = N^{-1} \textbf{1}$. 
% of these portfolios with (i) the equally weighted portfolio (EW), and (ii) the GMV estimator with gross exposure constraints proposed in \citet{fan2012vast} with gross exposure constraint $c=2$ (GEC). 
% and $c=1$ (i.e., no short exposure, GEC-2). 

% relative
In Table \ref{tab:GMV}, we report the out-of-sample variances of GMVs estimated using different covariance matrix estimators, with various combinations of $N$ and $K$. 
As expected, the sample covariance matrix yields the worst performance due to the error accumulation and the singularity of the covariance estimator. 
Consistent with the literature \citep{DING2020}, we find the factor model-based GMV estimators generally perform better. 
% than the plug-in estimator and also EW. 
Among the factor model-based GMV estimators, 
FMSP performs favorably in most cases. % compared with the other methods
Compared with PCA, we again observe the advantage of FMSP is particularly significant when $N$ is moderate,  which is also consistent with our theoretical results in Section \ref{theory}. 
As $K$ grows, we can observe a lower out-of-sample variance for most estimators (especially when $N$ is larger), which indicates the need for a larger number of factors to explain the more complex factor structure. 
Besides, we can also find that the relative advantage of FMSP is more significant when $K$ is larger, which can be interpreted as the ability of FMSP to approximate more complex latent factors. 
In contrast, OP and SPCA rely on several manually picked proxies, and they both show larger out-of-sample variances in these cases. 

In Table \ref{tab:GMV_diff_p}, we further demonstrate that, naively applying SPCA with a large number of covariates makes the performance deteriorate, as it is not designed for the high-dimensional setting. 
Specifically, we rerun the experiment under the same settings as above, but with an increasing number of covariates, from $K$ ($3$ or $5$) to $99$. 
For fairness, the Fama-French factors are always included, and the others are randomly chosen. 
As expected, when $p$ is large, SPCA over-fits the training data and hence the fitted values $\hat{\textbf{y}}_t$ degenerates to $\yt$, which makes the performance of SPCA degenerate to that of the vanilla PCA. 
In contrast, when $p = K$, the performance of FMSP is close to that of OP as expected, since both estimators rely on linear transformation of the observed factors; 
while when $p$ increases, FMSP demonstrates the ability of utilizing additional approximation information contained in the pool of factor proxies. 
In general, except when $p = K$, FMSP performs better than SPCA. 

% With $N$ growing, the relative advantage of FMSP increases, which we interpret as the need for more complex factors to explain the larger set of stocks as well as the availability of more data to estimate the relatively more complex model. 

% ,  and the risks of the resulting portofolios decrease 
% When $N$ is small ($N = 100$), OP performs better. 
% We interpret this as that the data is not enough for FMSP to stably learn the latent factors and the Fama-French factors are decent proxies. 
% The trend with $p$ supports our motivation of incorporating a sufficient number of candidates again. 
% When the number of candidates $p$ grows, the performance of FMSP gradually improves, which supports our motivation of incorporating a sufficient number of candidates. 
% But trend w.r.t. K is not clear?
% When $N=100$, the sample size may not be large enough for FMSP to learn factor models that are more accurate than those estimated using Fama-French factors.  

%%%%%%%%%%%%%%%%%%%%%%%%%%%%%%%%%%%%%%%%%%%%%%%%%%%%%%%%%%%%%%%%%%%%%%%%%%%%%%%%%%%%%%%%%%%%%%%%%%%%%%%%%
%%%%%%%%%%%%%%%%%%%%%%%%%%%%%%%%%%%%%%%%%%%%%%%%%%%%%%%%%%%%%%%%%%%%%%%%%%%%%%%%%%%%%%%%%%%%%%%%%%%%%%%%%
\begin{table}[h!]
\centering
\caption{The out-of-sample variance for returns (in $\%$) of GMVs estimated with different  estimators. % covariance
No experiment results for OP and SPCA with $K=7$ due to 
the lack of a standard choice of proxies. 
% the challenge of selecting appropriate proxies. 
The variance of the sample covariance matrix based estimator is higher than the others for orders of magnitude. Bold indicates the best.}
\begin{tabular}{|c|ccc|ccc|ccc|}
\hline
N & \multicolumn{3}{c|}{50} & \multicolumn{3}{c|}{300} & \multicolumn{3}{c|}{567}\\ \hline
K   & 3 & 5  & 7 & 3 & 5  & 7 & 3 & 5  & 7 \\ \hline

$S(\times 10^3$) & 0.01 & 0.01 & 0.01 & 20.09 & 20.09 & 20.09 & 18.45 & 18.45 & 18.45 \\ 

EW & 19.00 & 19.00 & 19.00 & 17.52 & 17.52 & 17.52 & 17.54 & 17.54 & 17.54 \\ 

PCA & 7.88 & 7.88 & 7.78 & 5.14 & 4.92 & 4.84 & 5.45 & 4.77 & 4.48 \\ 

OP & 5.81 & 5.72 & $\backslash$  & 5.11 & 4.81 & $\backslash$  & 5.18 & 4.80 & $\backslash$  \\ 

PCR & 6.78 & 6.51 & 6.39 & 7.10 & 6.60 & 6.40 & 7.36 & 6.94 & 6.76 \\ 

PLS & 6.35 & 6.25 & 6.17 & 6.42 & 6.11 & 6.00 & 6.72 & 6.49 & 6.42 \\ 

% SPCA & 5.84 & 6.00 & $\backslash$  & 5.18 & 5.00 & $\backslash$  & \textbf{5.13} & 4.91 & $\backslash$  \\ 
SPCA & 5.75 & 5.97 & $\backslash$  & 5.14 & 5.02 & $\backslash$  & \textbf{4.97} & 4.96 & $\backslash$  \\ 

FMSP & \textbf{5.54} & \textbf{5.69} & \textbf{5.89} & \textbf{5.04} & \textbf{4.66} & \textbf{4.41} & 5.14 & \textbf{4.59} & \textbf{4.12} \\ 
\hline
\end{tabular}\label{tab:GMV}
\end{table}

%%%%%%%%%%%%%%%%%%%%%%%%%%%%%%%%%%%%%%%%%%%%%%%%%%%%%%%%%%%%%%%%%%%%%%%%%%%%%%%%%%%%%%%%%%%%%%%%%%%%%%%%%
%%%%%%%%%%%%%%%%%%%%%%%%%%%%%%%%%%%%%%%%%%%%%%%%%%%%%%%%%%%%%%%%%%%%%%%%%%%%%%%%%%%%%%%%%%%%%%%%%%%%%%%%%
\begin{table}[h!]
\centering
\caption{The out-of-sample variance of returns (in $\%$) of GMVs estimated with different  covariance estimators. 
We focus on comparing FMSP and SPCA with different number of covariates. 
The lower the better. 
}
\begin{tabular}{|c|cccc|cccc|c|c|}
\hline
%  & \multicolumn{3}{c|}{567}
% & 3 & 5  & 7 
{Estimator} & \multicolumn{4}{c|}{SPCA} & \multicolumn{4}{c|}{FMSP} & PCA & OP \\ \hline
p   & K & 10 & 20  & 99 & K  & 10 & 20  & 99 & $\backslash$  & $\backslash$  \\ \hline

$N = 50, K = 3$ & 5.75 &	6.12 &	6.45 &	6.45 & 5.89 &	5.86 &	5.71 & 5.54 & 6.45 & 5.81 \\ 

$N = 50, K = 5$ & 5.97 &	6.01 &	6.37 &	6.37 & 5.78 &	5.73 &	5.58 &	5.69& 6.37 & 5.73\\ 
$N = 300, K = 3$ & 5.14&	5.17&	5.44 &	5.44	 & 5.19&  	4.80&	4.78 &	5.04& 5.44 & 5.10\\ 
$N = 300, K = 5$ & 5.02&	4.87&	5.20 &	5.13	 & 4.81&	3.92&	4.01 &	4.66 & 5.13 & 4.81
\\ 

\hline
\end{tabular}\label{tab:GMV_diff_p}
\end{table}

\section{Extension to Heavy-tailed Data}\label{sec:robust}
In some applications of factor models, for example, analysis of daily stock returns, the errors may exhibit heavy tails and hence non-robust procedures may fail  \citep{calzolari2018estimating}. 
In this section, we extend FMSP to accommodate these scenarios by proposing a novel penalized robust RRR method as a subroutine. 
% we further propose a novel penalized robust RRR method as a subroutine to extend FMSP to the heavy-tailed situations.  

%%%%%%%%%%%%%%%%%%%%%%%%%%%%%%%%%%%%%%%%%%%%%%%%%%%%%%
%%%%%%%%%%%%%%%%%%    Robust    %%%%%%%%%%%%%%%%
%%%%%%%%%%%%%%%%%%%%%%%%%%%%%%%%%%%%%%%%%%%%%%%%%%%%%%
\subsection{Robust factor model estimation}\label{sec: robust FMSP}
% consistency? this method? 
% In some applications of factor models, for example, analysis of daily stock returns, the errors may exhibit heavy tails, where 
% Recall that we use the least square loss in  \eqref{general optimization equation}. 
For heavy-tailed data, the least square loss used in  \eqref{general optimization equation} is not appropriate and a more robust loss function is needed. 
Some common choices include the Huber loss \citep{huber1992robust} and Tukey's Biweight loss \citep{hoaglin2000understanding}. 
However, as discussed in \cite{tan2018distributionally} and \cite{she2017robust}, it is computationally infeasible to directly apply these robust loss functions in RRR due to the non-convexity caused by the low-rank constraint. 
\cite{tan2018distributionally}  proposes to solve a convex relaxation with the nuclear norm penalty, which is only an approximate solution, is still computationally expensive, and loses the explicit control of the number of factors. 

% $K$.
% \textbf{Some explanations here: System} % More reason? why?% statistics
% why important? 

Instead, inspired by recent developments on robust structural learning \citep{fan2016shrinkage, zhu2017taming}, 
we propose a simple robust RRR method to use as a subroutine in our factor model estimation procedure: we first shrink the raw data adaptively to handle the heavy tails, and then apply square loss on the shrunk data. 
Specifically, with a given hyperparameter $\tau$, we define the shrinkage factor  $\taot = (||\yt||\wedge \tau)/||\yt||$, and then replace $\yt$ and  $\xt$ in the step $1$ with $\Tyt = \taot \yt$ and $\Txt = \taot \xt$, respectively. 
This approach is a modification of the method proposed in  \cite{fan2016shrinkage} to deal with the factor model structure in our setting.
% Alternatively, t
This robust RRR estimator can also be understood as the minimizer of the weighted square loss
%  which downweights 
% with applying the same shrinkage on $\xt$ to reduce bias.
% why shrink in this way? not confident... so sad.
%  \mathcal{L}_\tau(\yt, \MA\MBt\xt) = 
\begin{equation*}
   ||\Tyt - \MA\MBt\Txt||^2 = 
    (\frac{||\yt||\wedge \tau}{||\yt||})^2
    ||\yt - \MA\MBt\xt||^2.
    %\tau_t^2
\end{equation*}

% after modification, the results improved
% our shrinkage is a little different with that in \cite{fan2016shrinkage} on that we apply the same shrinkage $\taot$ on $\xot$. 
% This is because under the low-rank setting, only shrinking $\yt$ will results in non-degenerate bias, for example, when the large $||\yt||$ is caused by large $||\ft||$.
% Our modified method actually extends the applicability and theories of  \cite{fan2016shrinkage}. 

% Lu: It is not clear how this results are related to the proposed estimation method? Or it is just a general result for heavy tail data.
Some additional insights about how the shrinkage mitigates the heavy-tailed issue in factor model estimation can be gained from the following proposition, which implies that the adaptive shrinkage can force the idiosyncratic errors to behave like sub-Gaussian. 
\begin{proposition}\label{lemma: shrinkage}
Suppose the eigenvalues of $N^{-1}K\MA' \MA$ are bounded away from zero and infinity, 
and that $||\ft||_{\psi_2} \le M'_0$ for some $M'_0 >0$. 
% $, $M_2
Then when we choose $\tau \asymp \sqrt{N}$, there is a positive constant $M_{\epsilon}$ depending on N, such that $\taut \et$ is a sub-Gaussian vector:
\begin{equation*}
    ||\taot \et||_{\psi_2} \le M_{\epsilon}.
\end{equation*}
\end{proposition}

% while handling the heavy-tailness. 
% More explanation. 
% in Section \ref{method:non-robust}

% although it is a sub-Gaussian, but norm scales, because of the introduced bias? 
% can handle heavy-tail? but consistency?
To construct a robust factor model estimator, 
we simply replace step 1 of FMSP with the proposed robust RRR method, and keep  step 2 the same. 
Unlike using other robust RRR estimators, with this estimator as a subroutine, the factor model estimators can still be explicitly solved, in the same manner as in  Proposition 2. 
We refer to this variant as FMSP-R. 

%%%%%%%%%%%%%%%%% Heavy %%%%%%%%%%%%%%%%% 

\subsection{Simulation results}\label{simu_robust}
The theoretical property of FMSP-R relies on the novel robust RRR method. 
In the standard RRR literature, it is commonly assumed that the error term has independent entries \citep{chen2012sparse, she2017robust, tan2018distributionally}, 
which may not hold under the general factor model setting considered in this paper. 
This poses additional technical challenges to our proof. 
We leave the theoretical analysis of FMSP-R for future investigation and conduct simulation studies in this section to support its effectiveness.

% where $\MTh$ is a low-rank coefficient matrix and $\tilde{\boldsymbol{\epsilon}}_t$ is the error term which is typically assumed to have independent entries. 
% Our problem is different with the standard RRR as we allow $\tilde{\boldsymbol{\epsilon}}_t$ to have correlated entries or even some factor structures, and the regression is only used as an intermediate step to facilitate the factor model estimation. 
% 
% Unlike the typically assumed 
% The theoretical analysis is challenging  
% Next, we numerically examine FMSP-R, the robust extension proposed in Section \ref{sec: robust FMSP}.  
Specifically, we repeat the experiments in Section \ref{sec: simulation}, except that $\et$ is now generated from the re-scaled log-normal distribution $c_1{exp(1+1.2\vec{\xi})-c_2}$, where $\vec{\xi}$ follows $N(0,1)$ and $c_1$ and $c_2$ are constants such that each entry of $\et$ has mean $0$ and variance $3K$. 
The robust version of SPCA proposed in \cite{fan2016augmented} is used for comparison.  
% The robust version of FMSP-NP is denoted as FMSP-RNP.
% as introduced in Section \ref{sec: simulation}. 
Results under the serially dependent setting are presented in Tables \ref{tab: simu_heavy_TS} and \ref{tab: simu_heavy_mis_TS}. 
Results for the i.i.d. setting are deferred to the appendix and the findings are similar. 
Compared with the competing methods, FMSP-R still generally outperforms. 
In particular, compared with Tables \ref{tab:non_robust_TS}
 and \ref{tab:non_robust_sin_TS}, the relative performance of the non-robust baselines (PCA and OP) deteriorates significantly. 
The simulation results support the usefulness of the proposed robust extension. 
% In particular, the improvements over PCA is even more significant than those in . 
The other findings are similar to Section \ref{sec: simulation}. 

\begin{table}[h!]
\centering
\caption{ Relative values of the smallest canonical correlations with FMSP-R as the denominator,  when the noises are heavy-tailed, $\vec{g}(\vec{z}_t) = \vec{z}_t$, and under the serially dependent setting. A value below 1 implies the estimation is less accurate than FMSP-R. }
\begin{tabular}{|c|c|ccccc|ccccc|}
\hline
                      &            & \multicolumn{5}{c|}{Factors} & \multicolumn{5}{c|}{Loading} \\ \hline
$\omega_2$            & $\omega_1$ & PLS & PCR & PCA & SPCA & OP  & PLS & PCR & PCA & SPCA & OP \\ \hline
\multirow{3}{*}{0.7}   
 &  1  &  0.73 & 0.53 & 0.21 & 0.69 & 0.12 & 0.54 & 0.36 & 0.20 & 0.60 & 0.22 \\ 

 &  2  &  0.54 & 0.42 & 0.23 & 0.83 & 0.23 & 0.40 & 0.29 & 0.22 & 0.74 & 0.38 \\ 

 &  5  &  0.62 & 0.45 & 0.25 & 1.05 & 0.64 & 0.47 & 0.30 & 0.24 & 1.08 & 0.77 \\ 

\hline \multirow{3}{*}{0.95}
 &  1  &  1.08 & 0.76 & 0.18 & 0.74 & 0.18 & 0.90 & 0.58 & 0.16 & 0.74 & 0.31 \\ 

 &  2  &  0.62 & 0.51 & 0.19 & 1.00 & 0.44 & 0.51 & 0.39 & 0.20 & 1.02 & 0.60 \\ 

 &  5  &  0.76 & 0.55 & 0.18 & 1.17 & 1.06 & 0.62 & 0.40 & 0.17 & 1.34 & 1.01 \\ 
                      \hline
\end{tabular}\label{tab: simu_heavy_TS}

\caption{ Relative values of the smallest canonical correlations with FMSP-R as the denominator,   when the noises are heavy-tailed, $\vec{g}(\vec{z}_t) = 0.5 ( sin(\frac{1}{2}\vec{z}_t) +  \vec{z}_t)$, and under the serially dependent setting. A value below 1 implies the estimation is less accurate than FMSP-R. }
\begin{tabular}{|c|c|ccccc|ccccc|}
\hline
                      &            & \multicolumn{5}{c|}{Factors} & \multicolumn{5}{c|}{Loading} \\ \hline
$\omega_2$            & $\omega_1$ & PLS & PCR & PCA & SPCA & OP  & PLS & PCR & PCA & SPCA & OP \\ \hline
\multirow{3}{*}{0.7}   
 &  1  &  0.53 & 0.43 & 0.20 & 0.57 & 0.11 & 0.38 & 0.27 & 0.18 & 0.50 & 0.19 \\ 

 &  2  &  0.46 & 0.35 & 0.20 & 0.71 & 0.18 & 0.32 & 0.24 & 0.19 & 0.64 & 0.29 \\ 

 &  5  &  0.50 & 0.35 & 0.22 & 0.96 & 0.48 & 0.38 & 0.24 & 0.20 & 0.97 & 0.61 \\ 

\hline \multirow{3}{*}{0.95}
 &  1  &  1.06 & 0.78 & 0.19 & 0.65 & 0.20 & 0.83 & 0.57 & 0.18 & 0.67 & 0.31 \\ 

 &  2  &  0.65 & 0.53 & 0.18 & 0.87 & 0.46 & 0.50 & 0.39 & 0.20 & 0.92 & 0.57 \\ 

 &  5  &  0.69 & 0.55 & 0.18 & 1.27 & 1.08 & 0.57 & 0.42 & 0.19 & 1.50 & 1.04 \\ 
                      \hline
\end{tabular}\label{tab: simu_heavy_mis_TS}

\end{table}

% \input{tables/tab3.tex}

% ||\tilde{X}BA'||_F^2+tr(\tilde{Y}'\tilde{X}BA') + ||Y||_F^2

% In this loss function, we replace components of the square loss with their robustified counterparts. Specifically, while the covariates $\xt$ and the data $\yt$ are both heavy-tailed, for example, when $\xt$ are macroeconomic variables and $\yt$ are stock returns, we set $\Txt = (||\xt||_4\wedge c_1)\xt/||\xt||_4$ and $\Tyt = (||\yt||_4\wedge c_2)\yt/||\yt||_4$ with $c=(c_1,c_2)$; when only the data $\yt$ are heavy-tailed, we set $\Txt = \xt$ and $\Tyt = (||\yt||_2\wedge c)\yt/||\yt||_2$. 
% This simple shrinkage approach yield both superior numerical and theoretical performance in our problem.
% computational, tractable; more to say here
% discuss so much about our choice... why? so many fancy stuff?
% Another remaining issue is the identification problem
% Not identification enough!

%%%%%%%%%%%%%%%%%%%%%%%%%%%%%%%%%%%%%%%%%%%%%%%%%%%%%%    
%%%%%%%%%%%%%%%%    Discussion    %%%%%%%%%%%%%%%%%%%%%%%%%%
%%%%%%%%%%%%%%%%%%%%%%%%%%%%%%%%%%%%%%%%%%%%%%%%%%%%%%

\section{Discussion}\label{discussion}
Motivated by the "factor zoo" problem in finance, in this paper, we consider 
improving the accuracy and robustness of latent factor model estimation by combining information in both the target multivariate data $\yt$ and a large pool of factor proxies $\xt$. 
Towards this end, a two-step procedure based on penalized reduced rank regression and projection is designed. 
The proposed factor model estimator, FMSP, is statistically more accurate, is robust to the bias in the factor proxy selection, and is flexible to be applied across various datasets, dimensions, as well as the number of factors. 
We establish improved rates of convergence compared with PCA, and show superior numerical performance compared with competing methods. 
% baselines such as
% in terms of estimation accuracy and convergence rates

There are several meaningful future directions worthy of investigation. 
First, this paper focuses on the accurate estimation of latent factor models. 
Another desired property is the interpretability, which seems achievable with the large pool of factor proxies. 
Variable selection or group-wise transformation can be considered for extensions in this direction. 

Secondly, compared with the existing methods, the novel robust reduced rank regression proposed in Section \ref{sec: robust FMSP} is computationally more  efficient and is able to utilize explicit knowledge of both the low-rank structure as well as the rank $K$. 
We provide some numerical guarantee on its finite-sample performance. 
The theoretical analysis of this method is a meaningful future direction. 
% out of the scope of this paper and left for future research. 

% Finally, supervised factor number estimation is expected to be more accurate than unsupervised approaches and is a meaningful next step. 

{
Third, in this paper, we focus on aggregating information in observed factor proxies to approximate the latent factors that drive the covariance structure. 
    An accurate estimation of latent factor models has broad applications including risk management, portfolio optimization, portfolio hedging, etc.  
    Another related topic is estimating the risk premia of observed factor proxies and accurate asset pricing under the factor pricing model induced by the arbitrage-pricing theory \citep{ross2013arbitrage}. 
    Given that our methodology design is for estimating \textit{latent} factor models, 
    % On one hand, given that ; clear how to; Although g
    it is not straightforward to directly extend it to estimating the risk premia of \textit{observed} factors. % On the other hand
    However, similar to \citet{kozak2020shrinking} and \citet{lettau2020estimating}, we can consider a variant of the proposed procedure that incorporates the first moment information of the data via an additional squared loss term, and hence make sure the learned latent factors can explain the expected excess returns well (referred to as \textit{latent asset-pricing factors} in \citet{lettau2020estimating}). 
    The problem can be solved following a similar procedure as introduced in Section \ref{sec:meth}. 
    % our paper. 
    We leave this extension for future research. 
}

Fourth, the high dimensionality is one of the major challenges to utilizing the large pool of factor proxies. 
% factor zoo. 
In this paper, we choose not to assume a sparsity structure, mainly due to two reasons: % and we apply the ridge penalty
(i) the sparsity-induced estimator (e.g., Lasso) has been observed to perform not well when the regressors are correlated (as in our dataset)  \citep{zou2005regularization, kozak2020shrinking}, and 
(ii) other empirical studies have found that the stochastic discount factor has a dense dependency on observed factor proxies \citep{bryzgalova2019bayesian, kozak2020shrinking}, which has similar implications in our case. Without the sparsity assumption, it is generally challenging to establish the theory under $p \asymp T$. 
Extensions along this direction would be meaningful.

\clearpage

\bibliography{0_CAFE}

% % \printbibliography

%%%%%%%%%%%%%%%%%%%%%%%%%%%%%%%%%%%%%%%%%%%%%%%%%%%%%%    
%%%%%%%%%%%%%%%%    Appendix    %%%%%%%%%%%%%%%%%%%%%%%%%%
%%%%%%%%%%%%%%%%%%%%%%%%%%%%%%%%%%%%%%%%%%%%%%%%%%%%%%

% \newpage
% \appendix
% % %%%%%%%%%%%%%%%%%%%%%%%%%%%%%%%%%%%%%%%%%%%%%%%%%%%%%%%%%%%%%%%%%%%%%%%%%%%%%%%%%%%%%%%%%%%%%%%%%%%%%%%%%%%%
% % %%%%%%%%%%%%%%%%%%   Proof for Propositions   %%%%%%%%
% % %%%%%%%%%%%%%%%%%%%%%%%%%%%%%%%%%%%%%%%%%%%%%%%%%%%%%%
% \input{appendix_Numerical}

% \input{proof/proof_proposition.tex}
% \section{Proof of Theorems in Section \ref{theory}}
% \label{proof of main results}

% \input{proof/Sec_lemmas.tex}

% \input{proof/Proof_Theorem_1}

% % %%%%%%%%%%%%%%%%%%%%%%%%%%%
% % %%%%%%%  Theorem 2  %%%%%%%  
% % %%%%%%%%%%%%%%%%%%%%%%%%%%%

% \input{proof/Proof_Theorem_2}

% \input{proof/proof_lemmas.tex}

\end{document}

\begin{frontmatter}
\title{Supplement to ``\mbox{}
Mining the Factor Zoo: \\ 
Estimation of Latent Factor Models  with Sufficient Proxies
}

\end{frontmatter}

\appendix
\section{Additional Numerical Results}\label{sec:appendix_numerical}
In this section, we present some addition numerical results under similar settings as in Section \ref{sec: simulation} and \ref{sec:robust}. 
In Tables \ref{tab: simu_heavy} and \ref{tab: simu_heavy_mis}, we report results under the heavy-tailed setting introduced in Section \ref{sec:robust}, and we consider there is no auto-correlation structure.  The findings are consistent as discussed in the main text, that the robust version of FMSP performs favorably.

%%%%%%%%%%%%%%%%%%%%%%%%%%%%%%%%%%%%%%%%%%%%%%%%%%%%%%%%%%%%%%%%%%%%%%%%%%%%%%%%%%%%%%%%%%%%%%%%%%%%%%%%%
%%%%%%%%%%%%%%%%%%%%%%%%%%%%%%%%%%%%%%%%%%%%%%%%%%%%%%%%%%%%%%%%%%%%%%%%%%%%%%%%%%%%%%%%%%%%%%%%%%%%%%%%%
\begin{table}[h!]
\centering
\caption{ Relative values of the smallest canonical correlations with FMSP-R as the denominator,   when the noises are \textit{heavy-tailed}, $\vec{g}(\vec{z}_t) = \vec{z}_t$, and under the \textit{i.i.d.} setting. A value below 1 implies the estimation is less accurate than FMSP-R. }
\begin{tabular}{|c|c|ccccc|ccccc|}
\hline
                      &            & \multicolumn{5}{c|}{Factors} & \multicolumn{5}{c|}{Loading} \\ \hline
$\omega_2$            & $\omega_1$ &  PLS & PCR & PCA & SPCA & OP  & PLS & PCR & PCA & SPCA & OP \\ \hline
\multirow{3}{*}{0.7} 
 &  1  &  0.68 & 0.57 & 0.19 & 0.65 & 0.10 & 0.49 & 0.37 & 0.18 & 0.56 & 0.18 \\ 

 &  2  &  0.41 & 0.34 & 0.21 & 0.84 & 0.19 & 0.29 & 0.23 & 0.20 & 0.75 & 0.31 \\ 

 &  5  &  0.74 & 0.56 & 0.23 & 1.03 & 0.52 & 0.55 & 0.37 & 0.22 & 1.04 & 0.67 \\ 

\hline \multirow{3}{*}{0.95}
 &  1  &  1.18 & 0.99 & 0.19 & 0.68 & 0.20 & 0.97 & 0.76 & 0.19 & 0.72 & 0.31 \\ 

 &  2  &  0.47 & 0.42 & 0.20 & 1.03 & 0.39 & 0.38 & 0.32 & 0.19 & 1.12 & 0.56 \\ 

 &  5  &  0.89 & 0.67 & 0.17 & 1.12 & 0.91 & 0.74 & 0.51 & 0.16 & 1.30 & 0.90 \\ 
 \hline
\end{tabular}\label{tab: simu_heavy}
\caption{ Relative values of the smallest canonical correlations with FMSP-R as the denominator,   when the noises are \textit{heavy-tailed}, $\vec{g}(\vec{z}_t) = 0.5 ( sin(\frac{1}{2}\vec{z}_t) +  \vec{z}_t)$, and under the \textit{i.i.d.} setting. A value below 1 implies the estimation is less accurate than FMSP-R. }
\begin{tabular}{|c|c|ccccc|ccccc|}
\hline
                      &            & \multicolumn{5}{c|}{Factors} & \multicolumn{5}{c|}{Loading} \\ \hline
$\omega_2$            & $\omega_1$ & PLS & PCR & PCA & SPCA & OP  & PLS & PCR & PCA & SPCA & OP \\ \hline
\multirow{3}{*}{0.7}   
 &  1  &  0.53 & 0.43 & 0.17 & 0.54 & 0.09 & 0.36 & 0.26 & 0.17 & 0.47 & 0.15 \\ 

 &  2  &  0.38 & 0.32 & 0.18 & 0.66 & 0.14 & 0.25 & 0.20 & 0.17 & 0.56 & 0.22 \\ 

 &  5  &  0.57 & 0.43 & 0.20 & 0.90 & 0.37 & 0.40 & 0.27 & 0.20 & 0.86 & 0.48 \\ 

\hline \multirow{3}{*}{0.95}
 &  1  &  1.18 & 1.01 & 0.20 & 0.58 & 0.22 & 0.91 & 0.72 & 0.22 & 0.65 & 0.31 \\ 

 &  2  &  0.47 & 0.45 & 0.19 & 0.87 & 0.41 & 0.38 & 0.33 & 0.19 & 0.94 & 0.50 \\ 

 &  5  &  0.86 & 0.68 & 0.18 & 1.17 & 0.91 & 0.67 & 0.49 & 0.16 & 1.36 & 0.87 \\ 
                      \hline
\end{tabular}\label{tab: simu_heavy_mis}
\end{table}

% In this section, we present our results under the setting that $N = 100, T= 90$, and $K=5$,  which is consistent with our real data experiments. 
% $s$ is set as 5. 
% Additional results for other combinations can be found in the supplementary material and the findings are similar. 

Next, we present results under the same setting as in Table \ref{tab:non_robust_IID} in the main text, but with different combinations of $N$ and $T$. 
This set of results further supports that FMSP is generally applicable. 
Specifically, we present results from Table \ref{tab:other_setting_1} to \ref{tab:other_setting_4}, for $(N, T)$ equal to $(100, 60), (200, 60), (50, 150)$, and $(200, 150)$, respectively. 
Overall, the findings are consistent with the conclusions presented in the main text. 
As expected, with a large $N$ and $T$, the relative performance of PCA improves compared with FMSP. 
In contrast, when $N$ and $T$ are small, the relative performance of OP improves as it becomes increasingly challenging for data-driven methods such as FMSP to learn the latent structures with a smaller data size. 

\begin{table}[h!]
\centering
\caption{$N = 100$ and $T = 60$. Relative values of the smallest canonical correlations with FMSP as the denominator,  when $\vec{g}(\vec{z}_t) = 0.5 ( sin(\frac{1}{2}\vec{z}_t) +  \vec{z}_t)$ and under the \textit{i.i.d.} setting. A value lower than 1 implies the estimation is less accurate than FMSP. }
\begin{tabular}{|c|c|ccccc|ccccc|}
\hline
                      &            & \multicolumn{5}{c|}{Factors} & \multicolumn{5}{c|}{Loading} \\ \hline
$\omega_2$            & $\omega_1$ & PLS & PCR & PCA & SPCA & OP  & PLS & PCR & PCA & SPCA & OP \\ \hline
\multirow{3}{*}{0.7}   
 &  1  &  0.67 & 0.57 & 1.02 & 0.51 & 0.13 & 0.52 & 0.41 & 1.03 & 0.42 & 0.22 \\ 

 &  2  &  0.44 & 0.36 & 1.01 & 0.64 & 0.23 & 0.34 & 0.26 & 1.03 & 0.53 & 0.38 \\ 

 &  5  &  0.72 & 0.54 & 0.99 & 0.88 & 0.60 & 0.56 & 0.39 & 1.01 & 0.76 & 0.79 \\ 

\hline \multirow{3}{*}{0.95}
 &  1  &  1.19 & 1.02 & 0.61 & 0.49 & 0.23 & 0.99 & 0.81 & 0.65 & 0.44 & 0.37 \\ 

 &  2  &  0.47 & 0.41 & 0.66 & 0.68 & 0.47 & 0.43 & 0.35 & 0.71 & 0.61 & 0.67 \\ 

 &  5  &  0.89 & 0.66 & 0.69 & 1.04 & 1.06 & 0.76 & 0.53 & 0.73 & 0.96 & 1.06 \\ 
                      \hline
\end{tabular}\label{tab:other_setting_1}
\end{table}

\begin{table}[h!]
\centering
\caption{$N = 200$ and $T = 60$. Relative values of the smallest canonical correlations with FMSP as the denominator,  when $\vec{g}(\vec{z}_t) = 0.5 ( sin(\frac{1}{2}\vec{z}_t) +  \vec{z}_t)$ and under the \textit{i.i.d} setting. A value lower than 1 implies the estimation is less accurate than FMSP. }
\begin{tabular}{|c|c|ccccc|ccccc|}
\hline
                      &            & \multicolumn{5}{c|}{Factors} & \multicolumn{5}{c|}{Loading} \\ \hline
$\omega_2$            & $\omega_1$ & PLS & PCR & PCA & SPCA & OP  & PLS & PCR & PCA & SPCA & OP \\ \hline
\multirow{3}{*}{0.7}   
 &  1  &  0.81 & 0.74 & 1.04 & 0.53 & 0.09 & 0.66 & 0.58 & 1.10 & 0.41 & 0.24 \\ 

 &  2  &  0.38 & 0.30 & 1.03 & 0.72 & 0.21 & 0.28 & 0.22 & 1.07 & 0.58 & 0.47 \\ 

 &  5  &  0.84 & 0.65 & 1.00 & 0.96 & 0.55 & 0.71 & 0.49 & 1.04 & 0.87 & 0.87 \\ 

\hline \multirow{3}{*}{0.95}
 &  1  &  0.96 & 0.90 & 0.61 & 0.42 & 0.15 & 0.85 & 0.77 & 0.66 & 0.35 & 0.30 \\ 

 &  2  &  0.33 & 0.27 & 0.72 & 0.67 & 0.36 & 0.29 & 0.22 & 0.76 & 0.57 & 0.61 \\ 

 &  5  &  0.90 & 0.69 & 0.83 & 1.00 & 0.90 & 0.81 & 0.56 & 0.87 & 0.94 & 0.96 \\ 
                      \hline
\end{tabular}\label{tab:other_setting_2}
\end{table}

\begin{table}[h!]
\centering
\caption{$N = 50$ and $T = 150$. Relative values of the smallest canonical correlations with FMSP as the denominator,  when $\vec{g}(\vec{z}_t) = 0.5 ( sin(\frac{1}{2}\vec{z}_t) +  \vec{z}_t)$ and under the \textit{i.i.d} setting. A value lower than 1 implies the estimation is less accurate than FMSP. }
\begin{tabular}{|c|c|ccccc|ccccc|}
\hline
                      &            & \multicolumn{5}{c|}{Factors} & \multicolumn{5}{c|}{Loading} \\ \hline
$\omega_2$            & $\omega_1$ & PLS & PCR & PCA & SPCA & OP  & PLS & PCR & PCA & SPCA & OP \\ \hline
\multirow{3}{*}{0.7}   
 &  1  &  0.52 & 0.43 & 1.01 & 0.63 & 0.15 & 0.40 & 0.32 & 1.01 & 0.57 & 0.23 \\ 

 &  2  &  0.38 & 0.30 & 1.00 & 0.68 & 0.22 & 0.29 & 0.22 & 0.99 & 0.61 & 0.33 \\ 

 &  5  &  0.57 & 0.41 & 0.99 & 0.81 & 0.54 & 0.45 & 0.29 & 1.00 & 0.74 & 0.66 \\ 

\hline \multirow{3}{*}{0.95}
 &  1  &  1.23 & 0.98 & 0.76 & 0.68 & 0.31 & 0.97 & 0.78 & 0.75 & 0.66 & 0.46 \\ 

 &  2  &  0.57 & 0.50 & 0.80 & 0.79 & 0.48 & 0.47 & 0.41 & 0.83 & 0.75 & 0.66 \\ 

 &  5  &  0.80 & 0.61 & 0.80 & 0.95 & 1.07 & 0.67 & 0.48 & 0.82 & 0.88 & 1.09 \\ 
                      \hline
\end{tabular}\label{tab:other_setting_3}
\end{table}

\begin{table}[h!]
\centering
\caption{$N = 200$ and $T = 150$. Relative values of the smallest canonical correlations with FMSP as the denominator,  when $\vec{g}(\vec{z}_t) = 0.5 ( sin(\frac{1}{2}\vec{z}_t) +  \vec{z}_t)$ and under the \textit{i.i.d} setting. A value lower than 1 implies the estimation is less accurate than FMSP. }
\begin{tabular}{|c|c|ccccc|ccccc|}
\hline
                      &            & \multicolumn{5}{c|}{Factors} & \multicolumn{5}{c|}{Loading} \\ \hline
$\omega_2$            & $\omega_1$ & PLS & PCR & PCA & SPCA & OP  & PLS & PCR & PCA & SPCA & OP \\ \hline
\multirow{3}{*}{0.7}   
 &  1  &  0.79 & 0.72 & 1.02 & 0.69 & 0.08 & 0.60 & 0.52 & 1.08 & 0.53 & 0.20 \\ 

 &  2  &  0.36 & 0.30 & 1.01 & 0.77 & 0.17 & 0.24 & 0.19 & 1.06 & 0.61 & 0.40 \\ 

 &  5  &  0.78 & 0.64 & 1.01 & 0.95 & 0.47 & 0.62 & 0.44 & 1.04 & 0.84 & 0.80 \\ 

\hline \multirow{3}{*}{0.95}
 &  1  &  0.91 & 0.84 & 0.82 & 0.50 & 0.12 & 0.77 & 0.69 & 0.87 & 0.41 & 0.26 \\ 

 &  2  &  0.29 & 0.26 & 0.89 & 0.68 & 0.27 & 0.23 & 0.19 & 0.94 & 0.57 & 0.52 \\ 

 &  5  &  0.86 & 0.68 & 0.93 & 0.97 & 0.75 & 0.74 & 0.53 & 0.96 & 0.90 & 0.91 \\ 

                      \hline
\end{tabular}\label{tab:other_setting_4}
\end{table}

\section{Proof of Propositions}\label{sec:proof_proposition}
\subsection{Proof of Proposition \ref{proposition: exact_identification}}
\begin{proof}
The proof is similar with that of Theorem 2.1 in \cite{fan2016augmented}. 
Note that $E(\yt|\xt) = \MA\MB'\xt + \MA E(\ut|\xt)$, 
we have 
\begin{align*}
    E(E(\yt|\xt)E'(\yt|\xt)) 
    &= E((\MA\MB'\xt + \MA E(\ut|\xt))(\MA\MB'\xt + \MA E(\ut|\xt))') \\
    &= \MA E((\MB'\xt + E(\ut|\xt))(\MB'\xt + E(\ut|\xt))')\MA’\\
    &= \MA \Sigma_{f|x} \MA', 
\end{align*}
where $\Sigma_{f|x} = E((\MB'\xt + E(\ut|\xt))(\MB'\xt + E(\ut|\xt))')$. 
Recall the condition that $\sigma_{min}(E(E(\ft|\xt) E'(\ft|\xt))/K) > 0$, we have $r(E(E(\yt|\xt)E'(\yt|\xt))) = r(A) = K$, 
and according to the property of singular values decomposition, we can find an invertible matrix $\MH$ such that the eigenvectors of  $E(E(\yt|\xt) E'(\yt|\xt))$ corresponding to its nonzero eigenvalues 
are the columns of $\MA\MH$.  
% Let the singular values decomposition of $\Sigma_{f|x}$ be $\Sigma_{f|x} = \MV\Tilde{\Sigma}\MV'$, where $\Tilde{\Sigma}$ is a diagnoal matrix which consists of the eigenvalues of $\Sigma_{f|x}$, and the columns of $\MV$ are the eigenvectors of $\Sigma_{f|x}$. 
\end{proof}

\subsection{Proof of Proposition \ref{proof: Derivation}}
\begin{proof}
Recall the optimization problem is 
\begin{equation}\label{eqn:step2}
    \hA,\hB = \argmin_{\MA \in \mathcal{R}^{N\times K}, \MB \in \mathcal{R}^{p\times K}, \MA'\MA=NK^{-1}\mathbf{I}}\quad ||\MY - \MX\vec{B}\MA'||_F^2 +  \lambda ||\vec{B}||_F^2.
\end{equation}
We first show that one solution of (\ref{eqn:step2}) can be  obtained by solving the following problem: 
\begin{equation}\label{eqn:lemma_derivation}
   \hat{\MTh} = \argmin_{r(\MTh) \le k} ||\MY - \MX \MTh||_F^2 +  \frac{\lambda}{N} ||\MTh||_F^2.
\end{equation}
Notice that $||\MY - \MX \MTh||_F^2 +  (\lambda / N) ||\MTh||_F^2 = 
||\MY^* - \MX^* \MTh||_F^2$, where 
$\MY^* = (\MY',\mathbf{0})'$ and 
$\MX^* = (\MX ',\sqrt{\lambda /N }\mathbf{I})'$.
So  (\ref{eqn:lemma_derivation}) can be regarded as the standard reduced rank regression problem with respect to $\MY^*$ and $\MX^*$. 
% Similar with the proof in  \ref{proof:proposition1}, 
With the assumption that the largest $K$  singular values of
$\hat{\MY}^*$ are non-zero and  distinct, the solution of (\ref{eqn:lemma_derivation}) is unique as $\hMTh = \hB^*_{OLS} \MV_K^* \MV^{*'}_K$, 
where $\hB^*_{OLS} = ({\MX^*}' \MX^*)^{+}\MX^{*'}\MY^* = (\MX '\MX + (\lambda / N)\vec{I})^{-1}\MX '\MY$ 
and the $N \times K$ matrix $\MV^*_K$ is constructed by stacking the first $K$ principal axes of $\hat{\vec{Y}}^* = \MX^*\hB^*_{OLS}$. 
Define $\hA = \sqrt{N}\MV^*_K$ and $\hB = \hB^*_{OLS} \MV_K^* / \sqrt{N}$. 
For any matrix $\MA$ $\in \mathcal{R}^{N\times K}$ satisfying $\MA'\MA = N\vec{I}$ and matrix $\vec{B}$ $\in \mathcal{R}^{p\times K}$, the following inequality holds: 
 \begin{align*}
||\MY - \MX\vec{B}\MA'||_F^2 +  \lambda ||\vec{B}||_F^2 
&=
||\MY - \MX(\vec{B}\MA')||_F^2 +  \frac{\lambda}{N} ||(\vec{B}\MA')||_F^2 \\
&\le 
||\MY - \MX\hMTh||_F^2 +  \frac{\lambda}{N} ||\hMTh||_F^2 
 = ||\MY - \MX\vec{\hat{B}}\hA'||_F^2 +  \lambda ||\vec{\hat{B}}||_F^2 
 \end{align*}
Thus, $(\hA,\hB)$ is one solution of  (\ref{eqn:step2}). 
% Since $\hmth$ has \blue{k distinct non-zero singular values.}
% if we can obtain $\hat{\MTh}$, 
% we have that there exists an orthonormal matrix $\vec{Q}^*$ such that $\MV^*_K \vec{Q}^* = \MV$. This completes our proof.
% Finally, by writing
% $\hat{\MTh} = \vec{U}\vec{D}\MV' = \hB^*_{OLS} \MV_K^* \MV^{*'}_K$, 

% The uniqueness of the solution implies $\hMTh = \hMTh'$. 

% [Q: up to a rotation???]
% , there exists matrixes 
% $\vec{U}$ $\in \mathcal{R}^{p\times K}$,  $\MV$ $\in \mathcal{R}^{N\times K}$ and $\vec{D}$ $\in \mathcal{R}^{K\times K}$ such that  $\vec{U}\vec{D}\MV'$ is the 
% singular value decomposition of $\hat{\MTh}$, and 
% $\MV$ and $\Vec{U}\vec{D}$ are both unique (up to a rotation).

On the other hand, let $(\MA_1,\MB_1)$  be one solution of  (\ref{eqn:step2}), and denote $\MTh_1 = \MB_1\MA_1'$. 
The fact that $r(\MTh_1) \le K$ and the  optimality of $(\MA_1,\MB_1)$ implies the following inequality:

\begin{equation*}\label{eqn:lemma_derivation2}
 \begin{split}
 ||\MY - \MX \MTh_1||_F^2 +  \frac{\lambda}{N}  ||\MTh_1||_F^2 
&=
 ||\MY - \MX\MB_1 \MA_1'||_F^2 +  \lambda ||\MB_1||_F^2 \\
&\le 
||\MY - \MX\MB_1\MA_1'||_F^2 +  \lambda ||\MB_1||_F^2 
= ||\MY - \MX\MTh_1||_F^2 +  \frac{\lambda}{N} ||\MTh_1||_F^2.
 \end{split}
\end{equation*}
Therefore, $\MTh_1$ is also one solution of (\ref{eqn:lemma_derivation}), and thus $\MTh_1 = \hMTh$. 
Because of the condition that the largest $K$  singular values of
$\hat{\MY}^*$ are distinct, which implies the uniqueness (up to a rotation) of SVD of $\hMTh$, we have an orthonormal matrix $\Vec{Q}$ such that $\MA = \MA_1\MQ$ and $\MB = \MB_1\MQ$.
\end{proof}

\subsection{Proof of Proposition \ref{lemma: shrinkage} }
\begin{proof}
Note that according to the triangle inequality, it holds that
$||\yt|| \ge |(||\et|| - ||\MA\ft||)|$. 
Therefore, conditional on the event $\{||\et|| \ge \sqrt{N}t\}$, we have the relationship 
$\{||\MA\ft|| \le (1 -  \tau/(\sqrt{N}t)) ||\et||\} 
\supset \{||\yt|| \ge \tau/(\sqrt{N}t) ||\et||\}
\supset \{\taut \le \sqrt{N}t/||\et|| \} $. 
Hence, we have $P(||\et|| \ge \sqrt{N}t, ||\et||\taut \ge \sqrt{N}t) \le P(||\et|| \ge \sqrt{N}t, ||\MA\ft|| \ge (1 - \tau/(\sqrt{N}t) ) ||\et||)$.

Based on the assumptions that there are constants $c_1, c_2 >0$, such that
$||\MA|| \le \sqrt{N}c_1$ and $\tau \ge \sqrt{N}c_2$, we have for any $\vv \in \Sd{N}$ and sufficiently large $t$, it holds that 
\begin{align*}
    P(|\taot \vv'\et| \ge \sqrt{N} t) 
    &\le P(||\et|| \taot \ge \sqrt{N} t)\\
    &\le P(||\et|| \ge \sqrt{N}t, ||\MA \ft|| \ge (1-\frac{\tau}{\sqrt{N}t})||\et||)\\
    &\le P(\sqrt{N}||\ft|| \ge c_1 (1-\frac{\tau}{\sqrt{N}t})\sqrt{N}t)\\
    &\le P(||\ft|| \ge t - \frac{\tau}{\sqrt{N}})\\
    &\le  P(||\ft|| \ge c_1(t - c_2))
\end{align*}

Therefore, note that $\ft$ is a sub-Gaussian random vector, %???
we have that $\vv' \taut\et$ is a sub-Gaussian random variable.
By the arbitrariness of $\vv$, we conclude that there is a constant $M$ such that $||\taut \et||_\psit \le M$.
\end{proof}

\section{Proof of Theorems in Section \ref{theory}}
\label{proof of main results}

\subsection{Some Technical Results}

\begin{flushleft}
% Theorem \ref{rate:loading} and \ref{rate:factors} are both stated up to an invertible matrix $\MH$
Let $\lN = \lambda / N$ and $\wl = (\MX'\MX + \lN \Vec{I})^{-1}$. 
Recall Proposition \ref{proof: Derivation}, with $\hY = \MXs \wl \MX' \MY $ and $\hCOV = \hY'\hY / T$, the loading matrix estimator $\hA$ satisfies 
\begin{equation}\label{relationship:loading}
 \frac{1}{N} \hCOV \hA = \hA \TV,   
\end{equation}
 where $\TV$ is a $K \times K$ diagonal matrix with the $k$-th diagonal entry as the $k$-th largest eigenvalue of $\hCOV / N$, for $k=1,\dots,K$. 
\end{flushleft}

We first state several results which will be repeatedly used in the proof of both Theorem \ref{rate:loading} and Theorem \ref{rate:factors}. 
The proofs of these lemmas  are left to  \ref{appendix: lemmas}. 
To simplify the notations, we write 
$\MTh = \MB \MA'$, $\FX = \MX\MB$, 
$\MD = (\MTh' \MX ' \MY - \MTh' \MX ' \MX \MTh)/T$, and 
$\MDD = (\ME + \MU\MA')'(\XWX) (\ME + \MU\MA') / NT$.

%%%%%%%%%%%%%%%%%%%%%%%%%%%%%%%%%%%%%%%%%%%%%%%%%%%%%%%%%%%%%%%%%%%%%%%%%%%%%% 
%%%%%%%%%%%%%%%%%%%%%%%%%% Lemmas %%%%%%%%%%%%%%%%%%%%%%%%%%%%%%%%%%%%%%%%  
%%%%%%%%%%%%%%%%%%%%%%%%%%%%%%%%%%%%%%%%%%%%%%%%%%%%%%%%%%%%%%%%%%%%%%%%%%%%%% 
Under Assumptions 1-3, the following lemmas hold:

%%%%%%%%%%%%%%%%%%%%%%%%%%%%%%%%%%%%%%%%%%%%%%%%%%%%%%%%%%%%%%%%%%%%%%%%%%%%%% 
%%%%%%%%%%%%%%%%%%%%%%%%%% Lemma 1 %%%%%%%%%%%%%%%%%%%%%%%%%%%%%%%%%%%%%%%%  
%%%%%%%%%%%%%%%%%%%%%%%%%%%%%%%%%%%%%%%%%%%%%%%%%%%%%%%%%%%%%%%%%%%%%%%%%%%%%% 
\begin{lemma}\label{lemma: technical_lemmas}
\begin{enumerate}
\item 
For any matrix $\MB$ and any symmetric matrix $\MA$ of appropriate dimensions, it holds that $||\MB'\MA\MB|| \le ||\MB'\MB|| \times ||\MA||$; 
\item 
$T^{-1} \sumoT||\et||^2  = O_p(N)$
, $T^{-1} \sumoT||\ft||^2 =  O_p(K)$
, and $T^{-1} \sumoT||\MA'\et||^2 = O_p(N)$. 
\end{enumerate}

\end{lemma}

%%%%%%%%%%%%%%%%%%%%%%%%%%%%%%%%%%%%%%%%%%%%%%%%%%%%%%%%%%%%%%%%%%%%%%%%%%%%%% 
%%%%%%%%%%%%%%%%%%%%%%%%%% Lemma 2 %%%%%%%%%%%%%%%%%%%%%%%%%%%%%%%%%%%%%%%%  
%%%%%%%%%%%%%%%%%%%%%%%%%%%%%%%%%%%%%%%%%%%%%%%%%%%%%%%%%%%%%%%%%%%%%%%%%%%%%% 
\begin{lemma}\label{lemma: rate of wl and signal}
\begin{enumerate}
\item 
$||\wl|| \le \lN^{-1}$, $||\wl|| = O_p(T^{-1})$, and $||\MX\wl\MX'|| \le 1$;
%  = O_p(1)
\item 
$||\MTh||= O(\sqrt{N})$. 
% and $||\fxt||_{\psi_2} \le K^{1/2}M_5$ for $M_5 >0$.
\end{enumerate}
\end{lemma}
% , $||\wl|| = O_p(T^{-1})$, $||\MX\wl|| = O_p({T}^{-\frac{1}{2}})$ 
%%%%%%%%%%%%%%%%%%%%%%%%%%%%%%%%%%%%%%%%%%%%%%%%%%%%%%%%%%%%%%%%%%%%%%%%%%%%%% 
%%%%%%%%%%%%%%%%%%%%%%%%%% RATEs %%%%%%%%%%%%%%%%%%%%%%%%%%%%%%%%%%%%%%%%  
%%%%%%%%%%%%%%%%%%%%%%%%%%%%%%%%%%%%%%%%%%%%%%%%%%%%%%%%%%%%%%%%%%%%%%%%%%%%%% 
\newcommand{\rateofxuo}{-1/2-(\delo \wedge \delt)}
\newcommand{\rateofxut}{\sqrt{p}T^{-(1/2+\delta_1)} + T^{-(1/2+\delt)}}

\newcommand{\rateofxeo}{\sqrt{N/T}}
\newcommand{\Rateofxeo}{\sqrt{\frac{N}{T}}}
\newcommand{\rateofxet}{ \sqrt{N/T}} %(N/T) \vee 
\newcommand{\rateofxeth}{ \sqrt{N(p+K)(TK)^{-1}}}
\newcommand{\Rateofxeth}{ \sqrt{ \frac{N(p+K)}{TK} }}
\newcommand{\rateofxef}{ \sqrt{(N+p)/T} \vee (N/T)}

%%%%%%%%%%%%%%%%%%%%%%%%%%%%%%%%%  1 %%%%%%%%%%%%%%%%%%%%%%%%%% 
\newcommand{\covXE}{\MX'\ME / T}
\newcommand{\rateXE}{
\sqrt{(N+p)/T}
    % + \sqrt{N/T}(p^{\frac{1}{4}} \wedge \sqrt{N/T})
}
\newcommand{\RateXE}{
\sqrt{\frac{N+p}{T}}
    % + \sqrt{\frac{N}{T}}(p^{\frac{1}{4}} \wedge \sqrt{\frac{N}{T}})
}
    
\newcommand{\covXU}{\MX'\MU / T}
\newcommand{\rateXU}{\sqrt{(p+K)/T}u_2 + u_3}
\newcommand{\RateXU}{\sqrt{\frac{p+K}{T}}u_2 + u_3}

%%%%%%%%%%%%%%%%%%%%%%%%%%%%%%%%% 2 %%%%%%%%%%%%%%%%%%%%%%%%%% 

\newcommand{\covThetaXE}{\MTh'\MX'\ME / T}
\newcommand{\covThetaXU}{\MTh'\MX'\MU / T}
\newcommand{\rateThetaXE}{N/\sqrt{TK} + \sqrt{N/T}}
\newcommand{\RateThetaXE}{\frac{N}{\sqrt{TK}} + \sqrt{\frac{N}{T}}}
\newcommand{\rateThetaXU}{\sqrt{N/T}u_2 + \sqrt{N/K}u_3}

%%%%%%%%%%%%%%%%%%%%%%%%%%%%%%%%% 3 %%%%%%%%%%%%%%%%%%%%%%%%%% 
\newcommand{\covXEE}{\MX'\ME\ME' / T}
\newcommand{\rateXEE}{ N\sqrt{p/T} + \sqrt{Np}    }

\newcommand{\RateXEE}{ N\sqrt{\frac{p}{T}} + \sqrt{Np}    }

\newcommand{\covFXEE}{\FX'\ME\ME' / T}
\newcommand{\rateFXEE}{N \sqrt{K/T} + \sqrt{KN}}
\newcommand{\RateFXEE}{N \sqrt{\frac{K}{T}} + \sqrt{KN}}

%%%%%%%%%%%%%%%%%%%%%%%%%%%%%%%%% 4 %%%%%%%%%%%%%%%%%%%%%%%%%% 
\newcommand{\ratedelta}{
N(TK)^{-1/2}(u_2 + 1)  + N u_3 / K
}
% NT^{-1/2}(u_2 + 1) + Nu_3K^{-1/2}
% \sqrt{N}(\sqrt{N / T} + \sqrt{k/T}u_2 + u_3 )

\newcommand{\Ratedelta}{
\frac{N}{\sqrt{TK}}(u_2 + 1) + \frac{N}{K}u_3
}
% \frac{N}{\sqrt{T}}(u_2 + 1) + \frac{Nu_3}{\sqrt{K}}}
% \sqrt{N}(\sqrt{\frac{N}{T}} 
% + \sqrt{\frac{k}{T}}u_2 + u_3 )

\newcommand{\rateDD}{
T^{-1} + (NT)^{-1}p + (TK)^{-1}(p + K)u_2^2 + u_3^2 / K
}

\newcommand{\RateDD}{
\frac{1}{T} + \frac{p}{NT} +  \frac{p + K}{TK}u_2^2 + \frac{u_3^2}{K}
}

\newcommand{\rateXG}{
\sqrt{(N+p)/T}
    % + \sqrt{N/T}(p^{\frac{1}{4}} \wedge \sqrt{N/T})
+ \sqrt{N/K}(\sqrt{(p+K)/T}u_2 + u_3)
}

\newcommand{\RateXG}{
\sqrt{\frac{N+p}{T}}
    % + \sqrt{\frac{N}{T}}(p^{\frac{1}{4}} \wedge \sqrt{\frac{N}{T}})
+ \sqrt{\frac{N}{K}}(\sqrt{\frac{p+K}{T}}u_2 
+ u_3)
}

% \newcommand{\RateXG}{\frac{N+p}{T} + \sqrt{
% \frac{N+p}{T}} + \sqrt{\frac{Np}{T}}u_2 + \sqrt{N}u_3}
%%%%%%%%%%%%%%%%%%%%%%%%%%%%%%%%%%%%%%%%%%%%%%%%%%%%%%%%%%%%%%%%%%%%%%%%%%%%%% 
%%%%%%%%%%%%%%%%%%%%%%%%%% Lemma 3 %%%%%%%%%%%%%%%%%%%%%%%%%%%%%%%%%%%%%%%%  
%%%%%%%%%%%%%%%%%%%%%%%%%%%%%%%%%%%%%%%%%%%%%%%%%%%%%%%%%%%%%%%%%%%%%%%%%%%%%% 

\begin{lemma}\label{lemma: rate of CovMatrix}
\begin{enumerate}
\item  %%% i
% $||||\xt||_{\psit}^{-1} \covXE|| = \rateXE$ and 
% $||||\xt||_{\psit}^{-1} \covXU|| = O_p(\rateXU)$; 
$||\MDD||
= O_p(\rateDD)$
\item  %%% ii
$||\covThetaXE|| = O_p(\rateThetaXE)$
, $||\covThetaXU|| = O_p(\rateThetaXU)$,
$||\XEo/T|| = O_p(\rateofxeo)$, 
$|| ||\xt||_{\psit}^{-1} \XEth/T|| = O_p(\rateofxeth)$
, and $||\MD|| = O_p(\ratedelta)$
; 
\item  %%% iii
$|| ||\xt||_{\psit}^{-1} \covXEE||_F = O_p(\rateXEE)$ and $|| \covFXEE||_F = O_p(
\rateFXEE)$; 
% \item %%% iv
% $||\MD|| = O_p(\ratedelta)$
% and
% $|| ||\xt||_{\psit}^{-1} (\MX ' \MY - \MX ' \MX \MTh) / T|| = 
% O_p(\rateXG)$. 
\end{enumerate}
\end{lemma}

%%%%%%%%%%%%%%%%%%%%%%%%%%%%%%%%%%%%%%%%%%%%%%%%%%%%%%%%%%%%%%%%%%%%%%%%%%%%%% 
%%%%%%%%%%%%%%%%%%%%%%%%%% Lemma 4 %%%%%%%%%%%%%%%%%%%%%%%%%%%%%%%%%%%%%%%%  
%%%%%%%%%%%%%%%%%%%%%%%%%%%%%%%%%%%%%%%%%%%%%%%%%%%%%%%%%%%%%%%%%%%%%%%%%%%%%% 
% \item  $ || \ME'  \ME /T|| = O_p( N/T + \sqrt{N/T} + M_2)$
\begin{lemma}\label{lemma: rate of V}
$||\TV^{-1}||$ is bounded away from zero and infinity with probability approaching one.
\end{lemma}

\subsection{Proof of Theorem \ref{rate:loading}}

%%%%%%%%%%%%%%%%%%%%%%%%%%%
%%%%%%%  Theorem 1  %%%%%%%  
%%%%%%%%%%%%%%%%%%%%%%%%%%%
\begin{proof}
We begin by writing $\hY = \MXs \MTh + \C{1} + \C{2}$, where $\C{1} = -\lN \MXs \wl \MTh$ and $\C{2} = \MXs \wl (\MX '\hY - \MX ' \MX \MTh)$. Plugging this relationship into (\ref{relationship:loading}), we obtain a decomposition of the estimation error of $\MA$ as:

\begin{equation}\label{decomp_the1}
    \hA - \MA\MH = \sumoe \D{i}, \quad \MH = \frac{1}{TN}\MB'{\MXs}'\MXs\MTh\hA\TV^{-1}, 
\end{equation}
where $\{\D{i}\}_{i=1}^7$ are in the following forms:

% $ \MD = \frac{1}{T}(\MTh' \MX ' \MY - \MTh' \MX ' \MX \MTh) = \fot \MA\FX'\MG, \MG = \MU\MA' + \ME$

\begin{align*}
\D{1} &= -2 \lN \MTh'\MTh \MR  & \D{2} &= \MD \MRN    \\
\D{3} &=  \XYcovDiff' \wl \XYcovDiff \MR & \D{4} &=   \lN^2 \MTh' \wl \MTh \MR\\
\D{5} &=  -\lN \MTh' \wl \XYcovDiff \MR  & \D{6} &= \MD' \MRN\\
\D{7} &= -\lN \XYcovDiff' \wl \MTh \MR \\
\end{align*}
 
With   Lemma \ref{lemma: rate of wl and signal} - \ref{lemma: rate of V} established and with the sub-multiplicative property of the matrix operator norm $||\Vec{A_1}\Vec{A_2}|| \le ||\Vec{A_1}||||\Vec{A_2}||$, 
we can bound the convergence rates of each term in (\ref{decomp_the1}) as follows. 

% {WlamX interaction?}

\allowdisplaybreaks
\begin{align*}
    ||\D{1}|| 
    &= ||-2 \lN \MTh'\MTh \MR|| \\
    &= O_p(\lN||\MTh||^2 ||\MR||) \\
    &\leftstackrel{\text{(Lemma 4)}}{=} O_p(\lN||\MTh||^2\frac{1}{NT}\sqrt{\frac{N}{K}})\\
    &\leftstackrel{\text{(Lemma 2.ii)}}{=} 
    O_p(\lN\frac{\sqrt{N/K}}{T}) \\
    ||\D{2}|| 
    &= || \MD \MRN || \\
    &\leftstackrel{\text{(Lemma 4)}}{=} O_p(\frac{1}{\sqrt{NK}}||\MD||) \\
    &\leftstackrel{\text{(Lemma 3.ii)}}{=}
    O_p(\frac{1}{\sqrt{NK}} (\Ratedelta)) \\
    &= O_p(\sqrt{N}(\frac{u_2  + 1}{\sqrt{TK}} + \frac{u_3}{K} ))
    \\
    ||\D{3}|| 
    &= ||\MDD \MRclean|| \\
    % &\leftstackrel{\text{(Lemma 4)}}{=} O_p(\frac{1}{NT} \sqrt{\frac{N}{K}}
    % ||\XYcovDiff' \wl \XYcovDiff||) \\
    % &\leftstackrel{\text{(Lemma 1)}}{=} O_p(\frac{1}{NT} \sqrt{\frac{N}{K}}
    % ||\XYcovDiff||^2 ||\wl||)\\
    % &\leftstackrel{\text{(Lemma 2.i)}}{=}
    % O_p(
    % \frac{1}{\sqrt{N}}
    % ||\fot \xtphituatpe||^2  
    % )\\
    &\leftstackrel{\text{(Lemma 3.i)}}{=} \sqrt{\frac{N}{K}} (\RateDD)\\
    % &\leftstackrel{\text{(Lemma 1)}}{=}
    % O_p(\frac{1}{\sqrt{NK}}[\RateXG]^2)\\
    ||\D{4}|| 
    &= ||\lN^2 \MTh' \wl \MTh \MR|| \\
    &\leftstackrel{\text{(Lemma 1 and 4)}}{=} 
    O_p(\lN^2 \frac{1}{NT} \sqrt{\frac{N}{K}} ||\MTh||^2 ||\wl||) 
     \stackrel{\text{(Lemma 2)}}{=} O_p(\lN\frac{\sqrt{N/K}}{T})\\
    ||\D{5}|| 
    &= ||-\lN \MTh' \wl \XYcovDiff \MR|| \\
    % &\leftstackrel{\text{(Lemma 1 and 4)}}{=} 
    % O_p(\frac{\lN}{T} ||\frac{1}{T} \xtphituatpe||) \\
    &\leftstackrel{\text{(Lemma 1 and 4)}}{=} 
    O_p(\lN \sqrt{N} ||\wl^{1/2}|| \frac{1}{NT} \sqrt{\frac{N}{K}} ||\wl^{1/2} \xtphituatpe|| ) \\
    &\leftstackrel{\text{(Lemma 3)}}{=} 
    O_p(\lN \frac{\sqrt{N}}{T} \frac{1}{\sqrt{K}} (\frac{1}{\sqrt{T}} + \sqrt{\frac{p}{NT}} + \sqrt{\frac{p+K}{TK}} u_2 + \frac{u_3}{\sqrt{K}}) )\\
    % &\leftstackrel{\text{(Lemma 3)}}{=} 
    % O_p(\frac{\lN}{T} (\RateXG))\\
    ||\D{6}|| 
    &= O_p(||\D{2}||) \\% O_p(N^{-1/2}||\MD||) = 
    ||\D{7}|| &= O_p(||\D{5}||)
\end{align*}

Recall that $\hA - \MA\MH = \sum_{i=1}^{7}\D{i}$,
putting the above bounds together, by choosing $\lN = O(1 / \sqrt{T+N})$, we can achieve that
\begin{align*}
\frac{K}{N}||\hA -\MA\MH||^2 
&= O_p\Big(\frac{K}{N}\sum_{i=1}^7||\D{i}||^2\Big) \\
&= \frac{K}{N} O_p\Big([\lN\frac{\sqrt{N/K}}{T}]^2
+ [\sqrt{N}(\frac{u_2  + 1}{\sqrt{TK}} + \frac{u_3}{K} )]^2\\
&+ [\sqrt{\frac{N}{K}} (\RateDD)]^2
+ [\lN\frac{\sqrt{N/K}}{T}]^2\\
&+ [\lN \frac{\sqrt{N}}{T} \frac{1}{\sqrt{K}} (\frac{1}{\sqrt{T}} + \sqrt{\frac{p}{NT}} + \sqrt{\frac{p+K}{TK}} u_2 + \frac{u_3}{\sqrt{K}})]^2\\
&= \frac{K}{N} O_p\Big(\lN^2\frac{N}{KT^2}
+ N(\frac{u_2^2  + 1}{TK} + \frac{u_3^2}{K^2} )\\
&+ \frac{N}{K} [\frac{1}{T^2} + \frac{p^2}{N^2T^2} + \frac{p^2}{T^2K^2}u_2^4 + \frac{1}{T^2}u_2^4 + \frac{u_3^4}{K^2}]\\
&+ \lN^2 \frac{N}{T^2K}  [\frac{1}{T} + \frac{p}{NT} + \frac{p+K}{TK} u_2^2 + \frac{u_3^2}{K}]
\Big)\\
&= O_p\Big(\frac{1}{T} + \frac{u_2^2}{T} + \frac{u_3^2}{K} + \frac{p^2}{N^2T^2} + \frac{p^2u_2^4}{T^2K^2}
\Big)
\end{align*}

Finally, recall the relationship that, for any matrix $\MA$ of rank $d$, we have $||\MA||_F \le \sqrt{d} ||\MA||$. We can conclude that 
\begin{align*}
    &\frac{1}{N}||\hA -\MA\MH||_F^2\\
    &= O_p( \frac{K}{N}||\hA -\MA\MH||^2)\\
    &=  O_p\Big(\frac{1}{T} + \frac{u_2^2}{T} + \frac{u_3^2}{K} + \frac{p^2}{N^2T^2} + \frac{p^2u_2^4}{T^2K^2}
    \Big)\\
    &=  O_p\Big(\frac{1}{T} + \frac{u_2^2}{T} + \frac{u_3^2}{K} + 
    {\frac{1}{N^2T^{2\eta}}} + 
    {\frac{u_2^4}{T^{2\eta}K^2}}
    \Big).
\end{align*}

% \begin{align*}
% \frac{1}{N}||\hA -\MA\MH||_F^2 
% &= O_p(\frac{1}{N}\sum_{i=1}^7||\D{i}||^2) \\
% &= O_p(\lN^2\frac{1}{TK} + 
% \frac{(u_2  + 1)^2}{T} + \frac{u_3^2}{K}\\
% &+ 
% \frac{1}{N^2}[\RateXG]^4\\
% &+ 
% \frac{\lN^2}{T^2N} [\RateXG]^2 )\\
% &= O_p(\lN^2\frac{1}{TK} + 
% \frac{(u_2  + 1)^2}{T} + \frac{u_3^2}{K}\\
% &+ 
% \frac{1}{N^2}[\frac{N^2}{T^2} 
% + \frac{p^2}{T^2K} 
% % + \frac{N^2}{T^2} (p \wedge \frac{N^2}{T^2})
% + \frac{N^2}{K^2} (\frac{p^2}{T^2}u_2^4 + \frac{K^2}{T^2}u_2^4 + u_3^4)
% ]\\
% &+ 
% \frac{\lN^2}{T^2N} 
% [\frac{N}{T} + \frac{p}{T} 
% % + \frac{N}{T}(\sqrt{p} \wedge \frac{N}{T})
% + \frac{N}{K} (\frac{p}{T}u_2^2 + \frac{K}{T}u_2^2 + u_3^2)
% ])\\
% &=
% O_p(\ratetheoremo)
% \end{align*}

\end{proof}

%%%%%%%%%%%%%%%%%%%%%%%%%%%
%%%%%%%  Theorem 2  %%%%%%%  
%%%%%%%%%%%%%%%%%%%%%%%%%%%

\subsection{Proof of Theorem \ref{rate:factors}}

\begin{proof}
% \hl{where EA is explicit; why I mentioned it here?}
% Find where is the EA term, which makes things complex
Let $\EA = \hA - \MA \MH$. Recall the definition of $\hft$ as $\hft = N^{-1}K\hA'\yt $,  we have 
\begin{equation*}
    \begin{split}
         \hft - \MH^{-1}\ft 
    = \frac{K}{N} \Big[\EA'\et + \MH'\MA'\et   - \EA'\EA\MH^{-1}\ft - (\MA\MH)'\EA\MH^{-1}\ft \Big].
    \end{split}
\end{equation*}
Therefore, the convergence rate of the estimated factors can be decomposed as 
\begin{equation}\label{theory:factor:decomposition}
    \begin{split}
     \frac{1}{K}\oTsumoT ||\hft - \MH^{-1}\ft ||^2 
    &\le \frac{K}{N^2}
     \fot \sumoT \Big[||\EA'\et||^2 + ||\MH'\MA'\et||^2  \\
     &\quad + ||\EA'\EA\MH^{-1}\ft||^2 + ||(\MA\MH)'\EA\MH^{-1}\ft||^2 \Big].
    \end{split}
\end{equation}

% \flN (\MA\MH+\EA)'(\MA\MH+\EA)  = \MH' \ft + \flN \MH'\MA'\et + \flN \EA' \MA \ft + \flN \EA' \et

We first establish a bound for $||\MH||$. 
Recall $\MH = (TN)^{-1}\MB'{\MXs}'\MXs\MTh\hA\TV^{-1}$, $||(TK)^{-1}\MBt\MB|| =  ||(TN)^{-1}\MA\MBt\MB\MA'|| = (TN)^{-1}||\MTh||^2 = o(1)$, and the bound  $||(TK)^{-1} \FX'\FX || = O_p(1)$ obtained in the proof of Lemma \ref{lemma: rate of V}. We have 
\begin{equation}
    \begin{split}
            ||\MH|| &= ||N^{-1}K\MA'\hA\TV^{-1}|| \times || (TK)^{-1} (\FX' \FX + \lN \MBt\MB )|| \\
    &= O_p((TK)^{-1} (\FX' \FX) + (TK)^{-1} \lN \MBt\MB) = O_p(1)
    \end{split}
\end{equation}\label{rate:H}

% EA below: but decomposed already
We now begin to bound each term in \eqref{theory:factor:decomposition} separately. First of all, by the proof of Theorem \ref{rate:loading}, we know $\EA =  \sum_{i=1}^{7}\D{i}$. 
% Under the simplifying conditions imposed in Theorem \ref{rate:factors}, we have 
% $||\MDD|| = O_p(T^{-1}), ||\MD|| = O_p(N / \sqrt{T})$, and $KN^{-1}||\EA||^2 = O_p(T^{-1})$. 
% and $KN^{-1}||\EA||_F^2 = O_p(T^{-1})$. 

\textbf{Part 1.}
For the first term of \eqref{theory:factor:decomposition}, notice that 
\begin{equation*}
\oTsumoT ||\flN \EA'\et||^2 = O_p(  \sum_{i=1}^{7} \oTsumoT ||  \flN \D{i}' \et ||^2)    
\end{equation*}

For $i \neq 2, 3, 6$, by choosing $\lN = O(1 / \sqrt{T+N})$, we can simply use the inequality $||\D{i}'\et /N||^2 \le  || \D{i}||^2 ||\et||^2 /N^2$ to get that 
% {lamN here?}
\begin{align}\label{the2_1_0}
    \oTsumoT || \D{i}' \et /N ||^2 = O_p(||\D{i}||^2/N) = 
    O_p(\rate{N}
    ), i \neq 2, 3, 6
\end{align}

For $\D{2}$, we have 
\begin{equation}
\begin{split}\label{the2_1_1}
    \fot \sumoT ||\fln \D{2}' \et||^2 
    &\stackrel{(\text{Lemma 4})}{=}
     O_p( ||\frac{1}{N^2T}\hAt(\Yres)'  \FX||^2 \fot\sumoT||\MA'\et||^2 )\\
    &\leftstackrel{(\ref{rate: Aet})}{=}
     O_p( \frac{1}{N^2K} ||\fot \FX'  (\Yres)||^2) \\
    &=O_p( \frac{1}{N^3} ||\fot \MTh'\MX'  (\Yres)||^2) \\
    &= O_p( \frac{1}{N^3} ||\MD||^2)\\
    &= O_p(\frac{1}{N})
    % &\leftstackrel{(\text{Lemma 3.ii})}{=}
    % O_p(\frac{1}{N^3} (\Ratedelta)^2 ) 
    % = o_p(\fln)
\end{split}
\end{equation}

For $\D{6}$, we have 
\begin{equation}
\begin{split}\label{the2_1_2}
    \fot \sumoT ||\fln \D{6}' \et||^2 
    &=\fot \sumoT ||\fln \TV^{-1}\hAt \fln \fot (\MX\MTh)' (\MU\MA'+\ME) \et||^2\\
    &\leftstackrel{(\text{Lemma 4})}{=}
    O_p(\fot \frac{1}{N^3K}\sumoT || \fot \MTh' \MX' (\MU\MA' + \ME) \et ||^2)\\
    &= O_p(\frac{1}{N^3K}||\fot \MTh' \MX' \MU||^2\fot\sumoT||\MA'\et||^2) % \frac{1}{NT^{\frac{1}{2}+\delta_1}}
    + O_p(\frac{1}{TN^3K}||\fot \MTh' \MX' \ME \ME'||_F^2) \\ 
    &\leftstackrel{(\ref{rate: Aet} \text{ and Lemma 3)}}{=}
    O_p(\rate{N^2K}[\rateThetaXU]^2 + \rate{TN^2K^2}[\RateFXEE]^2)\\
    &= O_p(\frac{1}{NT} + \frac{1}{NK} +\frac{1}{T^2K})
\end{split}
\end{equation}

% &= O_p(\frac{1}{TN^2}(p +N \sqrt{p} + Np\frac{u_2^2}{K} + \frac{u_3^2NT}{K})(\frac{p}{T}(\frac{N}{T} + 1 + u_2^2) + u_3^2)))\\

Similarly, for $\D{3}$, let $\MG = \MY - \MX\MTh = \ME + \MU\MA'$, we have
\begin{equation}\label{the2_1_3}
\begin{split}
    \fot \sumoT ||\fln \D{3}' \et||^2 
    &= \fot \sumoT  ||\fln \big[\XYcovDiff' \wl \XYcovDiff \MR\big]'\et||^2 \\
    &= O_p \Big(||\MDD||^2 \frac{1}{K} [T ||\fot ||\xt||_{\psit}^{-1}\MX' \ME\ME'||^2 + T^2 ||\fot ||\xt||_{\psit}^{-1}\MX'\MU||^2 \fot \sumoT||\MA'\et||^2] \Big)\\
    &= O_p(\frac{1}{T^3K} + \frac{1}{T^2NK})
\end{split}
\end{equation}

Combining the results from (\ref{the2_1_0}) to (\ref{the2_1_3}), we can obtain the bound for the first term in equation (\ref{theory:factor:decomposition}) as 
\begin{equation}\label{the2: first}
    \frac{K}{TN^2}\sumoT ||\EA'\et||^2 =  \frac{K}{T}\sumoT ||\frac{1}{N} \EA'\et||^2 =
    O_p(\frac{1}{T^2} + \frac{K}{N} )
%     O_p(
% \frac{K}{N^2} + \frac{p^2u_2^2K}{N^2T^2}  + \frac{1}{T^2}
% + \frac{p^{3/2}K}{T^3} + \frac{p^2u_2^2}{T^3} 
% + \frac{u_3^2p}{T^2}
%     )
\end{equation}

% {the following two terms can  never be deleted}
\textbf{Part 2.}
For the second term in equation (\ref{theory:factor:decomposition}), we have 
\begin{equation}
     \frac{K}{T}\sumoT ||\frac{1}{N}\MH'\MA'\et||^2 
    =
 \frac{K}{N^2} \oTsumoT ||\MA'\et||^2 
    \stackrel{(\ref{rate: Aet})}{=}
    O_p(\frac{K}{N})
\end{equation}

% \hl{% EA below - already used full below (simply later)}
\textbf{Part 3.}
For the third term in equation (\ref{theory:factor:decomposition}), we can directly bound it by 
\begin{equation}
\begin{split}\label{the2_3}
    \frac{K}{T}\sumoT ||\frac{1}{N} \EA'\EA\MH^{-1}\ft||^2 
    &\le K||\frac{1}{\sqrt{N}} \EA||^4   \fot \sumoT ||\MH^{-1}\ft||^2  \\
    &= O_p(||\sqrt{\frac{K}{N}} \EA||^4)\\
    &= O_p\Big(
    \big[\frac{1}{T} + \frac{u_2^2}{T} + \frac{u_3^2}{K} + \frac{p^2}{N^2T^2} + \frac{p^2u_2^4}{T^2K^2}
\big]^2
    % \frac{1}{T^2}
    \Big)
    % &\leftstackrel{(\text{Theorem 1 and Lemma 2.ii})}{=}  
    % O_p(
    % (\ratetheoremo)^2K^2
    % )\\
    % &= O_p(
    % \frac{K^2}{T^2}
    % + \frac{p^4K^2}{T^4N^4}
    % + \frac{u_2^4K^2}{T^2}
    % + \frac{p^4u_2^8}{K^2T^4}
    % )
\end{split}
\end{equation}

\textbf{Part 4.}
Finally, for the last term in equation (\ref{theory:factor:decomposition}), we also can  decompose it as
\begin{equation*}
\oTsumoT ||\flN  (\MA\MH)'\EA\MH^{-1}\ft||^2 = O_p(  \sum_{i=1}^{7} \oTsumoT ||  \flN \MA'\D{i} \ft ||^2)    
\end{equation*}

For $i \neq 2, 3, 6$, we can simply bound the component by 
% O_p(||\D{i}||^2/N) = 
\begin{align*}
\oTsumoT||(\MA\MH)'\D{i} \ft /N||^2 
&= ||\MA\MH||^2 ||\D{i}||^2 \oTsumoT  ||\ft / N||^2 \\
&\leftstackrel{(\text{Lemma 2.ii})}{=} 
O_p(\flN ||\D{i}||^2) = O_p(\frac{1}{N}) %\\
% &\leftstackrel{(\text{Theorem 1's proof})}{=} 
% O_p(\rate{N})
\end{align*}

% % EA below , but even with the full form, still in this form
For $\D{2}$, we have
\begin{align*}
    % ||\flN \MH' \MA' \D{2}||^2
    \oTsumoT ||\fracon (\MA\MH)'\D{2} \ft ||^2 
    &= ||\fracon (\MA\MH)'\fot (\MTh' \MX ' \MY - \MTh' \MX ' \MX \MTh) \MRN ||^2 \times \oTsumoT ||\ft ||^2 \\
    &\leftstackrel{(\text{Lemma 2.2})}{=} 
    O_p(\rate{N^3} ||\fot \MTh'\MX'(\MU\MA' + \ME)\EA||^2) + 
    O_p(\rate{N^3} ||\fot \MTh'\MX'(\MU\MA' + \ME)\MA||^2)\\
    &\leftstackrel{(\text{Lemma 3})}{=} 
    O_p(\rate{N^2}[(\RateThetaXE)^2][\frac{1}{K} \frac{K}{N} ||\EA||^2])\\
    &\quad + O_p(\frac{1}{NK^2}(\rateThetaXU)^2 + \rate{N^2K}(\rateofxeo)^2) \\
    &=
    O_p([\frac{1}{TK^2} + \frac{1}{NTK}][\frac{K}{N} ||\EA||^2])
    + O_p(\frac{u_2^2}{TK} + \frac{u_3^2}{K^2} + \frac{1}{T^2K} + \frac{1}{NT}) \\
    &= O_p(\frac{u_2^2}{TK} + \frac{u_3^2}{K^2} + \frac{1}{T^2K} + \frac{1}{NT} + \frac{p^2}{N^2T^2K} + \frac{p^2u_2^4}{T^2K^3})\\
\end{align*}

% no EA here; just \hat{A}'s property. Can be improved
By similar argument, we can obtain that 
\begin{align*}
    &\oTsumoT ||\fracon (\MA\MH)'\D{6} \ft ||^2 \\
    &= \oTsumoT ||\frac{1}{NT} (\MA\MH)'(\MTh' \MX ' \MY - \MTh' \MX ' \MX \MTh)’\MRN  \ft ||^2 \\
    % EA, etc
    % &= \frac{1}{N^4T^2} ||\MA||^4  ||\MTh' \MX ' \MY - \MTh' \MX ' \MX \MTh||^2 \oTsumoT ||\ft||^2 \\
    % &= \frac{1}{N^2T^2K} ||\MTh' \MX ' \MY - \MTh' \MX ' \MX \MTh||^2  \\
    &\leftstackrel{(\text{Lemma 1 and 4})}{=}  O_p(\rate{NK^2} ||\fot\MTh'\MX'\MU||^2 + \rate{N^3}||\fot\MA'\ME'\MX\MTh||^2)\\
    &\leftstackrel{(\text{Lemma 3})}{=} 
    O_p( \rate{NK^2} (\rateThetaXU)^2 + \rate{N^3}(\rateofxeo \sqrt{N/K})^2 )\\
    &= O_p(\frac{u_2^2}{KT} + \frac{u_3^2}{K^2} + \frac{1}{NT} )
    % &= O_p(\frac{1}{T^{\newrateofxu}} + \frac{1}{NT})\\=
\end{align*}

For the term related with $\D{3}$, we can first decompose it as
 \begin{equation}\label{equ:factor_last_3}
     \begin{split}
          &\oTsumoT ||\fracon (\MA\MH)'\D{3} \ft ||^2 \\
 &=\oTsumoT ||\fracon (\MA\MH)' 
 \XYcovDiff' \wl \XYcovDiff \MR \ft ||^2 \\
&\le ||\fracon (\MA\MH)' 
 \XYcovDiff' \wl \XYcovDiff \MR ||^2 \times \oTsumoT||\ft||^2 \\
&\leftstackrel{(\text{Lemma 1 and 4})}{=}  O_p(K||\fracon \MA' 
 \XYcovDiff' \wl \XYcovDiff  \frac{1}{NT} (\EA + \MA \MH) ||^2) \\
&= \frac{K}{N^4} O_p(||\MA'(\Yres)'\MX\wl\MX'(\Yres)\frac{1}{T}
\MA||^2\\
    &+
    ||\MA'(\Yres)'\MX\wl\MX'(\Yres)\fot\EA||^2)
\end{split}
\end{equation}

For the first term of (\ref{equ:factor_last_3}), we have 
\begin{align*}
    &\frac{K}{N^4}O_p(||\MA'(\Yres)'\MX\wl\MX'(\Yres)\frac{1}{T}\MA||^2)\\
    &= O_p(\frac{K}{N^4}(\frac{N}{K})^4 ||\frac{1}{T} \MU'\MX\wl \MX'\MU||^2 
    + \frac{K}{N^4}||\frac{1}{T} \MA'\ME'\MX\wl\MX'\ME\MA||^2) \\
    &=  O_p(\frac{1}{K^3}(\sqrt{\frac{p+K}{T}}u_2 + u_3)^4 
    + \frac{K}{N^4}\frac{N^2(K^2 + p^2)}{T^2K^2})\\
    &= O_p(
    % \frac{K}{T^2N^2} + \frac{p^2}{T^2N^2K} 
    \frac{u_3^4}{K^3} 
    + \frac{p^2u_2^4}{K^3T^2}
    + \frac{u_2^4}{KT^2}
    + \frac{1}{N^2K}). 
\end{align*}
% where the second equality follows from the same arguments with the proof of \text{Lemma 3.i} and the last equality utilizes the simplifying conditions. 
% EA below; I used that 1/T; otherwise, too complex
% even no explicitly EA under those conditions, those are used in some other ways. 

% A'EA  can be simplified?
For the second term of (\ref{equ:factor_last_3}), 
% recall $N^{-1}||\EA||_F^2 = O_p(T^{-1})$, 
we have 
\begin{align*}
    &\frac{K}{N^4}O_p(||\MA'(\Yres)'\MX\wl\MX'(\Yres)\fot\EA||^2)\\
    &=O_p((\frac{1}{N^3} ||\frac{1}{T}\MA'(\Yres)'\MX||\xt||_{\psit}^{-1}||^2) ||\fot||\xt||_{\psit}^{-1}\MX'(\Yres)||^2  \frac{K}{N}||\EA||^2)\\
    &\leftstackrel{(\text{Lemma 3})}{=} 
    O_p\Big(\frac{1}{N^3} [\frac{N(p+K)}{TK} + \frac{N^2}{K^2}(\frac{p+K}{T} u_2^2 + u_3^2)]
    \times [(\frac{p+K}{T} u_2^2 + u_3^2) \frac{N}{K} + \frac{N+p}{T} ]
    \times \frac{K}{N}||\EA||^2
    \Big)\\
    &= 
    O_p\Big(
    \frac{1}{N} \big[\frac{(p+K)}{TK} + \frac{N}{K^2}(\frac{p+K}{T} u_2^2 + u_3^2)\big]
    \times \big[(\frac{p+K}{T} u_2^2 + u_3^2) \frac{1}{K} + \frac{1}{T} + \frac{p}{TN} \big] \\
    &\quad\quad\quad \times \big[\frac{1}{T} + \frac{u_2^2}{T} + \frac{u_3^2}{K} + \frac{p^2}{N^2T^2} + \frac{p^2u_2^4}{T^2K^2}\big]
    \Big)\\
    &= 
    O_p\Big(
    \frac{1}{N} + \frac{1}{T^2K} + \frac{1}{K}\big[\frac{u_2^2}{T} + \frac{u_3^2}{K} + \frac{p^2}{N^2T^2} 
    + \frac{p^2u_2^4}{T^2K^2}\big]
    \Big) 
\end{align*}

Combining the above two terms, we have 
\begin{align*}
  &\oTsumoT ||\fracon (\MA\MH)'\D{3} \ft ||^2 \\
  &=O_p\Big(
    \frac{1}{N} + \frac{1}{T^2K} + \frac{1}{K}\big[\frac{u_2^2}{T} + \frac{u_3^2}{K} + \frac{p^2}{N^2T^2} 
    + \frac{p^2u_2^4}{T^2K^2}\big]
    \Big) 
% &= O_p( \frac{1}{N} + \frac{K}{T^2} + 
% \frac{p^2u_2^4}{K^3T^2} +\frac{p^3u_2^4}{T^3K^2}
% +  \frac{p^2u_2^2}{T^3K}(\sqrt{p} \wedge \frac{N}{T}) 
)
\end{align*}

For the last term in equation (\ref{theory:factor:decomposition}), combining results above, we have 
\begin{equation}\label{the2: last}
\begin{split}
&\frac{K}{T}\sumoT ||\flN (\MA\MH)'\EA\MH^{-1}\ft||^2 \\
&= K O_p(\frac{u_2^2}{TK} + \frac{u_3^2}{K^2} + \frac{1}{T^2K} + \frac{1}{NT} + \frac{u_2^2}{KT} + \frac{u_3^2}{K^2} + \frac{1}{NT}  + \frac{K}{T^2N^2} + \frac{p^2}{T^2N^2K} + \frac{1}{T^2K} 
+ \frac{p^2u_2^4}{K^3T^2} + \frac{1}{NT}) \\
&= O_p(
\frac{u_2^2}{T} + \frac{u_3^2}{K}+ \frac{1}{T^2}
+ \frac{K}{TN} + \frac{p^2}{T^2N^2}
+ 
\frac{p^2u_2^4}{K^2T^2}
% +  \frac{p^2u_2^2}{T^3}(\sqrt{p} \wedge \frac{N}{T}) 
)
\end{split}
\end{equation}

Combining results for the four parts from Equation (\ref{the2: first}) - (\ref{the2: last}), we conclude with
\begin{equation*}
\begin{split}
    &\frac{1}{KT}\sumoT ||\hft - \MH^{-1}\ft ||^2 \\
    &\le K
     \Big[\fot \sumoT||\frac{1}{N}\EA'\et||^2 + 
     \fot \sumoT|| \frac{1}{N}\MH'\MA'\et||^2  \\
     &\quad + 
     \fot \sumoT||\frac{1}{N}\EA'\EA\MH^{-1}\ft||^2 + 
     \fot \sumoT|| \frac{1}{N}(\MA\MH)'\EA\MH^{-1}\ft||^2 \Big]\\
&=  
O_p(
\frac{u_2^2}{T} + \frac{u_3^2}{K}+ \frac{1}{T^2}
+ \frac{K}{N} + \frac{p^2}{T^2N^2}
+ 
\frac{p^2u_2^4}{K^2T^2}. 
% +  \frac{p^2u_2^2}{T^3}(\sqrt{p} \wedge \frac{N}{T}) 
)\\
&=  
{O_p(
\frac{u_2^2}{T} + \frac{u_3^2}{K}+ \frac{1}{T^2}
+ \frac{K}{N} + \frac{1}{T^{2\eta}N^2}
+ 
\frac{u_2^4}{K^2T^{2\eta}}. 
% +  \frac{p^2u_2^2}{T^3}(\sqrt{p} \wedge \frac{N}{T}) 
)
}
% &= O_p(
% \frac{K}{N} + \frac{K^2}{T^2}  + \frac{Ku_2^2}{T} + u_3^2
% + \frac{p^{3/2}K}{T^3} 
%     + \frac{p^2u_2^4}{K^2T^2} +\frac{p^3u_2^4}{T^3K}
% +  \frac{p^2u_2^2}{T^3}(\sqrt{p} \wedge \frac{N}{T}) )
            % \ratetheoremt
\end{split}
\end{equation*}
\end{proof}

\section{Proof of Lemmas in \ref{proof of main results}}\label{appendix: lemmas}

\subsection{Proof of Lemma \ref{lemma: technical_lemmas}}
\begin{proof}
(i) 
For any matrix $\MB$ and any symmetric matrix $\MA$ of appropriate dimensions, by the definition of matrix operator norm, we have $||\MB'\MA\MB|| = sup_{||\vx_1|| = 1} \vx_1'\MB'\MA\MB\vx_1 \le ||\MB||^2 sup_{||\vx_2|| = 1} \vx_2'\MA\vx_2
= ||\MB||^2||\MA|| = ||\MB'\MB||||\MA||$. 

(ii)
Note the fact that, for a $d$-dimensional sub-Gaussian vector $\zt$ with $||\zt||_\psit \le M_z$, it always holds that $|| (||\zt||^2) ||_\psio \le 2dM_z^2$. Therefore, recall Assumption \ref{asmp:tail}, we have 
\begin{equation*}
    (NT)^{-1} \sumoT||\et||^2 = O_p(E( N^{-1} ||\et||^2)) = O_p(1)
\end{equation*}

By similar arguments, we can obtain 
\begin{equation*}
    T^{-1} \sumoT||\ft||^2 =  O_p(K)
\end{equation*}

Finally, let $\MA = (\va_1,...,\va_K)$, and note that $\MA'\MA = NK^{-1}\I$, we have for any $i = 1, \dots, K$, $var(\va_i' \et) = \va_i' (E\et\et') \va_i \le ||\va_i||^2 \times ||E\et\et'|| \le NK^{-1}M_2$. 
By similar arguments, this implies 
\begin{equation}\label{rate: Aet}
    T^{-1} \sumoT||\MA'\et||^2 =  O_p(E(||\MA'\et||^2))  = O_p(N)
\end{equation}
\end{proof}

\subsection{Proof of Lemma \ref{lemma: rate of wl and signal}}

\begin{proof}
(i)
% any citations?
By the property of the matrix operator norm, for any invertable matrix $\boldsymbol{S}$, it holds that $||\boldsymbol{S}^{-1}|| = (\sigma_{min}(\boldsymbol{S}))^{-1}$. 
Therefore, we have 
$||\wl|| = (\sigma_{min}(\XX + \lN \I))^{-1} = (\sigma_{min}(\XX) + \lN)^{-1} \le \lN^{-1}$. 

In addition, recall the assumption $p = o(T)$. By Corollary 3.1 in \cite{yaskov2014lower}, when $T$ is sufficiently large, we have that $\sigma_{min}(\XX) \ge c_0 T$ with probability approaching one, where $c_0 >0$ is some constant, and thus $||\wl|| = (\sigma_{min}(\XX) + \lN)^{-1} = O_p(T^{-1})$. 

Finally, let the singular value decomposition of $\MX$ be $\MX = \MU\Sigma\MV'$. Then we have 
\begin{equation*}
    \begin{split}
        ||\MX\wl\MX'|| &= ||\MU\Sigma\MV'(\MV\Sigma^2\MV' + \lambda_N \I)^{-1} \MV\Sigma\MU'|| \\
            &= ||\MU\Sigma(\Sigma^2 + \lambda_N \I)^{-1} \Sigma\MU'||\\
            &= ||\Sigma(\Sigma^2 + \lambda_N \I)^{-1} \Sigma|| = max_{1 \le i \le p} [\sigma_i(\XX) (\sigma_i(\XX) + \lN)^{-1}] \le 1
    \end{split}
\end{equation*}

(ii)
By Assumption \ref{asmp:signal}, we have that 
$||E\MA\fxt\fxt'\MA'||  = K^{-1}N ||E\fxt\fxt'|| 
= N ||E(\ft-\ut)(\ft-\ut)' / K|| \le 2N (||E\ft\ft'|| + ||E\ut\ut'||)/ K
\le c_1 N$, where $c_1 = (2M_3 + 4M_4)$. 
% CONDITION: X's low correlation has been used here
Therefore, for any $\vv \in \Sd{N}$, we have $N c_1 \ge \vv' E\MA\fxt\fxt'\MA' \vv = \vv' \MTh' E(\xt \xt') \MTh \vv \ge m_5 ||\MTh \vv||^2$, which implies $||\MTh|| \le \sqrt{Nc_1 /m_5}$ and thus $||\MTh|| = O(\sqrt{N})$.
% It follows that 
% $\sqrt{NK^{-1}}   ||\fxt||_{\psi_2} = ||\MA \fxt||_{\psi_2}
% = ||\MTh \xt||_{\psit} \le ||\xt||_{\psi_2} ||\MTh||  \le \sqrt{N}M_5$ for some $M_5 >0$, and thus $K^{-1/2} ||\fxt||_{\psi_2} \le M_5$. 
\end{proof}

\subsection{Proof of Lemma \ref{lemma: rate of CovMatrix}}
\begin{proof}
Let $\Nd{N}$ be a 1/4-net on $\Sd{N}$. We recap Result (A.12) in \cite{fan2016shrinkage} here: for a $N_1 \times N_2$ matrix $\MA_1$, define $\Phi(\MA_1) = sup_{\vv \in \Nd{N_1}, \vu \in \Nd{N_2}} \vv' \MA_1 \vu$, then we have 
\begin{align}\label{covering argument}
    ||\MA_1|| \le \frac{16}{7} \Phi(\MA_1)
\end{align}

We also need to use the following Bernstein-type inequality for sub-exponential random variables: 

\begin{lemma}\label{Bernstein}
(Proposition 5.16 in \cite{vershynin2010introduction})
Let $X_1, \dots, X_N$ be independent centered sub-exponential random variables, and $K = max_i ||X_i||_{\psio}$. Then for every $a = (a_1,\dots,a_N) \in \Nd{N}$ and every $t \ge 0$, we have an absolute constant $c > 0$, such that
\begin{equation*}
    P(|\sum_{i=1}^N a_i X_i| \ge t) \le 2exp[-c min(\frac{t^2}{K^2||a||^2},\frac{t}{K||a||_{\infty}} )]. 
\end{equation*}
\end{lemma}

(i)
First of all, note that for any $\vu \in \Nd{K}$ and $\vv \in \Nd{N}$, 
we have $||\vu'||\xt||_{\psit}^{-1}\xt\ut'\vv||_\psio \le ||\vu'||\xt||_{\psit}^{-1}\xt||_\psit ||\ut'\vv||_\psit \le 2 ||\ut||_\psit \le 2 u_2$. By Lemma \ref{Bernstein},  for every $t \ge 0$, we have a constant $c_5 >0$ such that 
\begin{equation*} 
% P(|\fot \vu'\sumoT \xt\ut'\vv | > t) 
% = 
P(|\fot \vu'\sumoT ||\xt||_{\psit}^{-1}\xt\ut'\vv - E(\vu ||\xt||_{\psit}^{-1}\xt\ut'\vv')| > t u_2) 
\le 2exp(-c_5 T min(t^2,t)). 
\end{equation*}
Note that $E(\vu ||\xt||_{\psit}^{-1}\xt\ut'\vv') \le u_3$. 
% where we use the fact that $E(\vu \xt\ut'\vv') = \vu E(\xt\ut')\vv' = 0$. 
Therefore, by Equation (\ref{covering argument}) and the union bound over $\Nd{K}$, we have 
\begin{equation}\label{formula: XU}
||\fot ||\xt||_{\psit}^{-1} \MX'\MU|| = O_p(\sqrt{\frac{p + K}{T}}u_2 + u_3). 
\end{equation}

Next, by Lemma 3 in \citet{negahban2011estimation}, using similar covering arguments, we have 
\begin{equation}\label{formula: XE}
    || ||\xt||_{\psit}^{-1} \MX'\ME / T|| = O_p(\sqrt{(N+p)/T}). 
\end{equation}

Combining \eqref{formula: XU} and \eqref{formula: XE}, we get 
\begin{align*}
    &||\flnt (\ME + \MU\MA')'(\XWX) (\ME + \MU\MA')||\\
    &= O_p(||\flnt \ME'\XWX\ME|| + ||\flnt (\MU\MA')'(\XWX) (\MU\MA')||)\\
    &= O_p(||\flnt (||\xt||_{\psit}^{-1}\ME'\MX) (||\xt||_{\psit}^2 \wl) (||\xt||_{\psit}^{-1}\MX'\ME||) + O_p(\frac{p + K}{TK}u_2^2 + \frac{u_3^2}{K})\\
    &= O_p(\frac{1}{T} + \frac{p}{NT} + \frac{p + K}{TK}u_2^2 + \frac{u_3^2}{K})
\end{align*}

% $P(||\sumoT \xt\ut' /T|| > t) \le 2exp((N + p)log8-c_5Tmin(t,t^2))$, 
% which implies 
% $||\MX'\ME / T|| = O_p((N+p)/T) $ when  $T = o(N)$ and
% $O_p(\sqrt{(N+p)/T})$ when $N = O(T)$, and thus
% \begin{equation}\label{formula: XE}
% ||\fot \MX'\ME|| = O_p((N+p)/T + \sqrt{(N+p)/T}).     
% \end{equation}
% We will improve this bound in the proof of Lemma \ref{lemma: rate of CovMatrix}.(iii). 

% Next, 
% note that $||\vu'\xt\ut'\vv||_\psio \le ||\vu'\xt||_\psit ||\ut'\vv||_\psit \le  2M_0 u_2$ holds for any $\vu \in \Nd{p}$ and $\vv \in \Nd{K}$,  by similar arguments, we have 
% \begin{equation*}
% ||\fot \MX'\MU|| = O_p(\sqrt{\frac{p + K}{T}}u_2 + u_3) 
% \end{equation*}

(ii) Let $\mathcal{M}_B$ be the row space of $\MB$. Recall in the proof of Lemma \ref{lemma: rate of wl and signal} we have $||\MTh|| \le \sqrt{Nc_1 /m_5}$,  which implies that, 
for any $\vv \in \Nd{N}$, it holds that $||\sqrt{m_5 / Nc_1}\vv'\MTh' \MX' \ME|| = ||\vv_1 \MX' \ME||$, where $\vv_1 = \sqrt{m_5 / Nc_1}\vv_2 \in \mathcal{M}_B$ with some $\vv_2$ satisfying $||\vv_2|| \le 1$. Therefore, by similar arguments with proof of  Lemma 3.i
% those for the rate of $||\MX'\MU/T||$
, we can obtain that 
\begin{equation}\label{temp2}
||\MTh'\MX'\ME/T|| = O_p(\sqrt{\frac{N}{K}}(\sqrt{\frac{N+K}{T}}))
= O_p(\frac{N}{\sqrt{TK}} + \sqrt{\frac{N}{T}})
= O_p(\frac{N}{\sqrt{TK}})
% \frac{N+K}{T} + 
% \frac{N}{T} + 
% \sqrt{N}(\sqrt{\frac{N}{T}})
\end{equation}

Likewise, we can derive the following results:
\begin{gather*}
||\MTh' \MX'\MU/T|| = O_p(\sqrt{\frac{N}{T}}u_2 + \sqrt{\frac{N}{K}}u_3), \\
||\XEo/T|| = O_p(\Rateofxeo),  \\
|| ||\xt||_{\psit}^{-1} \XEth/T|| = O_p(\Rateofxeth). 
\end{gather*}
 
% CONDITION: where do we use these results? correct?
\newcommand{\tildeepsto}{\tilde{\boldsymbol{\epsilon}}_{t_1}}

Combining the above results, we can get 
\begin{align*}
||\MD|| 
= ||\frac{1}{T}(\MTh' \MX ' \MY - \MTh' \MX ' \MX \MTh)||
&\le ||\fot \MTh' \MX'\ME|| + 
|| \fot \MTh' \MX'\MU\MA'|| \\
% & = O_p(\frac{N}{\sqrt{T}}(u_2 + 1) + \frac{Nu_3}{\sqrt{K}} )\\
& = O_p(\frac{N}{\sqrt{TK}}(u_2 + 1) + \sqrt{\frac{N}{T}} + \frac{N}{K}u_3)\\
& = O_p(\frac{N}{\sqrt{TK}}(u_2 + 1) + \frac{N}{K}u_3)
\end{align*}

(iii)
For $||\FX'\ME\ME' / T||_F^2$, we first  decompose it into two parts using Cauchy-Schwarz inequality: 
\begin{align}\label{bound: FXEE_decom}
    ||\fot \FX'\ME\ME' ||_F^2 
    \le T^{-2}\sumoT ||\fxt\et'\et||^2 + \sumto||\fot \sumnto \fxt\et' \eto||^2. 
\end{align}
First, note that by Cauchy-Schwarz inequality, we have
\begin{align*}
    E(||\fxt\et'\et||^2) &= E(||\et||^4 ||\fxt||^2) 
    \le \sqrt{E(||\et ||^2)^4} \sqrt{E(||\fxt||^2)^2}
    = O(KN^2),
\end{align*}
where we use the fact that for a $d$-dimensional vector $\zt$ satisfying $||\zt||_\psit \le M_z$, it holds that $|| (||\zt||^2) ||_\psio \le 2dM_z^2$. By the law of large numbers, It follows that
\begin{align*}
    T^{-2}\sumoT ||\fxt\et'\et||^2 = O_p(\frac{KN^2}{T}).
\end{align*}
Next, by Proposition 5.16 in \cite{vershynin2010introduction}, for any $T$, $t_1 \in \{1,\dots,T\}$, $\vv \in \Sd{k}$ and $\tildeepsto \in \mathcal{R}^N$, we have
\begin{align*}
    P(|\frac{1}{T-1} \sumnto(\vv'\fxt\et')\frac{\tildeepsto}{||\tildeepsto||} - E(\vv'\fxt\et'\frac{\tildeepsto}{||\tildeepsto||})| > K^{1/2}t |\eto=\tildeepsto) \\
    \le 2 exp[-c ((T-1)min(t^2,t)],
\end{align*}
where $c$ is some universal constant because $||K^{-1/2}\fxt||_\psit$ and $||\et||_\psit$ are bounded. 
Because $E(\vv'\fxt\et'\tildeepsto/||\tildeepsto||)=0$ and $\eto$ is independent with $(\fxt, \et)$ since  $t \neq t_1$,  we have 
$P(||(T-1)^{-1} \sumnto(\fxt\et')(\eto/||\eto||) || 
\ge \sqrt{Klog(\epsilon/2)/(-c(T-1))}) \le \epsilon$, for any small enough $\epsilon$. 
Note that $||\eto|| = O_p(\sqrt{N})$, it follows that % from these two results
\begin{align*}
    \sumto||\fot \sumnto \fxt\et' \eto||^2 = O_p(NK). 
\end{align*}
Therefore, combining results for the two parts of (\ref{bound: FXEE_decom}), we obtain that 
\begin{equation}\label{bound: FXEE}
    ||\fot \FX'\ME\ME'||_F = O_p( N \sqrt{\frac{K}{T}} + \sqrt{KN}
    )    
\end{equation}

By similar arguments, we can derive 
\begin{align*}
    ||\fot ||\xt||_{\psit}^{-1} \MX'\ME\ME'||_F = O_p(
    N\sqrt{\frac{p}{T}} + \sqrt{Np}
    )
\end{align*}

\end{proof}

% Lemma \ref{lemma: rate of wl and signal} and 

\subsection{Proof of Lemma \ref{lemma: rate of V}}
\begin{proof}
Denote $\bsigma_0 = \MA \MA'$. 
$\MA'\MA = K^{-1}N\boldsymbol{I}$ implies that the eigenvalues of $KN^{-1}\bsigma_0$ are bounded away from zero and infinity. 
It then follows from Assumption \ref{asmp:signal} that the eigenvalues of $\bsigma_1 =  \MTh'E(\xt\xt')\MTh/N = \MA E(\fxt\fxt')\MA' /N$ are bounded away from zero and infinity.
Since $K = o(T)$, we know the sample covariance matrix $T^{-1}\sum_{t=1}^{T}(\fxt\fxt'/K)$ is a consistent estimator of $E(\fxt\fxt'/K)$ under the operator norm \citep{vershynin2010introduction}. 
Therefore
\begin{align*}
    ||\bsigma_1 - \flnt \MTh'\MX'\MX\MTh|| 
    &= ||\frac{1}{N} \MA(E(\fxt\fxt') - \fot \sum_{t=1}^{T}\fxt\fxt')\MA'||\\
    &= ||\frac{K}{N}\MA\MA'|| ||E(\frac{1}{K}\fxt\fxt') - \frac{1}{KT} \sum_{t=1}^{T}\fxt\fxt'||
    = o_p(1)
\end{align*}

In addition, notice that 
\begin{align*}
||\flnt [(\MX\MTh)'(\XWX)(\MX\MTh) - (\MX\MTh)'(\MX\MTh)]||
&= ||\flnt (\MX\MTh)' (\XWX - \Vec{I}) (\MX\MTh)|| \\
&\le ||\frac{1}{T} \MX' (\XWX - \Vec{I}) \MX|| \\
&= \frac{1}{T} max_{1\le i \le p} \frac{\lN \sigma_i(\XX)}{\lN + \sigma_i(\XX)}\\
&\le \frac{1}{T} \lN = o_p(1)
\end{align*}

Moreover, write $\MG = \MY - \MX\MTh = \ME + \MU\MA'$, we have
\begin{align*}
&\flnt \MY'(\XWX)\MY - \flnt (\MX\MTh)'(\XWX)(\MX\MTh)\\
& =  \flnt \MG'(\XWX)\MX\MTh + 
\flnt (\MG'(\XWX)\MX\MTh)' + 
\flnt \MG'(\XWX)\MG
\end{align*}

According to Lemma \ref{lemma: rate of CovMatrix}, we have 
\begin{align*}
||\flnt \MG'(\XWX)\MG||
= 
||\MDD||
% \frac{1}{N} ||\fot\MX' \MG||^2 
= o_p(1). 
\end{align*}
Besides, by using similar arguments with the proof of Lemma \ref{lemma: rate of CovMatrix}, we have  $||\flnt \MG'(\XWX)\MX\MTh|| = ||\flnt (\MG'\MX) \wl (\MX'\FX) \MA'|| = o_p(1)$. 
% Besides, note $||\wl\XX|| = max_i [\sigma_i(\XX)  (\sigma_i(\XX) + \lN)^{-1}] \le 1$, and 
% the fact that $||\MG' \MX / T|| = O_p(\rateXG)
% = o_p(\sqrt{N})$, we have 
% \begin{align*}
% ||\flnt (\MG'(\XWX)\MX\MTh)'|| = ||\flnt \MG'(\XWX)\MX\MTh||
% &\le \frac{1}{\sqrt{N}} ||\fot \MG' \MX|| ||\wl\XX||\\
% &= o_p(1)
% \end{align*}

% It follows from $||\wl|| = O_p(T^{-1})$ that 

% I try to avoid that? not necessary. we can do a good job here. But very weird??? I am not clear... not good results. But should easily be strict.  [??? good or not?]

Therefore, putting the above results together, we have that $||(NT)^{-1} \MY'(\XWX)\MY  - (NT)^{-1}(\MX\MTh)'(\MX\MTh)|| = o_p(1)$, which implies
$(NT)^{-1}\MY'(\MX\wl\MX')\MY $ is also a consistent estimator of $\bsigma_1$ under the operator norm. 
Finally, the well-known Weyl's theorem completes our proof, which we cite here: 
\begin{lemma}(Weyl's theorem)
For two $N\times N$ symmetric matrices $\boldsymbol{\Sigma}$ and $\hCOV$, let $\{\sigma_i\}_{i=1}^N$ and $\{\hat{\sigma}_i\}_{i=1}^N$ be their corresponding eigenvalues in descending order. Then we have for i = 1, \dots, N,
\begin{equation*}
    |\hat{\sigma}_i - \sigma_i| \le ||\bsigma  - \hCOV||.
\end{equation*}
\end{lemma}
With $\sigma_i$ as the $i$-th largest eigenvalue of $\bsigma_1$ and $\hat{\sigma}_i$ as the $i$-th largest eigenvalue of $(NT)^{-1}\hCOV / N = \MY'(\MX\wl\MX')\MY$, we have $|\hat{\sigma}_i - \sigma_i| = o_p(1)$, and thus all $\hat{\sigma}_i$ are bounded away from zero and infinity with probability approaching one, which completes the proof.
\end{proof}

\bibliography{0_CAFE}